\documentclass[extra,mreferee]{gji}
\usepackage{timet}
\usepackage{amsmath,amssymb,amsfonts,latexsym}[mathlines] 
\usepackage{mathrsfs}
\usepackage{graphicx}
\usepackage{color}
\usepackage[normalem]{ulem}

\usepackage{lineno}
\let\oldequation\equation
\let\oldendequation\endequation
\renewenvironment{equation}
  {\linenomathNonumbers\oldequation}
  {\oldendequation\endlinenomath}
\let\oldflalign\flalign
\let\oldendflalign\endflalign
\renewenvironment{flalign}
  {\linenomathNonumbers\oldflalign}
  {\oldendflalign\endlinenomath}

\title[Reduction in fully Bayesian inversions]
  {Appropriate reduction of the posterior distribution in fully Bayesian inversions}
\author[D. Sato et al.
]
  {Dye SK Sato$^{1*}$, 
  Yukitoshi Fukahata$^{1}$
  \& Yohei Nozue$^2$
\\
  $^1$ 
  Disaster Prevention Research Institute, Kyoto University, Gokasho, Uji, Kyoto 611-0011, Japan. 
  \\
  $^2$ 
  Division of Earth and Planetary Sciences, Graduate School of Science, Kyoto University, 
  \\\,\, Gokasho, Uji, Kyoto 611-0011, Japan. 
  \\$^*$ Correspondence: sato.daisuke.44p@st.kyoto-u.ac.jp
  }
\date{Received 20xx Xxxx xx; in original form 20xx Xxxx xx}
\pagerange{\pageref{firstpage}--\pageref{lastpage}}
\volume{xxx}
\pubyear{xxxx}

\begin{document}

\label{firstpage}

\maketitle


\begin{summary}
Bayesian inversion generates a posterior distribution of model parameters from an observation equation and prior information both weighted by hyperparameters. 
The prior is also introduced for the hyperparameters in fully Bayesian inversions and enables us to evaluate both the model parameters and hyperparameters probabilistically by the joint posterior. 
However, even in a linear inverse problem, it is unsolved how we should extract useful information on the model parameters from the joint posterior. 
This study presents a theoretical exploration into the appropriate dimensionality reduction of the joint posterior in the fully Bayesian inversion. 
We classify the ways of probability reduction into the following three categories focused on the marginalisation of the joint posterior: (1) using the joint posterior without marginalisation, (2) using the marginal posterior of the model parameters and (3) using the marginal posterior of the hyperparameters. 
First, we derive several analytical results that characterise these categories. 
One is a suite of semianalytic representations of the probability maximisation estimators for respective categories in the linear inverse problem. 
The mode estimators of categories (1) and (2) are found asymptotically identical for a large number of data and model parameters. 
We also prove the asymptotic distributions of categories (2) and (3) delta-functionally concentrate on their probability peaks, which predicts two distinct optimal estimates of the model parameters. 
Second, we conduct a synthetic test and find an appropriate reduction is realised by category (3), typified by Akaike's Bayesian information criterion (ABIC). The other reduction categories are shown inappropriate for the case of many model parameters, 
where the probability concentration of the marginal posterior of the model parameters is found no longer to mean the central limit theorem. 
The main cause of these results is that the joint posterior peaks sharply at an underfitted or overfitted solution as the number of model parameters increases. 
The exponential growth of the probability space in the model-parameter dimension makes almost-zero-probability events finitely contribute to the posterior mean and distributions of categories (1) and (2) be pathological. 
One remedy for this pathology is counting all model-parameter realisations by integrating the joint posterior over the model-parameter space of exponential multiplicity. 
Hence, the marginal posterior of the hyperparameters for categories (3) becomes appropriate and can conform to the law of large numbers even with numerous model parameters. 
The exponential rarity of the posterior mean and ABIC estimates implies the exponential time complexity of ordinary Monte Carlo methods in population mean and ABIC computations. 
We also present a geophysical application to estimate a continuous strain-rate field from spatially discrete Global Navigation Satellite System (GNSS) data, demonstrating denser basis function expansions of the model-parameter field lead to oversmoothed estimates in naive fully Bayesian approaches, while detailed fields are resolved with convergence by the reduction of category (3). 
We usually naively believe a good solution can be constructed from a finite number of samples with high probabilities, but the high-probability domain could be inappropriate, and exponentially many samples become necessary for generating appropriate estimates in the high-dimensional fully Bayesian posterior probability space. 
\end{summary}



\section{Introduction}

Typical geophysical inverse problems first set a model, termed an observation equation, which describes a theoretical relationship between observed data and model parameters to be estimated~\citep{jackson1972interpretation}. 
From the observation equation, we obtain the likelihood of the model parameters for an observed data set. 
The maximum likelihood principle provides the model parameters that best fit the data. 
However, the maximum likelihood estimate is often 
not unique~\citep[giving an ill-posed problem;][]{backus1967numerical} and unstable~\citep[overfitting observation noises;][]{tarantola1982generalized} for numerous model parameters. 
Hence it is hard to discuss the details of finely-resolved inverted results when using only the likelihood, even with densely distributed model parameters.

Bayesian inversion combines the observation equation with a prior~\citep{jackson1985bayesian,yabuki1992geodetic,matsu2007geodetic}, 
which represents a priori information on the model parameters~\citep{jackson1979use}, typically taking the form of damping, smoothing and sparsity constraints.  
The solution of the Bayesian inversion is associated with regularisation techniques such as the regularised least-square method~\citep{jackson1985bayesian} that can stably solve the problem even considering a number of model parameters. 
In return, Bayesian approaches and regularisation techniques require tuning hyperparameters that weight the prior against the observed data. 
The determination of the hyperparameters in the regularised least-square estimation is often subjective or based on the optimisation functions that lack a firm theoretical basis, complicating interpretation~\citep{minson2013bayesian}. 
For example, the use of the trade-off curve between the regularisation and data misfit is criticised by~\citet{fukuda2008fully}, as it can define the optimal value arbitrarily by changing the coordinate axes of the trade-off curve plot, thus ill-defined. 
This major drawback is removed in the Bayesian inversion using Akaike's Bayesian information criterion \citep[ABIC;][]{akaike1980likelihood,yabuki1992geodetic}. ABIC determines the optimal values of the hyperparameters by maximising the marginal likelihood of the hyperparameters with integrating out the model parameters. 
Meanwhile, the use of ABIC is referred to as empirical Bayes in statistical literature and is criticised for the point estimation of the hyperparameters~\citep[p.104]{gelman2013bayesian}. 

In the fully Bayesian inversion, the hyperparameters are also random variables accompanied by their own priors~\citep[hyperpriors;][]{fukuda2008fully}. 
We then have an observation equation, prior of the model parameters and hyperprior, which generate the joint posterior of the model parameters and hyperparameters. The joint posterior allows us to evaluate the optimal set of the model parameters and hyperparameters with uncertainties~\citep{minson2013bayesian,kubo2016development,amey2018bayesian}. 
The fully Bayesian inversion is usually regarded as a non-approximated version of ABIC~\citep{malinverno2004expanded,gelman2013bayesian}. 
We encounter, however, many difficulties in evaluating the joint posterior, such as non-Gaussianity even with a linear observation equation, and as discussed later, inherent incapability of projecting a unique model-parameter distribution. 
Coping with these problems has refined the numerical methods of sampling the joint posterior, typified by Markov-chain Monte Carlo (MCMC) methods~\citep{hastings1970monte,gamerman2006markov}.

The fully Bayesian inversion commonly involves routines to extract useful information from the joint posterior~\citep{sen2013global}. 
The operations include determining the optimal values of the model parameters and hyperparameters. 
We may eliminate the hyperparameters by integration (marginalisation) from the joint posterior for evaluating the model-parameter distribution~\citep{fukuda2008fully}. 
Projecting the joint posterior onto low-dimensional profiles is usual in MCMC implementations~\citep{duputel2014accounting,amey2018bayesian,bagnardi2018inversion}. 
Operations of reshaping the joint posterior into these tractable forms can all be regarded as instances of the dimensionality reduction of the joint posterior. 

We can classify these reduction methods in the fully Bayesian inversion broadly into the following three categories in view of the marginalisation entailed in the reduction~\citep{malinverno2004expanded}. 
One is the ABIC type~\citep{good1965estimation,akaike1980likelihood}, 
dividing the inversion of the model parameters and hyperparameters into two inference stages (detailed later in the next section). 
By contrast, most of the fully Bayesian approaches adopt direct sampling of the joint posterior by the MCMCs~\citep[e.g.][]{minson2013bayesian,livermore2014core,kubo2016development}, which offers two different categories of reduction. 
One of them is a straightforward use of the joint posterior, where the probability value is evaluated for a set of the model parameters and hyperparameters, not for the model parameters alone. 
A representative (point) estimation of this reduction is the maximisation of the joint posterior, called maximum a posteriori (MAP). 
The other way of reduction is to use the marginal posterior of the model parameters by integrating out the hyperparameters from the joint posterior~\citep{fukuda2008fully}; as we see later, this reduction is performed implicitly whenever only the model parameters are output from the joint posterior. 
A synthetic test of \citet{fukuda2008fully} suggests that the marginal posterior of the model parameters gives a result close to the ABIC estimate. However, it is unclear whether the closeness of these solutions holds in general, and as we will see later, it is not true.

This study treats the above issue: how we should perform the reduction of the joint posterior in the fully Bayesian inversions to obtain an appropriate distribution of the model parameters. 
First, we derive a series of semianalytic solutions of the reductions in a linear inverse problem with relatively generic hyperpriors, which shows that any measurable estimate is asymptotically equivalent to the MAP or ABIC estimate. There are, so to speak, intrinsically only two choices: MAP and ABIC. 
The analysis also illustrates a distinctive difference between these estimates for a large number of model parameters, the vast probability space of which was beyond the reach of numerical techniques. 
Second, we investigate the asymptotic property of the joint posterior for a high-dimensional model-parameter space with the aid of synthetic tests, which elucidates that and why the two-stage inference of ABIC is a rather appropriate reduction. 
This conclusion is supported by a geophysical application to estimate a strain-rate field from GNSS data. 
We will also identify the statistical character of the high-dimensional probability space and pose a sampling problem hidden in the fully Bayesian approaches.


\section{Framework}
We first set the fully Bayesian inverse problem analysed in this study. We next categorise the methods for the reduction of the joint posterior and introduce the reduction problem. 

\subsection{Fully Bayesian inversion}
\label{sec:FBIframework}
We consider inference of a model parameter field $a(\xi)$ over a coordinate space of $\xi$ from observation data $d(x_n)$, where $x_n$ represents the location of observation point $n=1,2,...,N$. 
Here $x_n$ and $\xi$ may belong to different coordinate systems, as in seismic tomography that inverts data recorded on the Earth's surface ($x_n$) to a slowness field across space ($\xi$). 

Suppose the data $d(x)$ at $x$ is described by an integral equation that convolves the model parameter field $a(\xi)$ and integral kernel (Green's function) $H(x,\xi)$ over $\xi$, plus an observation error $e(x)$: 
\begin{equation}
d(x)=\int d\xi H(x;\xi)a(\xi) +e(x).
\label{eq:obseq}
\end{equation}
In this study, $H$ is assumed to be error-free. 

We discretise the model parameter field $a(\xi)$ by superposing a finite number of basis functions $X_m(\xi)$ $(m=1,2,...,M)$:
\begin{equation}
a(\xi)\approx\sum_{m=1}^{M}a_mX_m(\xi).
\label{eq:expansionruleofa}
\end{equation}
Equation~(\ref{eq:expansionruleofa}) rewrites eq.~(\ref{eq:obseq}) into the following matrix-vector equation:
\begin{equation}
{\bf d}={\bf Ha}+{\bf e},
\label{eq:discreteobseq}
\end{equation}
with 
\begin{equation}
H_{nm}=\int d\xi H(x_n,\xi)X_m(\xi).
\label{eq:defofdiscretekernel}
\end{equation} 
Here the ($n$, $m$)-entry of ${\bf H}$ is $H_{nm}$, 
the $m$-component of ${\bf a}$ is $a_m$, 
and the $n$-components of ${\bf d}$ and ${\bf e}$ are $d(x_n)$ and $e(x_n)$, respectively ($n=1,...,N$; $m=1,...,M$).
Our inverse problem is to estimate the model parameter vector ${\bf a}$ from observed data ${\bf d}$. 

We assume the error ${\bf e}$ in eq.~(\ref{eq:discreteobseq}) follows a Gaussian distribution of zero mean and covariance $\sigma^2 {\bf E}$: 
\begin{equation}
{\bf e}\sim \mathcal N ({\bf 0},\sigma^2 {\bf E}),
\label{eq:defofobserror}
\end{equation}
where $\sigma^2(>0)$ is a hyperparameter 
that scales the variance of the error ${\bf e}$,
and ${\bf E}$ is the normalised covariance of ${\bf e}$. 
We presume the positive-definiteness of symmetric matrix ${\bf E}$ (and ${\bf E}^{-1}$). 
Equations~(\ref{eq:discreteobseq}) and (\ref{eq:defofobserror}) 
assign a probability density function (pdf) of the 
data ${\bf d}$ given the model parameters ${\bf a}$ and the hyperparameter $\sigma^2$: 
\begin{equation}
P({\bf d}|{\bf a},\sigma^2)=e^{-U({\bf d},{\bf a})/\sigma^2 +F_{\rm obs}(\sigma^2)}
\label{eq:obseqprob}
\end{equation}
with
\begin{flalign}
U({\bf d},{\bf a})&=\frac 1 2 ({\bf H}{\bf a}-{\bf d})^{\rm T}{\bf E}^{-1}({\bf H}{\bf a}-{\bf d})
\label{eq:defofU}
\\
F_{\rm obs}(\sigma^2)&=-\frac{N}{2} \ln (2\pi\sigma^2)-\frac 1 2 \ln |{\bf E}|
\end{flalign}
where $|\cdot|$ denotes the determinant for a matrix, and the superscript ${}^{\rm T}$ denotes the transpose. 
Equation~(\ref{eq:obseqprob}) is a likelihood function of ${\bf a}$ given ${\bf d}$ and $\sigma^2$. 
Hereafter, we omit the ${\bf d}$-dependence of $U({\bf d},{\bf a})$, considering ${\bf d}$ constant as in usual inversion analyses. 

Besides the observation equation, 
Bayesian formulation introduces 
a priori information on the model parameters in a probabilistic form. 
We assume the following prior: 
\begin{equation}
P({\bf a}|\rho^2)= e^{-V({\bf a})/\rho^2+F_{\rm pri}(\rho^2)}
\label{eq:conjpriorobs}
\end{equation}
with 
\begin{flalign}
V({\bf a})&=\frac 1 2 {\bf a}^{\rm T}{\bf G}{\bf a}.
\label{eq:defofV}
\\
F_{\rm pri}(\rho^2)
&=c-\frac{P}{2} \ln (2\pi\rho^2)+\frac 1 2 \ln |\boldsymbol\Lambda_G|,
\label{eq:FpriforrankdeficientG}
\end{flalign}
where $\rho^2>0$ is a hyperparameter that represents the rigour of the prior constraint, 
${\bf G}$ denotes a positive-semidefinite symmetric matrix of rank $P$, 
and 
$|\boldsymbol\Lambda_G|$ is the product of the nonzero eigenvalues of ${\bf G}$.
When ${\bf G}$ is rank deficient, a normalisation factor $c$ is required in practice~\citep{fukahata2012inversion}.

Equations~(\ref{eq:obseqprob}) and (\ref{eq:conjpriorobs}) represent a standard linear inversion that encompasses the regularised least-square method. 
We mainly treat this problem setting throughout the paper. 
$P({\bf d}|{\bf a},\sigma^2)$ and $P({\bf a}|\rho^2)$ in this problem belong to a special class of pdfs called the exponential family~\citep{gelman2013bayesian}. 
$F_{\rm obs}$ and $F_{\rm pri}$ are the normalisation factors that depend on $\sigma^2$ and $\rho^2$, respectively, but now independent of ${\bf a}$; 
functions $U$ and $V$ are the cost functions of the above distributions~\citep[Gibbs distributions;][]{landau1994statistical}. 
The exponential family [taking the same form as eqs.~(\ref{eq:obseqprob}) and (\ref{eq:conjpriorobs})] is comprehensive and includes distributions of nonlinear inversions [s.t. ${\bf d}={\bf F}({\bf a})+{\bf e}$ assuming a nonlinear function ${\bf F}$ of ${\bf a}$] and non-Gaussian errors. 
The linear inverse problem constitutes a simplest class contained in it. 

The fully Bayesian inference, where hyperparameters are also random variables, further introduces a hyperprior $P(\sigma^2,\rho^2)$ for $\sigma^2$ and $\rho^2$. 
We primarily consider the following uniform hyperprior over $0<\sigma^2<\infty,0<\rho^2<\infty$, which represents we know nothing about the hyperparameters a priori:
\begin{equation}
P(\sigma^2,\rho^2)\propto const.
\label{eq:noninfohyperprior}
\end{equation}
Lack of knowledge is expressed by several noninformative priors, such as Jefferey's noninformative prior~\citep{jeffreys1998theory}, 
which is a logarithmically uniform prior $P(\sigma^2)\propto1/\sigma^2$ and $P(\rho^2)\propto1/\rho^2$ for variances of Gaussian distributions $\sigma^2$ and $\rho^2$~\citep[the logarithmic prior;][]{carlin2008bayesian}. 
Including both the uniform $(n_{\sigma^2}=n_{\rho^2}=0)$ and logarithmic $(n_{\sigma^2}=n_{\rho^2}=-1)$ hyperpriors, 
we also treat $n_{\sigma^2}$- and $n_{\rho^2}$-th power functions of $\sigma^2$ and $\rho^2$: 
\begin{equation}
\begin{aligned}
P(\sigma^2)&=c(\sigma^2)^{n_{\sigma^2}}
\\
P(\rho^2)&=c(\rho^2)^{n_{\rho^2}},
\end{aligned}
\label{eq:slightgenhyperpriors}
\end{equation}
where $c$ denotes normalisation constants. 
In this paper, we generally use ``$c$'' as normalisation constants of probabilities, and each ``$c$'' may have different values as in the top and bottom parts of eqs.~(\ref{eq:slightgenhyperpriors}). 
We note the uniform prior over an infinite (or a semi-infinite) range is the improper prior that is not normalisable~\citep{gelman2013bayesian}, while it can be regarded as a limit of a (normalised) uniform prior over a sufficiently wide range~\citep{ulrych2001bayes,fukahata2012inversion}. 


Using Bayes' theorem, we incorporate the data distribution (eq.~\ref{eq:obseqprob}) 
with the priors of the model parameters (eq.~\ref{eq:conjpriorobs}) and hyperparameters (eq.~\ref{eq:noninfohyperprior}) 
into the joint posterior of the fully Bayesisan inverse problems: 
\begin{equation}
P({\bf a},\sigma^2,\rho^2|{\bf d})=
\frac{P({\bf d}|{\bf a},\sigma^2)P({\bf a}|\rho^2)P(\sigma^2,\rho^2)}{P({\bf d})},
\label{eq:Bayestheoremforposterior}
\end{equation}
specifically, 
\begin{equation}
\begin{aligned}
    &P({\bf a},\sigma^2,\rho^2|{\bf d})\\
    &=c \exp\left[-\frac{U({\bf a})+\alpha^2 V({\bf a})}{\sigma^2}+F_{\rm obs}(\sigma^2)+F_{\rm pri}(\rho^2)\right], 
\end{aligned}
\label{eq:posterior4uniformhyperprior}
\end{equation}
where $\alpha^2$ $(:=\sigma^2/\rho^2)$ controls the relative weight of the two cost functions $U$ and $V$. When eqs.~(\ref{eq:slightgenhyperpriors}) replace eq.~(\ref{eq:noninfohyperprior}) as a hyperprior, 
the joint posterior becomes 
\begin{equation}
\begin{aligned}
    P({\bf a},\sigma^2,\rho^2|{\bf d})=&
    c
    (2\pi\sigma^2)^{-N^\prime/2}
    (2\pi\rho^2)^{-P^\prime/2}
    |{\bf E}|^{-1/2}|\boldsymbol\Lambda_G|^{1/2}
    \\&\times
    \exp\left[-\frac{U({\bf a})+\alpha^2 V({\bf a})}{\sigma^2}\right]
\end{aligned}
\label{eq:posterior4slightgenhyperpriors}
\end{equation}
with 
\begin{equation}
\begin{cases}
N^\prime&=N-2n_{\sigma^2}
\\
P^\prime&=P-2n_{\rho^2}  
\end{cases}
\label{eq:conversionrule_uniform2slightgenhyperpriors}
\end{equation}
Equation~(\ref{eq:posterior4slightgenhyperpriors}) shows we can transform the joint posterior (eq.~\ref{eq:posterior4uniformhyperprior}) with the uniform hyperprior 
into the one (eq.~\ref{eq:posterior4slightgenhyperpriors}) with a more general hyperprior
through the conversion rule specified by eq.~(\ref{eq:conversionrule_uniform2slightgenhyperpriors}), although we must be aware that unchanged $N$, $M$ and $P$ are implicitly included as sizes and ranks of vectors and matrices in eq.~(\ref{eq:posterior4slightgenhyperpriors}) as in eq.~(\ref{eq:posterior4uniformhyperprior}). 
Therefore, we obtain analytic results for eqs.~(\ref{eq:slightgenhyperpriors}) from those for eq.~(\ref{eq:noninfohyperprior})
by converting explicit $N$ and $P$ into $N^\prime$ and $P^\prime$ through eq.~(\ref{eq:conversionrule_uniform2slightgenhyperpriors}) while keeping the implicit $N$ and $P$ dependence of 
$U[=\mathcal O(N)]$ and $V[=\mathcal O(P)]$ invariant. 
The property of the joint posterior can generally vary for different $n_{\sigma^2},n_{\rho^2}$ values, 
but most of our subsequent asymptotic results for the uniform hyperprior are applicable to the nonuniform ones of
$n_{\sigma^2},n_{\rho^2}=\mathcal O(1)$ (e.g. the logarithmically uniform hyperprior) such that  $N^\prime\sim N$ and $P^\prime\sim P $ for large $N$ and $P$.


\subsection{Candidates of appropriate reduction}
The joint posterior $P({\bf a},\sigma^2,\rho^2|{\bf d})$ eq.~(\ref{eq:posterior4uniformhyperprior}) or (\ref{eq:posterior4slightgenhyperpriors})
is the formal solution of the fully Bayesian inversion defined in the previous section because the joint posterior includes all the information on the observed data, model parameters and hyperparameters~\citep{matsu1991development,carlin2008bayesian,sen2013global}. 
However, the joint posterior is commonly not as simple as the posterior in the linear inverse problem of the fixed hyperparameters. 

For the given hyperparameters $\sigma^2$ and $\rho^2$, 
the posterior is assigned to the model parameters ${\bf a}$ as $P({\bf a}|{\bf d},\sigma^2,\rho^2)$. 
Maximising Gaussian $P({\bf a}|{\bf d},\sigma^2,\rho^2)$, we obtain 
the mean ${\bf a}_*$ of $P({\bf a}|{\bf d},\sigma^2,\rho^2)$ as the regularised least-square estimate~\citep[e.g.][]{yabuki1992geodetic}: 
\begin{equation}
{\bf a}_*(\alpha^2)=({\bf H}^{\rm T}{\bf E}^{-1}{\bf H} +\alpha^2{\bf G})^{-1}{\bf H}^{\rm T}{\bf E}^{-1}{\bf d}.
\label{eq:leastsquarewithpenalty}
\end{equation}
Note the mean ${\bf a}_*$ is also the mode (and median) in the Gaussian $P({\bf a}|{\bf d},\sigma^2,\rho^2)$.  
The covariance ${\bf C}_{{\bf a}_*}$ of $P({\bf a}|{\bf d},\sigma^2,\rho^2)$ is given as
\begin{equation}
{\bf C}_{{\bf a}_*}(\sigma^2,\alpha^2)
=\sigma^2({\bf H}^{\rm T}{\bf E}^{-1}{\bf H} +\alpha^2{\bf G})^{-1},
\label{eq:leastsquarewithpenalty_variance}
\end{equation}
which is hereafter supposed positive definite, thus having full rank. 
When the hyperparameters are fixed, we can fully parametrise the probability profile by ${\bf a}_*$ and ${\bf C}_{{\bf a}_*}$ and reasonably regard ${\bf a}_*$ as the optimal solution of ${\bf a}$. 

On the other hand, the joint posterior $P({\bf a},\sigma^2,\rho^2|{\bf d})$ takes a non-Gaussian profile, which is asymmetric in terms of the hyperparameters [$\ln P({\bf a},\sigma^2,\rho^2|{\bf d})\propto -U/\sigma^2-V/\rho^2$] and, as seen later, could be multimodal in terms of ${\bf a}$, $\sigma^2$ and $\rho^2$. 
Such $P({\bf a},\sigma^2,\rho^2|{\bf d})$ does not equate
various point estimates (mean, mode, median and so on) unlike Gaussian cases. 
Even worse, 
there is no unique projection rule of $P({\bf a},\sigma^2,\rho^2|{\bf d})$ onto a probability profile of ${\bf a}$; 
in one method, we can integrate out (marginalise out) the hyperparameters~\citep{fukuda2008fully} and obtain $P({\bf a}|{\bf d})$ as
\begin{equation}
P({\bf a}|{\bf d})=\int d\sigma^2\int d\rho^2 P({\bf a},\sigma^2,\rho^2|{\bf d}),
\label{eq:defofmarginalized}
\end{equation}
while in another, 
decomposing the joint posterior as
\begin{equation}
P({\bf a},\sigma^2,\rho^2|{\bf d})=
P({\bf a}|\sigma^2,\rho^2,{\bf d})
P(\sigma^2,\rho^2|{\bf d}),
\label{eq:twostage}
\end{equation}
we can project $P({\bf a},\sigma^2,\rho^2|{\bf d})$ onto $P({\bf a}|\sigma^2,\rho^2,{\bf d})$ with reasonable hyperparameter values ~\citep[e.g. probability peaks;][]{akaike1980likelihood} inferred from their marginal posterior: 
\begin{equation}
P(\sigma^2,\rho^2|{\bf d})=\int d{\bf a}  P({\bf a},\sigma^2,\rho^2|{\bf d}).
\label{eq:marginalposofhyper}
\end{equation}
Due to the non-Gaussianity of the joint posterior, these two projection rules generally do not conclude an equivalent pdf of ${\bf a}$, and as seen later, resultant pdfs are not necessarily well-behaved.

Reduction of information earns a practical significance as above in evaluating the joint posterior. 
We here investigate it, and our focus is on marginalisation of the joint posterior. 
The following three categories are considered: (1) no marginalisation, (2) marginalisation with respect to the hyperparameters, 
and (3) marginalisation with respect to the model parameters. 

Category (1) evaluates the probability value of
paired values of the model parameters and hyperparameters (${\bf a}$, $\sigma^2$ and $\rho^2$) directly from the joint-posterior $P({\bf a},\sigma^2,\rho^2|{\bf d})$. 
A representative point estimator of category (1) is the mode of the joint posterior (maximum a posteriori, MAP). 
The suite of the MAP estimates, $\hat {\bf a}_{\rm MAP},\hat \sigma^2_{\rm MAP}$ and $\hat \rho^2_{\rm MAP}$ is defined as
\begin{equation}
(\hat {\bf a}_{\rm MAP},\hat \sigma^2_{\rm MAP},\hat \rho^2_{\rm MAP})
:=\mbox{argmax}_{{\bf a},\sigma^2,\rho^2} P({\bf a},\sigma^2,\rho^2|{\bf d}),
\label{eq:defofMAP}
\end{equation}
where $\hat{\cdot}$ represents the optimal value, 
$:=$ denotes that the left-hand side is defined by the right-hand side, 
and 
$\mbox{argmax}_{\cdot}(\cdot)$ is a functional such that $\mbox{argmax}_{y}(f(y))$ returns $y$ maximising the function $f$ of $y$. 
The performance of the MAP is recognised as not necessarily high in the statistical literature, 
in both the Bayesian inference without hyperparameters~\citep{lin2006loss} and 
the fully Bayesian inference~\citep{iba1996learning}. 
Meanwhile, the MAP is also considered a generalisation of the maximum likelihood estimation~\citep[termed generalised maximum likelihood estimation;][]{carlin2008bayesian} with many practical applications~\citep{carlin2008bayesian,amey2018bayesian,goto2019delayed}. 

In category (2), the hyperparameters are regarded as secondary in the model-parameter estimation and marginalised out from the joint posterior~\citep{carlin2008bayesian,fukuda2008fully}. 
It 
leads to the marginal posterior of the model parameters $P({\bf a}|{\bf d})$ 
and 
includes a family of ordinary point estimators. 
The simplest estimator in category (2) would be the mode of $P({\bf a}|{\bf d})$~\citep[e.g.][]{amey2018bayesian}, here we call the maximum of the marginal posterior of the model parameters (the MMPM): 
\begin{equation}
\hat {\bf a}_{\rm MMPM}
:=
\mbox{argmax}_{{\bf a}}P({\bf a}|{\bf d}).
\end{equation}
The most popular estimator may be the posterior mean (expected a posteriori, EAP): 
\begin{equation}
\hat {\bf a}_{\rm EAP}:=
\langle
{\bf a}
\rangle_{{\bf a},\sigma^2,\rho^2|{\bf d}},
\label{eq:defofmodelEAP}
\end{equation}
where $\langle f\rangle_{y|z}:=
\int dy f P(y|z)$ represents the probability mean of a function $f$ over $y$ given $z$, where $y$ and $z$ may be vectors. 
That is, 
\begin{equation}
\hat {\bf a}_{\rm EAP}=
\int d{\bf a}\int d\sigma^2\int d\rho^2 P({\bf a},\sigma^2,\rho^2|{\bf d}){\bf a}=\int d{\bf a}P({\bf a}|{\bf d}){\bf a}.
\label{eq:elementarymath}
\end{equation}
Similar to the EAP estimate $\hat {\bf a}_{\rm EAP}$, quantities computable from $P({\bf a}|{\bf d})$ only are classified into category (2) [i.e. functions $f({\bf a})$ of ${\bf a}$, for which $\langle
f({\bf a})
\rangle_{{\bf a},\sigma^2,\rho^2|{\bf d}}=\int d{\bf a}P({\bf a}|{\bf d})f({\bf a})$]. 
Note there is no loss of information with respect to ${\bf a}$ in the transform from $P({\bf a},\sigma^2,\rho^2|{\bf d})$ to $P({\bf a}|{\bf d})$, and category (2) drops only the information on the hyperparameters contained in category (1), which is redundant for the description of the model parameters~\citep{fukuda2008fully}. 
To summarise, when only the model-parameter values are variables of interest as in eq.~(\ref{eq:elementarymath}), 
the information on the hyperparameters are integrated (marginalised) automatically, and the marginal posterior of the model parameters contains identical model-parameter information to the joint posterior. Category (2) represents this implicit reduction of the joint posterior erasing the information on the hyperparameters. 
Statistics in category (2) have similar but slightly different asymptotic properties from ones in category (1) as next seen in \S\ref{sec:analyticsolution}, except the after-mentioned intricacy of the EAP. 

In category (3), a two-stage inference is performed in accordance with the decomposition of the joint posterior expressed by eq.~(\ref{eq:twostage}); 
the first stage is the confidence evaluation for the hyperparameters using $P(\sigma^2,\rho^2|{\bf d})$,
and the second stage is 
that for the model parameters using $P({\bf a}|\sigma^2,\rho^2,{\bf d})$ based on the result of the first stage. 
A representative of this category is
ABIC~\citep{akaike1980likelihood}, which is nearly identical to the maximisation of Type II likelihood~\citep{good1965estimation} and evidence (the use of them is called empirical Bayes\citep{malinverno2004expanded,gelman2013bayesian}). 
In ABIC, 
the optimum set of the hyperparameters $(\hat \sigma^2_{\rm ABIC},\hat \rho^2_{\rm ABIC})$ is obtained through the maximisation of $P(\sigma^2,\rho^2|{\bf d})$:
\begin{equation}
(\hat \sigma^2_{\rm ABIC},\hat \rho^2_{\rm ABIC})
:=\mbox{argmax}_{(\sigma^2,\rho^2)} P(\sigma^2,\rho^2|{\bf d}).
\label{eq:pointestimationABIC_hyperparameter}
\end{equation}
Using these values, 
ABIC gives its optimum values of the model parameters ${\bf a}$ from $P({\bf a}|\sigma^2,\rho^2,{\bf d})$, where the mode estimate ${\bf a}_*$ is reasonably optimal:
\begin{equation}
\hat {\bf a}_{\rm ABIC}:=\mbox{argmax}_{{\bf a}} P({\bf a}|{\bf d},\hat\sigma^2_{\rm ABIC},\hat\rho^2_{\rm ABIC}).
\label{eq:pointestimationABIC_modelparameter}
\end{equation}
As mentioned in the Introduction, 
ABIC has sometimes been criticised for point estimation of the hyperparameters. 
However, 
we locate ABIC in the context of the joint posterior reduction in this study,
and 
the term `ABIC' in this paper does not imply such point estimation (eqs.~\ref{eq:pointestimationABIC_hyperparameter} and \ref{eq:pointestimationABIC_modelparameter}), but rather involves the uncertainty evaluation of the hyperparameters, analytic expressions of which are shown in \S\ref{sec:analyticsolution}. 
We also note the following considers maximising their marginal posterior $P(\sigma^2,\rho^2|{\bf d})$, 
although ABIC originally maximises the marginalised likelihood $P({\bf d}|\sigma^2,\rho^2)$ of the hyperparameters [formally corresponding to $P(\sigma^2,\rho^2|{\bf d})$ of the uniform prior eq.~(\ref{eq:noninfohyperprior})]~\citep{akaike1980likelihood}. 
In perspective on the reduction, as discussed later, the key of ABIC is in the two-stage inference of the model parameters and hyperparameters expressed by eq.~(\ref{eq:twostage}). 

\subsection{Semianalytic representations of estimates and asymptotic forms of the posteriors}
\label{sec:analyticsolution}

\subsubsection{ABIC and the marginal posterior of the hyperparameters}
We first present semianalytic representations of the ABIC estimates by extending their point-estimate expressions \citep{akaike1980likelihood,yabuki1992geodetic}, reproducing which is instructive for understanding the following derivation associated with categories (1) and (2). 

For the present linear inversion, 
we can analytically marginalise out the model parameters from the joint posterior [as eq.~(\ref{eq:marginalposofhyper})] and obtain 
the marginal posterior/likelihood $P(\sigma^2,\rho^2|{\bf d})$ of the hyperparameters~\citep{akaike1980likelihood,yabuki1992geodetic}: 
\begin{equation}
\begin{aligned}
\ln P(\sigma^2,\rho^2|{\bf d})=&
c-\frac{N-M+P}{2}\ln \sigma^2 +\frac P 2 \ln \alpha^2
\\&
-\frac 1 2 \ln |{\bf H}^{\rm T}{\bf E}^{-1}{\bf H}+\alpha^2{\bf G}|
-\frac{s({\bf a}_*(\alpha^2),\alpha^2)}{2\sigma^2} 
\end{aligned}
\label{eq:lnmarginalposofhyperforlinearinverse}
\end{equation}
with
\begin{equation}
\begin{aligned}
s({\bf a},\alpha^2) :=&2[U({\bf a})+\alpha^2V({\bf a})]
\\
=&({\bf d}-{\bf Ha})^{\rm T}{\bf E}^{-1}({\bf d}-{\bf Ha})
+\alpha^2{\bf a}^{\rm T}{\bf G}{\bf a},
\end{aligned}
\label{eq:defofs}
\end{equation}
where we use $\sigma^2,\alpha^2$ instead of $\sigma^2,\rho^2$. 
Then, we eliminate $\sigma^2$ from eq.~(\ref{eq:lnmarginalposofhyperforlinearinverse}) using the extremum condition with respect to $\sigma^2$ while fixing $\alpha^2$, 
which 
yields 
\begin{equation}
\tilde \sigma^2_{\rm ABIC}(\alpha^2):=\frac{s({\bf a}_*(\alpha^2),\alpha^2)}{N+P-M}.
\label{eq:sigmaatsaddleofmarginalposterior}
\end{equation}
Substituting eq.~(\ref{eq:sigmaatsaddleofmarginalposterior}) into eq.~(\ref{eq:lnmarginalposofhyperforlinearinverse}), the maximisation condition of $P(\sigma^2,\rho^2|{\bf d})$ provides the following representation of the optimal $\alpha^2$ value in the ABIC estimate: 
\begin{equation}
\begin{aligned}
\hat \alpha^2_{\rm ABIC}=\mbox{argmin}_{\alpha^2}[
&(N+P-M)\ln s({\bf a}_*(\alpha^2),\alpha^2) -P\ln (\alpha^2) 
\\&+ \ln | {\bf H}^{\rm T}{\bf E}^{-1}{\bf H}+\alpha^2 {\bf G}| ],
\end{aligned}
\label{eq:ABICalpha}
\end{equation}
where $\mbox{argmin}_{\cdot}(\cdot)$ is the functional such that $\mbox{argmin}_{y}(f(y))$ returns $y$ minimising the function $f$ of $y$. 
Note the extremum search rewrites eq.~(\ref{eq:ABICalpha}) for $\alpha^2=\alpha^2_{\rm ABIC}$ as follows (Appendix~\ref{sec:ABICextremal}):
\begin{equation}
P=
\alpha^2 \mbox{Tr}[({\bf H}^{\rm T}{\bf E}^{-1}{\bf H}+\alpha^2  {\bf G})^{-1}{\bf G}]+\frac{2\alpha^2V({\bf a}_*(\alpha^2))}{
\tilde \sigma^2_{\rm ABIC}(\alpha^2)},
\label{eq:ABICalpha_explicitextremal}
\end{equation} 
where $\mbox{Tr}(\cdot)$ denotes the trace of the matrix. 
Equation~\ref{eq:ABICalpha_explicitextremal}) 
is not linearly solvable, 
and 
we conduct a direct numerical search for the $\alpha^2$ value that meets eq.~(\ref{eq:ABICalpha}). 
The optimal $\alpha^2$ obtained from eq.~(\ref{eq:ABICalpha})
determines 
the optimal $\sigma^2$ using eq.~(\ref{eq:sigmaatsaddleofmarginalposterior}) of the extremum condition with respect to $\sigma^2$:
\begin{equation}
\hat \sigma^2_{\rm ABIC}= \tilde \sigma^2_{\rm ABIC}(\hat \alpha^2_{\rm ABIC}).
\end{equation}

For the point estimation of the hyperparameters, 
the distribution of the model parameters ${\bf a}$ for ABIC is simplified to a Gaussian form
${\bf a}\sim 
P({\bf a}|\hat \sigma^2_{\rm ABIC},\hat \rho^2_{\rm ABIC},{\bf d})=
\mathcal N(\hat {\bf a}_{\rm ABIC},{\bf C}_{\hat {\bf a}_{\rm ABIC},{\rm point}})$, 
where $\hat \rho^2_{\rm ABIC}=\hat\sigma^2_{\rm ABIC}/\hat\alpha^2_{\rm ABIC}$ holds, 
and 
the optimal model parameters $\hat {\bf a}_{\rm ABIC}$ for ABIC and 
the posterior covariance ${\bf C}_{\hat {\bf a}_{\rm ABIC},{\rm point}}$ of ${\bf a}$ are given as functions of the optimal hyperparameters; using eqs.~(\ref{eq:leastsquarewithpenalty}) and (\ref{eq:leastsquarewithpenalty_variance}) that describe the mean and covariance of $P({\bf a}|\sigma^2,\rho^2,{\bf d})$, we have
\begin{equation}
\hat {\bf a}_{\rm ABIC}={\bf a}_*(\hat\alpha^2_{\rm ABIC})
\label{eq:ABICmodel}
\end{equation}
with 
$ 
{\bf C}_{\hat {\bf a}_{\rm ABIC},{\rm point}}=
{\bf C}_{{\bf a}_*}(\hat \sigma^2_{\rm ABIC},\hat \alpha^2_{\rm ABIC}).
$ 

For more precise two-stage inferences, we should also evaluate the distribution of the hyperparameters. 
As described in Appendix~\ref{sec:ABIChyperparametervariance}, 
we can evaluate the associated covariance
${\bf C}_{\hat {\bf h}_{\rm ABIC}}$ of the hyperparameters as
\begin{equation} 
\begin{aligned}
{\bf C}_{\hat {\bf h}_{\rm ABIC}}=&\mathcal O[\min(1/N,1/P)]
\\
&+\left(\begin{array}{cc}0&0\\0&\mathcal O[(N/P)\min(1/N,1/P)]\end{array}\right),
\end{aligned}
\label{eq:negligiblevarianceofhyperparametersABIC}
\end{equation} 
where ${\bf h}=(\sigma^2,\rho^2)^{\rm T}$. 
The first term and then the variance of $\sigma^2$ and cross correlation between $\sigma^2$ and $\rho^2$  
vanish for large $N$ or large $P$, while the second term and also the variance of $\rho^2$ are $\mathcal O(1/P)$ for large $N$ and cancel only for large $P$. 
This strange order of the $\rho^2$ variation is probably because 
the influence of the prior to the estimates is negligible from the beginning for large $N$ [e.g. $\mathcal O(P/N)$ in the regularised least-square solution], which can weaken the constraint on $\rho^2$ (or equivalently, on $\alpha^2$) for $N\gg P$; meanwhile, the other covariance components are well constrained both for large $N$ and for large $P$. 
Equation~(\ref{eq:negligiblevarianceofhyperparametersABIC}) indicates the non-point estimation of the hyperparameters in ABIC 
generates 
the associated model-parameter distribution as 
${\bf a}\sim \mathcal N(\hat {\bf a}_{\rm ABIC},{\bf C}_{\hat {\bf a}_{\rm ABIC},{\rm point}}
+\mathcal O(1/ P)
)$ for large $P$ (the exact expression of the error propagation is given in Supplement 1, 
using the specific form of the covariance ${\bf C}_{\hat {\bf h}_{\rm ABIC}}$ shown in Appendix~\ref{sec:ABIChyperparametervariance}), which approaches to the point-estimate one asymptotically. 
Equation~(\ref{eq:negligiblevarianceofhyperparametersABIC}) also means the smallness of the standard deviations of the hyperparameters $\sigma^2$ and $\rho^2$ for large $N$ or $P$. As it suggests, the marginal posterior of the hyperparameters asymptotically approaches to a delta function as $N$ or $P$ increases (Appendix~\ref{sec:ABIChyperparametervariance}): 
for large $P$, 
\begin{equation}
P(\sigma^2,\rho^2|{\bf d})\to
\delta(\sigma^2-\hat\sigma^2_{\rm ABIC})
\delta(\rho^2-\hat\rho^2_{\rm ABIC}),
\label{eq:deltafunctionalconvergence_hyperparameter}
\end{equation}
and for large $N$, 
\begin{equation}
P(\sigma^2,\rho^2|{\bf d})\to
\delta(\sigma^2-\hat\sigma^2_{\rm ABIC})\psi(\rho^2),
\label{eq:deltafunctionalconvergence_hyperparameter_added}
\end{equation}
where $\psi$ is a pdf of $\rho^2$, peaking at $\hat \rho^2_{\rm ABIC}$. 
The inverse-square-root standard deviations are analogous to the central limit theorem,
and the delta-functional concentrations to the law of large numbers.

\subsubsection{MAP}
Below, we derive a semianalytic representation of the MAP estimate that satisfies the maximisation condition of the joint posterior $P({\bf a},\sigma^2,\rho^2|{\bf d})$, eq.~(\ref{eq:defofMAP}).
We use (${\bf a}$, $\sigma^2$, $\alpha^2$) as a set of independent variables instead of (${\bf a}$, $\sigma^2$, $\rho^2$). 
This conversion rewrites the MAP estimate as follows:
\begin{equation}
(\hat {\bf a}_{\rm MAP},\hat \sigma^2_{\rm MAP},\hat \alpha^2_{\rm MAP}) = \mbox{argmax}_{({\bf a},\sigma^2,\alpha^2)} P({\bf a},\sigma^2,\rho^2|{\bf d}),
\label{eq:MAPwithalpha}
\end{equation}
where $\hat \alpha^2_{\rm MAP}=\hat \sigma^2_{\rm MAP}/\hat \rho^2_{\rm MAP}$. 
We note that the maximisation function is still the joint posterior of ${\bf a}$, $\sigma^2$ and $\rho^2$, yet treated as a function of  
${\bf a}$, $\sigma^2$ and $\alpha^2$ 
through the relation $\rho^2=\sigma^2/\alpha^2$. 

We first eliminate the model parameters ${\bf a}$ from the joint posterior by applying the extremum condition with respect to ${\bf a}$ while fixing $\sigma^2$ and $\alpha^2$. 
It 
is equivalent to the extremum condition of $\ln P({\bf a}|\sigma^2,\rho^2,{\bf d})$ given the decomposition of the joint posterior eq.~(\ref{eq:twostage}),
thus now yielding 
$ 
{\bf a}={\bf a}_*(\alpha^2)
$ (eq.~\ref{eq:leastsquarewithpenalty}). 
We also consider the extremum condition of the joint posterior eq.~(\ref{eq:posterior4uniformhyperprior})
with respect to $\sigma^2$ while fixing ${\bf a}$ and $\alpha^2$, 
which leads to 
\begin{equation}
\tilde \sigma^2_{\rm MAP}({\bf a},\alpha^2):=\frac{s({\bf a},\alpha^2)}{N+P},
\label{eq:sigmaatsaddleofposterior}
\end{equation}
where $s({\bf a}, \alpha^2)$ is defined by eq.~(\ref{eq:defofs}). 
By substituting ${\bf a}={\bf a}_*$ and $\sigma^2=\tilde \sigma^2_{\rm MAP}$ into the joint posterior eq.~(\ref{eq:posterior4uniformhyperprior}), we derive the following representation of the optimal $\alpha^2$ value in the MAP estimate: 
\begin{equation}
\hat \alpha^2_{\rm MAP}:=\mbox{argmin}_{\alpha^2}
\left[ (N+P) \ln s({\bf a}_*(\alpha^2),\alpha^2)- P \ln \alpha^2\right].
\label{eq:MAPalphadeterminationcondition}
\end{equation} 
Here we multiplied the log joint posterior by $-2$ as in the ABIC estimate for comparison. 
Equation~(\ref{eq:MAPalphadeterminationcondition}) is a one-dimensional search problem of $\alpha^2$ as eq.~(\ref{eq:ABICalpha}) for the ABIC estimate. 

Once we obtain the $\alpha^2$ value of the MAP estimate, $\hat \alpha^2_{\rm MAP}$, we also have the 
MAP estimates of ${\bf a}$:
\begin{flalign}
\hat {\bf a}_{\rm MAP}
={\bf a}_*(\hat\alpha^2_{\rm MAP})
.
\label{eq:MAPmodel}
\end{flalign} 
The MAP estimate of $\sigma^2$ is also derived as
$
\hat \sigma^2_{\rm MAP}
=\tilde \sigma^2_{\rm MAP}(\hat {\bf a}_{\rm MAP},\hat \alpha^2_{\rm MAP})
$. 
Given the decomposition of the joint posterior eq.~(\ref{eq:twostage}), 
we obtain the second-order moment ${\bf C}_{\hat {\bf a}_{\rm MAP}}$ of ${\bf a}$ around $\hat{\bf a}_{\rm MAP}$ by substituting $\sigma^2=\hat\sigma^2_{\rm MAP}$ and $\rho^2=\hat\sigma^2_{\rm MAP}/\hat\alpha^2_{\rm MAP}$ into $P({\bf a}|\sigma^2,\rho^2,{\bf d})$ (eq.~\ref{eq:leastsquarewithpenalty_variance}): 
$ 
{\bf C}_{\hat {\bf a}_{\rm MAP}}=
{\bf C}_{{\bf a}_*}(\hat \sigma^2_{\rm MAP},\hat \alpha^2_{\rm MAP}). 
$ 

The above maximisation condition (eq.~\ref{eq:MAPalphadeterminationcondition}) for $\alpha^2$ also provides the following extremum condition, distinctively different from the requirement of the ABIC estimate eq.~(\ref{eq:ABICalpha_explicitextremal}):
\begin{equation}
\hat \alpha^2_{\rm MAP}=\frac{U({\bf a}_*(\hat \alpha^2_{\rm MAP}))/N}{V({\bf a}_*(\hat \alpha^2_{\rm MAP}))/P},
\label{eq:MAPalpha}
\end{equation}
where we used $\partial s({\bf a},\alpha^2)/\partial {\bf a}|_{{\bf a}={\bf a}_*}={\bf 0}$. 
Equation~(\ref{eq:MAPalpha}) indicates $\hat \alpha^2_{\rm MAP}$ is determined such that $U/(N\sigma^2)=V/(P\rho^2)$. That means the MAP estimate balances the normalised cost function of data fitting ($U/\sigma^2$) per degree of freedom $(U/\sigma^2)/N$ with that of the model-parameter prior ($V/\rho^2)/P$. 

Since the model-parameter estimate $\hat {\bf a}_{\rm MAP}$ of the MAP takes the same functional form ${\bf a}_*$ (eq.~\ref{eq:leastsquarewithpenalty}) as $\hat {\bf a}_{\rm ABIC}$ of ABIC (eq.~\ref{eq:ABICmodel}),
their difference in ${\bf a}$ is all ascribed to that in their maximisation functions of $\alpha^2$ [eqs.~(\ref{eq:MAPalphadeterminationcondition}) and (\ref{eq:ABICalpha}) for the MAP and ABIC, respectively]: $-M\ln s({\bf a}_*(\alpha^2))+ \ln | {\bf H}^{\rm T}{\bf E}^{-1}{\bf H}+\alpha^2 {\bf G}|$; 
one may notice
$ 
\hat {\bf a}_{\rm MAP}=\mbox{argmax}_{{\bf a}} P({\bf a}|{\bf d},\hat\sigma^2_{\rm MAP},\hat\rho^2_{\rm MAP}),
$ 
deduced from eqs~(\ref{eq:twostage}) and (\ref{eq:MAPwithalpha}), identical to eq.~(\ref{eq:pointestimationABIC_modelparameter}) of $\hat {\bf a}_{\rm ABIC}$. 
Regarding their $\sigma^2$ estimates, 
the difference exists also in the denominators besides their $\alpha^2$ values
[$N+P$ in the MAP (eq.~\ref{eq:sigmaatsaddleofposterior}) and $N-(M-P)$ in ABIC (eq.~\ref{eq:sigmaatsaddleofmarginalposterior})]. 

\subsubsection{The MMPM estimate and model-parameter values sampled with finite probabilities}

The marginal posterior of the model parameters $P({\bf a}|{\bf d})$ eq.~(\ref{eq:defofmarginalized}) is written as follows via eq.~(\ref{eq:posterior4uniformhyperprior}):
\begin{equation}
\begin{aligned}
&P({\bf a}|{\bf d})
=
\\
&c\int^\infty_0 d(\sigma^2)\int^\infty_0 d(\rho^2) 
(\sigma^2)^{-\frac N 2 }(\rho^2)^{-\frac P 2}
\exp \left[-\frac U {\sigma^2} - \frac V {\rho^2} \right].
\end{aligned}
\label{eq:explicitintegralformofmarginalposteriorofa}
\end{equation}
For $U>0$ and $V>0$,
converting the integration variables from $\sigma^2$ and $\rho^2$ to
$X=\sigma^2/U$ and $Y=\sigma^2/V$,  
we reduce the integral of 
eq.~(\ref{eq:explicitintegralformofmarginalposteriorofa}) 
to the Gamma functions $\Gamma(z)(:=\int^\infty_0 dt t^{z-1}e^{-t})$ as
\begin{equation}
\begin{aligned}
P({\bf a}|{\bf d})
=&
cU^{-N/2+1}
V^{-P/2+1}
\\&\times
\int^\infty_0 dXX^{-N/2}
e^{-1/X}
\int^\infty_0 dYY^{-P/2}
e^{-1/Y}
\\
=&
c U^{-N/2+1} 
V^{-P/2+1} \Gamma(N/2 -1)\Gamma(P/2 -1).
\end{aligned}
\label{eq:marginalizedpostprob_withGamma}
\end{equation}
Equation~(\ref{eq:marginalizedpostprob_withGamma}) states
\begin{equation}
P({\bf a}|{\bf d})
=c
U^{-N/2+1} V^{-P/2+1}. 
\label{eq:marginalizedpostprob}
\end{equation}
Equation~(\ref{eq:marginalizedpostprob}) is also valid for $U=0$ or $V=0$, where both hands of eq.~(\ref{eq:marginalizedpostprob}) are infinite.

We calculate the extremum condition of $P({\bf a}|{\bf d})$ 
with eq.~(\ref{eq:marginalizedpostprob}) and obtain the mode $\hat {\bf a}_{\rm MMPM}$ of the marginal posterior $P({\bf a}|{\bf d})$ of ${\bf a}$: 
\begin{equation}
\hat {\bf a}_{\rm MMPM}={\bf a}_*(\check \alpha^2_{\rm MMPM})
,
\label{eq:modelparammode}
\end{equation} 
with 
a scalar function of $\hat {\bf a}_{\rm MMPM}$: 
\begin{equation}
    \check \alpha^2_{\rm MMPM} := \frac{U(\hat {\bf a}_{\rm MMPM})}{V(\hat {\bf a}_{\rm MMPM})}\frac{P-2}{N-2}
    .
\label{eq:checkalphamode}
\end{equation}
Equations~(\ref{eq:modelparammode}) and (\ref{eq:checkalphamode}) 
yield a self-consistent equation of $\check \alpha^2_{\rm MMPM}$: 
\begin{equation}
\check\alpha^2_{\rm MMPM}=
\frac{U({\bf a}_*(\check \alpha^2_{\rm MMPM}))/(N-2)}{V({\bf a}_*(\check \alpha^2_{\rm MMPM}))/(P-2)}=\hat\alpha^2_{\rm MAP}+\mathcal O(1/N,1/P).
\label{eq:checkalphamode_selfconsistent}
\end{equation}
As above, we obtain the $\alpha^2$ value that gives the optimal ${\bf a}$ from a one-dimensional search of eq.~(\ref{eq:checkalphamode_selfconsistent}), analogous to eq.~(\ref{eq:MAPalpha}) in the MAP estimate. 
Although $\check \alpha^2_{\rm MMPM}$ is a virtual hyperparameter because the original hyperparameters are marginalised out to evaluate the marginal posterior of the model parameters, 
the value of $\check \alpha^2_{\rm MMPM}$ (eq.~\ref{eq:checkalphamode_selfconsistent}) is consistent with the associated MAP estimate $\hat \alpha^2_{\rm MAP}$ (eq.~\ref{eq:MAPalpha}) after converting $P-2$ and $N-2$ to $P$ and $N$, respectively. 
Thus, the mode of $P({\bf a}|{\bf d})$ is asymptotically consistent with the MAP estimate of the model parameters for large $N$ and $P$: 
\begin{equation}
\hat {\bf a}_{\rm MMPM}=\hat {\bf a}_{\rm MAP}+\mathcal O(1/N)+\mathcal O(1/P).
\label{eq:MMAPsimMAP}
\end{equation}

Equation~(\ref{eq:marginalizedpostprob}) 
also shows 
$P({\bf a}|{\bf d})$ concentrates as $U=\mathcal O(N)$ or $V=\mathcal O(P)$ increases, and indeed 
$P({\bf a}|{\bf d})$ asymptotically approaches to the following delta function as $N$ or $P$ increases (Appendix~\ref{sec:FBmarginilisation}): 
\begin{equation}
P({\bf a}|{\bf d})\to c\delta(X({\bf a})-X(\hat {\bf a}_{\rm MMPM})-0),
\label{eq:deltafunctionalconvergence_modelparameter}
\end{equation}
with
\begin{equation}
X({\bf a}):=
\begin{cases}
    \dfrac{U({\bf a})}{N}\left[
\dfrac{V({\bf a})}{P}
\right]^{\frac{P-2}{N-2}} & (N/P\geq1) \\
    \dfrac{V({\bf a})}{P}\left[
\dfrac{U({\bf a})}{N}
\right]^{\frac{N-2}{P-2}} & (N/P<1)
\end{cases}
\label{eq:univariableofPad}
\end{equation}
Because the distribution $P({\bf a}|{\bf d})$ converges to a delta function, 
arbitrary model parameter values with finite probabilities in $P({\bf a}|{\bf d})$ are asymptotically consistent.

Evaluating the second-order moment ${\bf C}_{\hat {\bf a}_{\rm MMPM}}$ of $P({\bf a}|{\bf d})$ around its peak $\hat {\bf a}_{\rm MMPM}$ up to the second-order deviation, 
we find 
$ 
{\bf C}_{\hat {\bf a}_{\rm MMPM}}
=
{\bf C}_{{\bf a}_*}(\check\sigma^2_{\rm MMPM},\check\alpha^2_{\rm MMPM})
+\delta {\bf C}_{\hat {\bf a}_{\rm MMPM}}
$  (Appendix~\ref{sec:FBmarginilisation}) 
with 
\begin{equation}
\check\sigma^2_{\rm MMPM}=
\frac{s({\bf a}_*(\check \alpha^2_{\rm MMPM}))}{N+P-4} 
=\hat\sigma^2_{\rm MAP}+\mathcal O(1/N)+\mathcal O(1/P),
\end{equation}
where the explicit form of $\delta {\bf C}_{\hat {\bf a}_{\rm MMPM}}$ is given in Appendix~\ref{sec:FBmarginilisation}. 
The first term of ${\bf C}_{\hat {\bf a}_{\rm MMPM}}$ is asymptotically consistent with ${\bf C}_{\hat {\bf a}_{\rm MAP}}$ of the MAP estimate, 
and thus the rather complicated second term $\delta {\bf C}_{\hat {\bf a}_{\rm MMPM}}$ may be interpreted as propagation of uncertainty from the hyperparameters involved with the marginalisation of the hyperparameters.


It may also be noteworthy that 
the mode of $P({\bf a}|{\bf d})$ for the logarithmically uniform hyperprior [given by eq.~(\ref{eq:checkalphamode_selfconsistent}) with converting $N\to N+2$ and $P\to P+2$ for the change in the hyperprior eq.~(\ref{eq:conversionrule_uniform2slightgenhyperpriors})] is identical to the model-parameter estimate of the MAP for the uniform hyperprior [given by eq.~(\ref{eq:MAPalpha})]; the difference between the MAP and MMPM estimates is such small.

The characteristics of the EAP estimate are quite complicated, and then we will explain them later in detail in \S\ref{sec:intepretationofprob}. 
In Table~\ref{tab:semianalyticresultsummary}, we summarise the analytical results derived in this subsection.


\begin{table*}
    \centering
    \caption{Semianalytic representations of the MAP (and MMPM) and ABIC estimates for the hyperparameters $\alpha^2$ and $\sigma^2$ and the model parameters ${\bf a}$. 
    Here, $s_*:=s_*({\bf a}_*(\alpha^2),\alpha^2)$. 
    Read $N$ as $N-2$ and $P$ as $P-2$ for the MMPM. 
    The second-order moments around the model-parameter estimates are also shown as ${\bf C}_{\hat {\bf a}}$ for the MAP and ABIC (not for the MMPM). 
    Note $\hat {\bf a}$ and ${\bf C}_{\hat{\bf a}}$ of the MAP and ABIC have different arguments ($\hat\alpha^2$, $\hat\sigma^2$) in the same formulas. 
    Associated marginal posteriors are further shown in asymptotic forms (using $\approx$) for $P\to\infty$.}
    \begin{tabular}{ccc}
        \hline
        &MAP (MMPM) &ABIC\\
        \hline
        $\hat\alpha^2$ &${\rm argmin}[(N+P)\ln s_*-P\ln\alpha^2]$ & ${\rm argmin}[(N+P-M)\ln s_*-P\ln\alpha^2+\ln|{\bf H}^T{\bf E}^{-1}{\bf H}+\alpha^2{\bf G}|]$  \\
        $\hat\sigma^2$ & $s_*/(N+P)$&$s_*/(N+P-M)$  \\
        $\hat{\bf a}$ & ${\bf a}_*(\hat\alpha^2)$ & ${\bf a}_*(\hat\alpha^2)$ \\
        ${\bf C}_{\hat {\bf a}}$ & ${\bf C}_*(\hat\sigma^2,\hat\alpha^2)$& ${\bf C}_*(\hat\sigma^2,\hat\alpha^2)$\\
        $P(\cdot|{\bf d})$ & $\approx\delta(X({\bf a})-X(\hat{\bf a}))$& $\approx\delta(\sigma^2-\hat\sigma^2)\delta(\rho^2-\hat\rho^2)$  \\ \hline
    \end{tabular}
    \label{tab:semianalyticresultsummary}
\end{table*}

\section{Synthetic tests} 
The analytic solutions derived in the previous section clarify 
the MAP estimate and estimates obtained from the marginal posterior $P({\bf a}|{\bf d})$ of the model parameters with finite probabilities, typified by the MMPM, asymptotically converge to the same solution for $N\to\infty$ or $P\to\infty$, while the ABIC estimate does not ($\hat {\bf a}_{\rm MAP}\approx\hat {\bf a}_{\rm MMPM}\neq\hat {\bf a}_{\rm ABIC}$), where the approximate sign $\approx$ expresses the asymptotic equality at $N\to\infty$ or $P\to\infty$. 
This marked difference between the ABIC solution and the others derives from a fundamental gap between the asymptotic convergence of the marginal posterior of the hyperparameters 
to the ABIC estimate (eq.~\ref{eq:deltafunctionalconvergence_hyperparameter})
and 
the asymptotic convergence of the marginal posterior of the model parameters
to the MAP estimate 
(eqs.~\ref{eq:MMAPsimMAP} and \ref{eq:deltafunctionalconvergence_modelparameter}). 
In brief, when $\hat{\bf a}_{\rm ABIC}\neq\hat {\bf a}_{\rm MAP}$, 
the ABIC estimate $\hat{\bf a}_{\rm ABIC}$ of the model parameters has asymptotically zero probability $P(\hat{\bf a}_{\rm ABIC}|{\bf d})\to0$ for large degrees of freedom, while the MAP estimate of the hyperparameters has an asymptotically zero probability in the marginal posterior of the hyperparameters,  $P(\hat\sigma^2_{\rm MAP},\hat\rho^2_{\rm MAP}|{\bf d})\to0$. 
Meanwhile, when the number of data $N$ is large enough for a given number of model parameters $M$ ($N\gg M$), since the role of priors is negligible, categories (1)--(3) set mostly the same reductions, as known in the literature~\citep[e.g.][]{gelman2013bayesian}. 
Hereafter, $P$ and $M$ are assumed to be of the same order for simplicity. 
Hence, the discrepancy between the analytic solutions in the previous section is considered to have captured certain asymptotic characteristics of the joint posterior for a relatively large number of model parameters ($N\lesssim M$). 

In this section, we focus on the two asymptotically exclusively measurable solutions, the ABIC and MAP estimates, and perform synthetic tests with examining their dependence on the number of model parameters $M$. 
It allows us to investigate whether the two-stage inference of ABIC or the conventional one-stage fully Bayesian approach is the more appropriate reduction and to study why such inconsistency arises.

\subsection{Model setting}
We consider a problem of reconstructing a continuous crustal deformation field $a(x)$ from discrete displacement data $d(x_n)$ of observation points $n(=1,...,N)$. 
A single displacement component is treated in a one-dimensional coordinate $x$ (e.g. along an east-west survey line) for simplicity. 
This model can be regarded as a simplest example of the observation equation eq.~(\ref{eq:obseq}), where the integral kernel is a delta function. 
Delta-functional kernels are employed in the observation equations of 
\citet{fukahata1996crustal}
and 
\citet{okazaki2021consistent} to 
invert levelling and Global Navigation Satellite System (GNSS) data, respectively. 
The identical one-dimensional scalar-field inference also appears in an inversion of 
the annual variation in the teleseismic detection capability of a station~\citep{iwata2015quantitative}. 
The error-free nature of the model is ideal for investigating the difference between the MAP and ABIC. 

For $H(x;\xi)=\delta(x-\xi)$, 
the observation equation eq.~(\ref{eq:obseq}) is reduced to
\begin{equation}
d(x_n)= a(x_n)+ e(x_n).
\label{eq:examplecase}
\end{equation}
The inference is performed in an interval $0<x<L$, where the true displacement $a_0$ is given by certain functions (specified later), 
and observation locations $x_n$ are randomly selected from a uniform distribution over $0<x<L$ in an uncorrelated manner. 
Observation errors $e(x_n)$ are assumed to follow a Gaussian distribution as ${\bf e}\sim \mathcal N(0,\sigma^2_0 {\bf E})$. 
We take $N=100$, $L=100$, $\sigma_0=0.15$ ($\sigma^2_0=0.0225$) and ${\bf E}={\bf I}$ (the unit matrix) in the following synthetic tests. 

We discretise the problem by expanding the model-parameter field $a(\xi)$ by the normalised
cubic B-spline function with an equally spaced local support $X_m(\xi)$ 
centred at 
$\xi_m:=(m-1)\Delta \xi$, where $\Delta \xi:=L/M$.
Substituting $H(x;\xi)=\delta(x-\xi)$ into eq.~(\ref{eq:defofdiscretekernel}), 
we obtain the $n,m$ entry $H_{nm}$ of the discretised integral kernel ${\bf H}$ as
\begin{equation}
H_{nm}=X_m(x_n).
\end{equation} 
The data distribution $P({\bf d}|{\bf a},\sigma^2)=\mathcal N({\bf H a},\sigma^2 {\bf E})$ (eq.~\ref{eq:obseqprob}) is then obtained.

We also impose the Laplacian smoothing as the prior of ${\bf a}$,  
which regulates the spatial integral of the second derivative $a^{\prime\prime}$ of the model-parameter field~\citep{inoue1986least,yabuki1992geodetic}  
and sets $V=\int d\xi [a^{\prime\prime}(\xi)]^2/2$ in eq.~(\ref{eq:conjpriorobs}). 
Discretising $V$ by eq.~(\ref{eq:expansionruleofa}), 
we have
the prior $P({\bf a}|\rho^2)=\mathcal N(0,\rho^2 {\bf G})$ 
with 
the following $n,m$ entries $G_{nm}$ of ${\bf G}$: 
\begin{equation}
    G_{nm}=\int d\xi\frac{d^2 X_n(\xi)}{d \xi^2}\frac{d^2 X_m(\xi)}{d \xi^2}.
    \label{eq:defofG_ex}
\end{equation}
Hereafter, we normalise $r$ and $G$ with respect to $\xi$ by the grid size $\Delta \xi$. 
Besides, we use the uniform hyperprior of $\sigma^2$ and $\rho^2$ (eq.~\ref{eq:noninfohyperprior}).

\subsection{Measures and a benchmark for evaluating the estimates}
As shown in \S\ref{sec:analyticsolution}, 
the model-parameter estimates are the same regularised least squares ${\bf a}_*$ for both ABIC 
(eq.~\ref{eq:ABICmodel})
and the MAP 
(eq.~\ref{eq:MAPmodel})
apart from the difference in their optimal estimates of $\alpha^2$. 
To evaluate the goodness of these estimates, 
we define the following measures and a benchmark.

The inferred field $a(\xi)$ is expected to be close to the given true field $a_0(\xi)$. 
Hence, we introduce the squared misfit between the true field $a_0(\xi)$ and the estimated one $\hat a(\xi)$ as a measure to evaluate the estimates.
We refer to it as true misfit on sources (TMS):
\begin{equation}
\mbox{TMS}
=
\int d\xi [a_0(\xi)-\hat a(\xi)]^2, 
\label{eq:defofTMS}
\end{equation}
where $\hat a(\xi):=\sum^{M}_{m=1}\hat a_mX_m(\xi)$, and $\hat a_j$ is the $j$-th component of the optimal model-parameter vector $\hat {\bf a}$, given by eq.~(\ref{eq:ABICmodel}) for ABIC and by eq.~(\ref{eq:MAPmodel}) for the MAP. 
Note the TMS can be measurable only in the synthetic tests because the true solution is unknown in actual inverse problems. 

Another measure more directly related to the observed data is the squared misfit of the data estimate ${\bf H}\hat{\bf a}$ from the synthetic data ${\bf d}$, here called ``data misfit'' (DM):
\begin{equation}
\mbox{DM}=
({\bf d}-{\bf H}\hat {\bf a})^{\rm T}{\bf E}^{-1}({\bf d}-{\bf H}\hat {\bf a}).
\label{eq:defofDataMisfit}
\end{equation}
This is exactly the minimisation function of the least-square method. 
However, as the data ${\bf d}$ contains the observation errors ${\bf e}$, minimising the data misfit does not necessarily minimise the misfit from the true solution (e.g. overfitting may occur). 
Original expectation to ${\bf H}\hat{\bf a}$ would be the proximity to ${\bf d}_0:={\bf d}-{\bf e}$ rather than to ${\bf d}$. 
We then define the squared misfit of ${\bf H}\hat{\bf a}$ from ${\bf d}_0$ as true misfit on receivers (TMR): 
\begin{equation}
\mbox{TMR}=
({\bf d}_0-{\bf H}\hat{\bf a})^{\rm T}{\bf E}^{-1}({\bf d}_0-{\bf H}\hat {\bf a}). 
\label{eq:defofTMR}
\end{equation}
The TMR represents only the difference between $\hat a(\xi)$ and $a_0(\xi)$ at the data points, and therefore the TMR is generally not equivalent to the TMS. 
We can relate the TMR to statistical quantities (Supplement 2), such as the cross entropy common in optimisations~\citep[e.g.][]{friedman2001elements}. 

The optimal estimates $\hat \sigma^2$ of $\sigma^2$ are also expected to be close to the true value $\sigma^2_0$ that generates the observation error $e(x_n)$. 
Although discrete inversion 
evaluates $\sigma^2_0$ in the continuous space plus the discretisation error as $\sigma^2$, 
the discretisation error is negligible for a sufficiently fine grid compared to the characteristic length $l$ of the true model-parameter field ($\Delta \xi\ll l$). 

The other hyperparameter, $\rho^2$ (or $\alpha^2$) does not possess the true value in this synthetic test, where the prior of the model parameters is not related to the generating process of ${\bf a}_0$. 
A deterministic ${\bf a}_0$ generator (adopted in this study, given in the next subsection) models practical cases where the absence of the true $\rho^2$ is common for regularisation techniques and priors such that ${\bf a}\sim \mathcal N(0,\rho^2 {\bf G})$, although it is also technically possible in a synthetic test to generate ${\bf a}_0$ probabilistically from the prior. 
Even though there is no true $\alpha^2$, 
we can define a reference value of $\alpha^2$ in the present case where 
the MAP and ABIC estimates are expressed as the regularised least-square solution ${\bf a}_*$ that are fully determined by the $\alpha^2$ values. 
We introduce a benchmark for the $\alpha^2$ estimates, $\alpha_{\rm min.~TMR}^2$, such that the regularised least-square solution minimises the TMR: 
\begin{equation}
\alpha^2_{\rm min.~TMR}:=\mbox{argmin}_{\alpha^2}
\{
[{\bf d}_0-{\bf H}{\bf a}_*(\alpha^2)]^{\rm T}{\bf E}^{-1}[{\bf d}_0-{\bf H}{\bf a}_*(\alpha^2)]
\}.
\label{eq:defofbenchmarkalpha}
\end{equation}
In the synthetic tests, 
we can consider $\alpha^2_{\rm min.~TMR}$ an ideal $\alpha^2$ value in the discrete inverse problem, by numerically solving eq.~(\ref{eq:defofbenchmarkalpha}) in the same manner as for $\alpha^2$ of the ABIC and MAP estimates, although $\alpha^2_{\rm min.~TMR}$ does not necessarily minimise the TMS in the continuous field. 
%
Note
${\bf H a}_*$ does not completely fit to ${\bf d}_0$ such that TMR $=0$ 
even if we use this ideal $\alpha^2_{\rm min.~TMR}$, since ${\bf a}_*$ is inferred from data ${\bf d}$ containing observation errors ${\bf e}$ (eq.~\ref{eq:leastsquarewithpenalty}) and is 
affected by ${\bf e}$. 
It should also be noted that $\alpha^2_{\rm min.~TMR}$ can be an observable only in a synthetic test since true ${\bf d}_0$ is an unknown in practice. 

\subsection{Results}

The following synthetic tests treat two kinds of model parameter fields: a long-wavelength cosine curve and a long-wavelength exponential plus a short-wavelength sinusoid. 
Each of these two kinds of model-parameter fields generates 10 data sets. 
Each data set may have different observation locations ($x_n$).
We conduct the inversion with varying the number of model parameters $M$, while the number of data $N$ is the same for all the data sets.
We discuss statistical characteristics of the reductions by
averaging the results over 10 data sets when necessary. 
We are directed toward the continuous limit of large $M$. 
We will see later that the behaviour of the estimates depending on $M$ is what highlights the difference between the MAP and ABIC. 

The grid search is performed for determining $\alpha^2$ ($\hat \alpha^2_{\rm ABIC}$, $\hat \alpha^2_{\rm MAP}$ and $\alpha^2_{\rm min.~TMR}$) on a closed interval $\alpha^2\in [10^{-4}:10^4]$. This corresponds to recasting a uniform hyperprior eq.~(\ref{eq:noninfohyperprior}) for the closed interval. 
Note the preceding maximisation functions of $\alpha^2$ for the MAP and ABIC estimates (eqs.~\ref{eq:ABICalpha} and \ref{eq:MAPalphadeterminationcondition}) are applicable to finite $\alpha^2$ intervals without any correction. 
Utilising the bounded range of $\alpha^2$ is to grasp the after-mentioned multimodality of the joint posterior, and we later discuss the infinite interval of $\alpha^2$. 
Obviously inappropriate overfitted ($\alpha^2\to0$) and underfitted (oversmoothed, $\alpha^2\to\infty$) solutions are thereby excluded from the results in this section in advance.

\subsubsection{Estimation of model parameters for a sinusoidal displacement field}
The first example is the following normalised sinusoidal model-parameter field: 
\begin{equation}
a_0(x)=\cos(2\pi x/l),
\label{eq:inversionex1}
\end{equation}
where 
$l$ is the characteristic wavelength, taken to be 50 ($l/L=1/2$). 

\begin{figure*}
	\includegraphics[width=152mm]{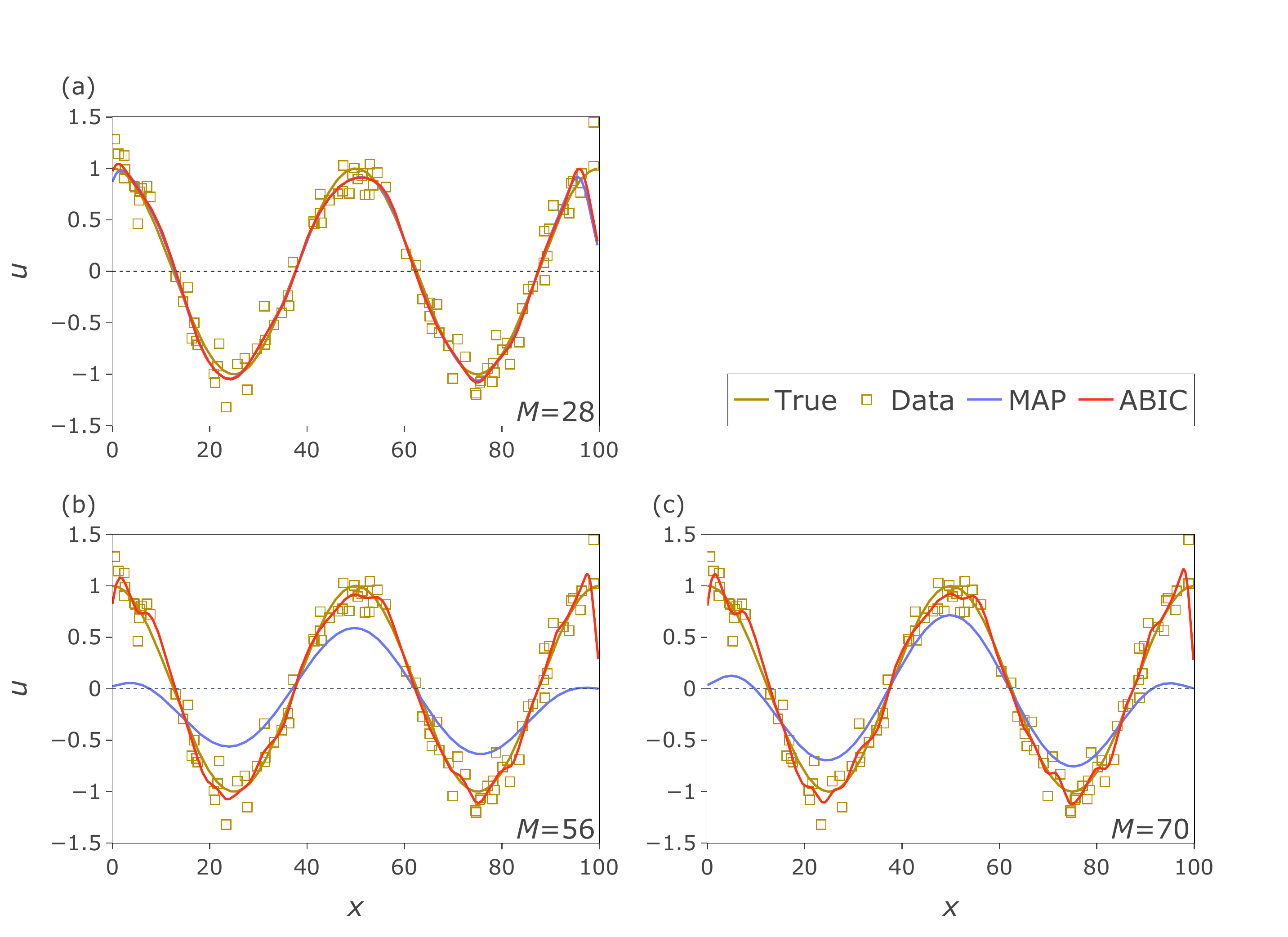}
 \caption{
Estimation examples for a sinusoidal displacement field. 
Synthetic data (yellow squares) are generated by
adding random noise to the true solution (yellow lines), which is inferred by the MAP (blue) and ABIC (red). 
The number of model parameters $M$ varies as 
28 (a), 56 (b) and 70 (c). 
The number of data $N$ is fixed to 100.
 }
\label{fig:demo}
\end{figure*}

Figure~\ref{fig:demo} illustrates examples of displacement inference, which are obtained for a particular synthetic data set with random noise. 
The estimated displacement fields reproduce the true displacement field well for both the MAP and ABIC, when the number of model parameters $M$ is relatively smaller compared to the number of data $N(=100$) [$M=28$, Fig.~\ref{fig:demo}(a)]. 
The difference is hardly observed between the MAP and ABIC in Fig.~\ref{fig:demo}(a), consistent with the report by \citet{fukuda2008fully} that the MAP and ABIC present similar results. 
However, 
the MAP estimate oversmooths 
for relatively large $M$ [$M=56$, Fig.~\ref{fig:demo}(b)]. 
This trend is maintained for even larger $M$ [$M=70$, Fig.~\ref{fig:demo}(c)]. 
By contrast, the ABIC estimate consistently well reproduces the true displacement field. 

\begin{figure}
	\includegraphics[width=90mm]{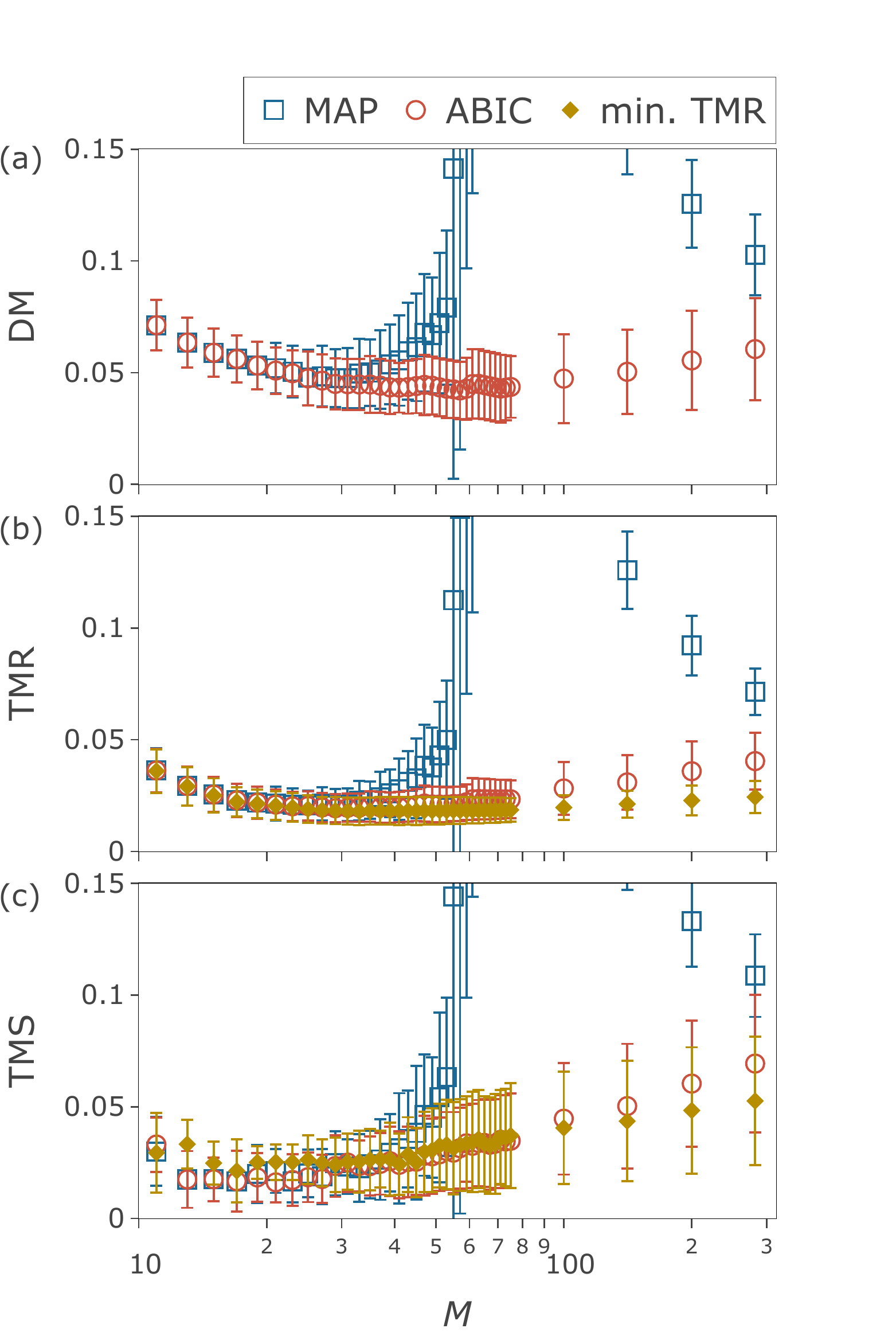} \caption{
Measures of misfits, 
DM (a), TMR (b) and TMS (c), defined in eqs.~(\ref{eq:defofTMS})--(\ref{eq:defofTMR}), plotted as functions of the number of model parameters $M$. 
The TMR and TMS of the min.~TMR estimate are evaluated with the regularised least-square estimate using $\alpha^2_{\rm min.~TMR}$ (eq.~\ref{eq:defofbenchmarkalpha}).
The true displacement field is given by $a_0(x)=\cos(2\pi x/l)$ with $l=50$, and random noise is added to generate synthetic data of the $N(=100)$ components. 
The mean and standard deviation of inversion results for 10 different data sets (one of which is drawn in Fig.~\ref{fig:demo}) are indicated for each estimate. 
The horizontal axis is taken on the log scale. 
 }
\label{fig:misfit}
\end{figure}

Since the results for each data set (e.g. drawn in Fig.~\ref{fig:demo}) are affected by the added random noise, 
in the following, 
we average the results over 10 different data sets and further study the systematic $M$-dependence of the estimates. 
Figure~\ref{fig:misfit} displays the measures of misfit, 
DM, TMR and TMS (eqs.~\ref{eq:defofTMS}-\ref{eq:defofTMR}), normalised by
${\bf d}^{\rm T}{\bf E}^{-1}{\bf d}$,
$({\bf Ha}_0)^{\rm T}{\bf E}^{-1}({\bf H a}_0)$ and
$\int d\xi a_0^2$, respectively. 
Here min.~TMR denotes $\alpha^2_{\rm min.~TMR}$ (eq.~\ref{eq:defofbenchmarkalpha}) and the regularised least-square solutions of the model parameters using $\alpha^2_{\rm min.~TMR}$. 
For relatively small $M\lesssim30$, both the MAP and ABIC estimates are close to each other in all the measures (DM, TMR and TMS) and to the min.~TMR estimate. 
For larger $M$, however, 
all the measures of the MAP estimates record systematically larger values than those of ABIC and 
are consistent with the oversmoothed model-parameter estimates of the MAP in Fig.~\ref{fig:demo} for large $M$. 
The misfits of the MAP estimate explosively increase around $M\simeq 55$. 
We can also notice the misfit decrease for even larger $M\gtrsim N$, but we will later explain this characteristic depends on the applied search interval of $\alpha^2$, thus insignificant. 
On the other hand, 
the ABIC estimate is consistently close to the min.~TMR estimate with regard to both the TMS and TMR. 

The min.~TMR estimate works as a benchmark by recording the minimum TMR by definition (Fig.~\ref{fig:misfit}b). 
The TMS values of the MAP and ABIC estimates are frequently below that of the min.~TMR estimate for relatively small $M\lesssim30$, but
the TMS of the min.~TMR estimate is always the minimum within the standard deviation (Fig.~\ref{fig:misfit}c). 
It may also be noted that the TMR of the min.~TMR estimate (i.e. the lower bound of the TMR for the regularised least-square estimates) tends to increase with $M$, although the increase is marginal within our observation. 

\begin{figure}
	\includegraphics[width=90mm]{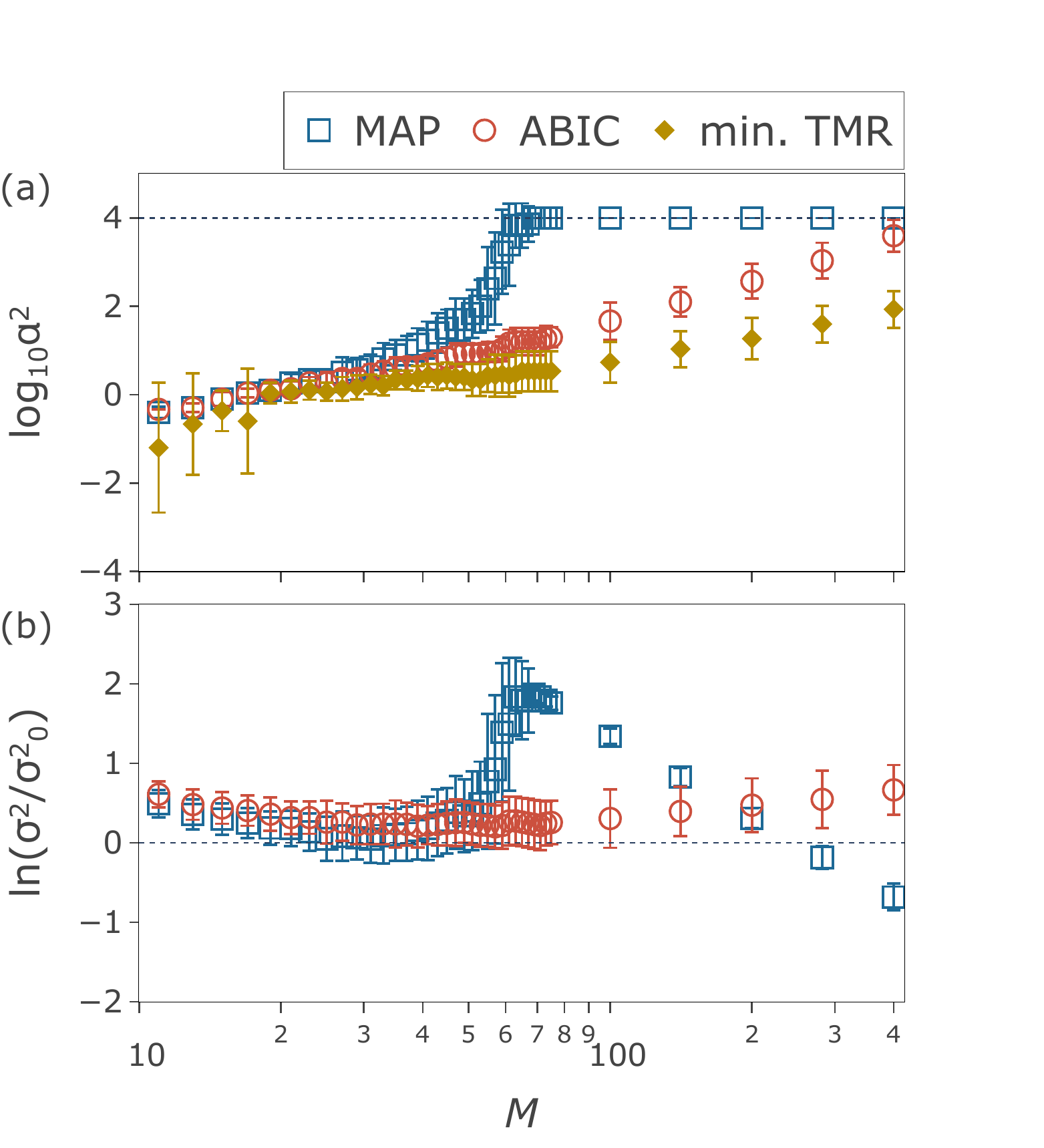}
 \caption{
Inverted $\alpha^2$ (a) and $\sigma^2$ (b) values on the log scale, plotted as functions of the number of model parameters $M$. 
The solved problem and visualisation method are the same as Fig.~\ref{fig:misfit}. 
The broken lines indicate the upper bound of the search interval for $\alpha^2=10^4$ in (a) 
and the true value of $\sigma^2$ in (b).
 }
\label{fig:alphasigma}
\end{figure}

Figure~\ref{fig:alphasigma}(a) shows the optimal $\alpha^2$ values and also captures the oversmoothing tendency of the MAP estimate for relatively large $M$. 
For relatively small $M\lesssim30$, 
the MAP and ABIC estimates return almost the same $\alpha^2$. 
Accompanying the increase in $M$, however, $\alpha^2$ of the MAP gradually becomes larger than that of ABIC and indicates the oversmooth tendency as in Figs.~\ref{fig:demo}(b) and (c). The growth of $\alpha^2$ accelerates as $M$ increases. 
After diverging around $M\simeq55$, $\alpha^2$ of the MAP estimate stays on the upper limit of the search interval. 
Meanwhile, although considerably smaller than that of the MAP, 
the $\alpha^2$ values of the ABIC and min.~TMR estimates also increase with $M$, then slowly approaching to the upper limit of the search interval and thus $\alpha^2$ of the MAP for relatively large $M$. 
In brief, the aforementioned misfit decrease of the MAP estimate for large $M\gtrsim N$ (Fig.~\ref{fig:misfit}) can be ascribed to the limited search interval of $\alpha^2$. 
If the search interval of $\alpha^2$ is not narrowed, there is no misfit decrease of the MAP estimate for large $M\gtrsim N$. 
Even worse, in this case, the oversmooth solution is always selected as the global minimum solution for the MAP estimation. This problem is described in \S\ref{sec:erasingoutsiders}. 

Figure~\ref{fig:alphasigma}(b) plots the $\sigma^2$ estimates normalised by the true value $\sigma^2_0$. 
For relatively small $M\lesssim 30$, where the $\alpha^2$ estimates are nearly identical between the MAP and ABIC (Fig.~\ref{fig:alphasigma}a), both the MAP and ABIC produce the $\sigma^2$ values that are overvalued but approximately coincide with the true value $\sigma^2_0$ within 30--50\% accuracy, and approach to $\sigma^2_0$ with an increase in $M$. 
For $30\lesssim M\lesssim 55$, however, $\sigma^2$ of the MAP estimate explosively increases, while initially slightly closer to $\sigma^2_0$ than that of the ABIC estimate. 
The sharp increase of $\sigma^2$ corresponds to the oversmoothed solution in Figs.~\ref{fig:demo}(b) and (c), where the estimated $\alpha^2$ value also bursts (Fig.~\ref{fig:alphasigma}a). 
Although we also observe $\sigma^2$ of the MAP estimate decreases for even larger $M\gtrsim70$, it is caused by the limitation of the search interval of $\alpha^2$ previously explained, and 
$\sigma^2$ of the MAP estimate also becomes larger when larger $\alpha^2$ is allowed. 
In contrast, although $\sigma^2$ of the ABIC estimate slightly increases with $M\gtrsim70$ as its TMR (Fig.~\ref{fig:misfit}b), 
it roughly agrees with the true value over the entire investigated range of $M$.

\subsubsection{Estimation of model parameters for a mixed-wavelength field}
In the previous example of a simple sinusoidal displacement field eq.~(\ref{eq:inversionex1}), we confirmed that the oversmoothed solution was obtained as the MAP estimate for a large number of model parameters $M$, while the ABIC estimate stably infers reasonable solutions even for such large $M$ (Figs.~\ref{fig:demo}-\ref{fig:alphasigma}). As seen in this example, the MAP estimate has a serious defect, but readers may also notice that we can obtain a reasonably good solution for a smaller number of $M$, as shown in Fig.~\ref{fig:demo}(a), even if we use the MAP estimate. However, this is not the case always. The previous example may be too simple.

In this subsection, we consider a displacement field with mixed wavelength components: exponential decrease with a sinusoidal perturbation. We specifically treat the following functional form:
\begin{equation}
a_0(x)=e^{-(x/l)^2}+\frac 1 4 \cos\left( \frac{2\pi x}{l/4}\right),
\label{eq:secondex}
\end{equation}
where $l$ is 50. 
The first term contains long-wavelength components in a wide wavenumber range, while the second term does a single short-wavelength Fourier component. 

\begin{figure*}
	\includegraphics[width=165mm]{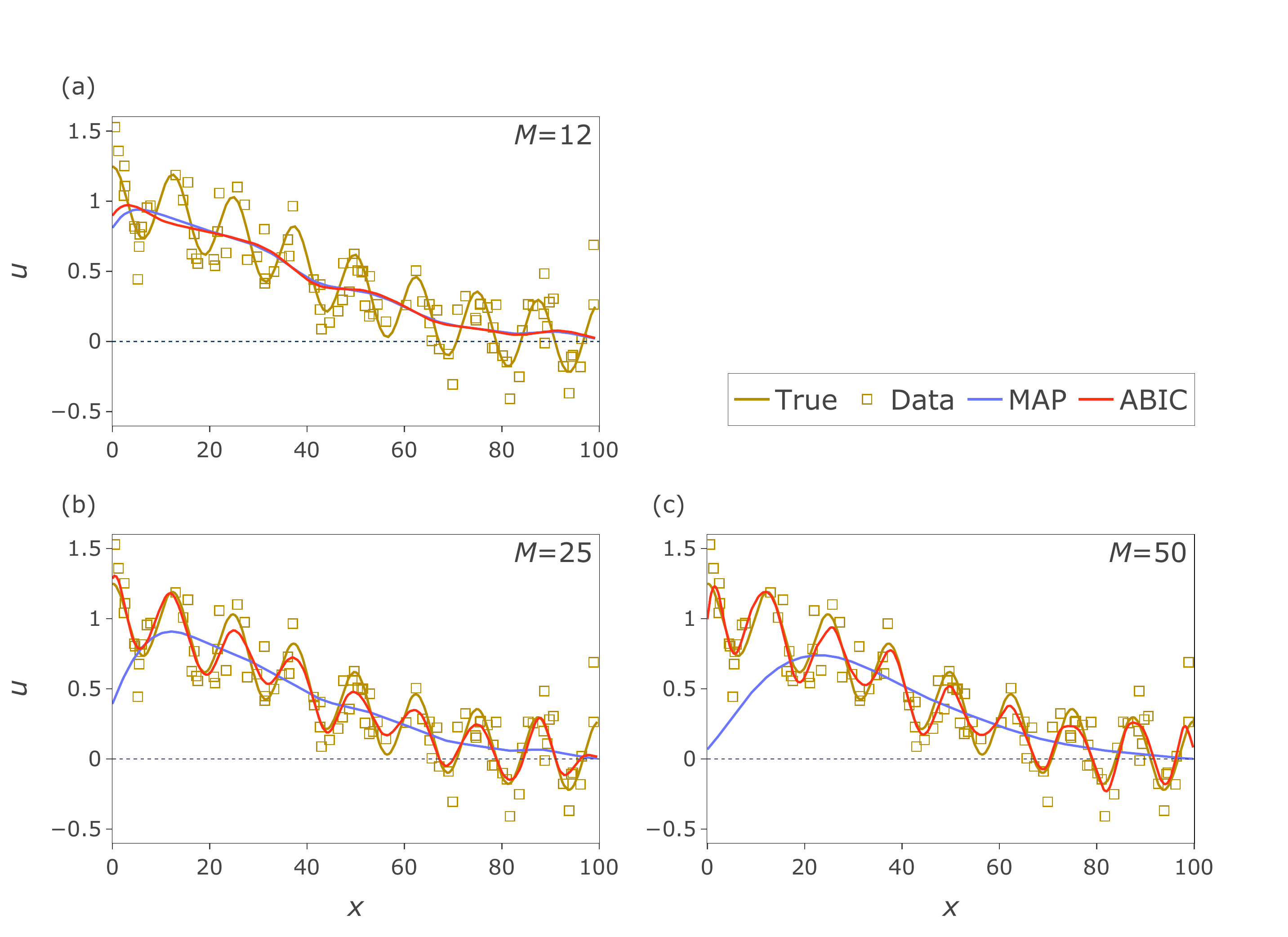}
 \caption{
Estimation examples for a mixed-wavelength field (eq.~\ref{eq:secondex}). 
Synthetic data (yellow squares) generated as
random noise plus the true field (yellow lines) are inverted by the MAP (blue) and ABIC (red). 
The number of model parameters $M$ varies as 
12 (a), 25 (b) and 50 (c). 
The number of data $N$ is fixed to 100.
 }
\label{fig:demo2}
\end{figure*}

Figure~\ref{fig:demo2} draws estimation examples for different $M$. 
When $M=12$ (Fig.~\ref{fig:demo2}a), 
both the MAP and ABIC estimates reproduce only the long-wavelength term of $\exp[-(x/l)^2]$, ascribed to the lack of degrees of freedom to reproduce the short-wavelength variation. 
Even for larger numbers of model parameters $M=25$ and $50$ (Figs.~\ref{fig:demo2}b and c), however, 
the MAP estimate still neglects the short-wavelength character. 
This is considered another instance of the oversmooth tendency of the MAP estimate. By contrast, the ABIC estimate is successful as reproduces both the exponential decrease and sinusoidal oscillation for $M = 25$ and $50$. It also means $M = 25$ and $50$ are enough degrees of freedom to recover the short-wavelength sinusoid and confirms the MAP estimate for $M = 25$ and $50$ is actually oversmooth with large $\alpha^2$. 
Incidentally, remarkable poor fitting of the MAP estimate around $x\sim0$ is not solely attributed to the oversmoothness for large $M$, but also to the boundary condition; nonzero displacement is now allowed only in the model regions covered by the basis functions, so the displacement is forced to be zero just outside of it (at $x=-2\Delta x,L+\Delta x$).

\begin{figure}
	\includegraphics[width=90mm]{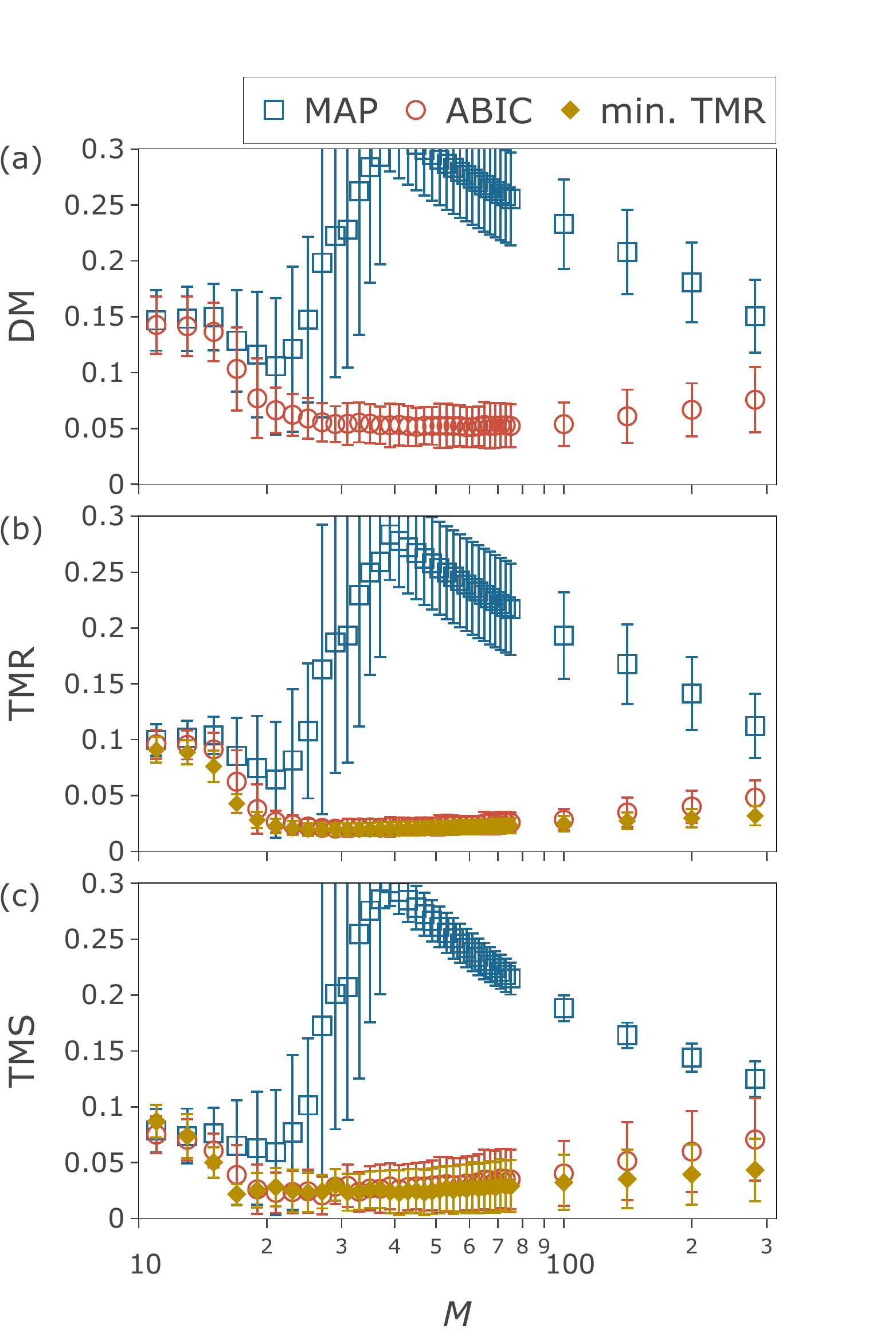}
 \caption{
Measures of misfits,
DM (a), TMR (b) and TMS (c), defined in eqs.~(\ref{eq:defofTMS})--(\ref{eq:defofTMR}), plotted as functions of the number of model parameters $M$. 
The TMR and TMS of the min.~TMR estimate are evaluated with the regularised least-square estimate using $\alpha^2_{\rm min.~TMR}$ (eq.~\ref{eq:defofbenchmarkalpha}).
The true displacement field is given by $a_0(x)=\exp[-(x/l)^2]+(1/4)\cos[2\pi x/(l/4)]$, and random noise is added to generate synthetic data of the $N(=100)$ components. 
The mean and standard deviation of inversion results for 10 different data sets (one of which is drawn in Fig.~\ref{fig:demo2}) are indicated for each estimate. 
The horizontal axis is taken on the log scale. 
 }
\label{fig:misfit_cosexp}
\end{figure}

\begin{figure}
	\includegraphics[width=90mm]{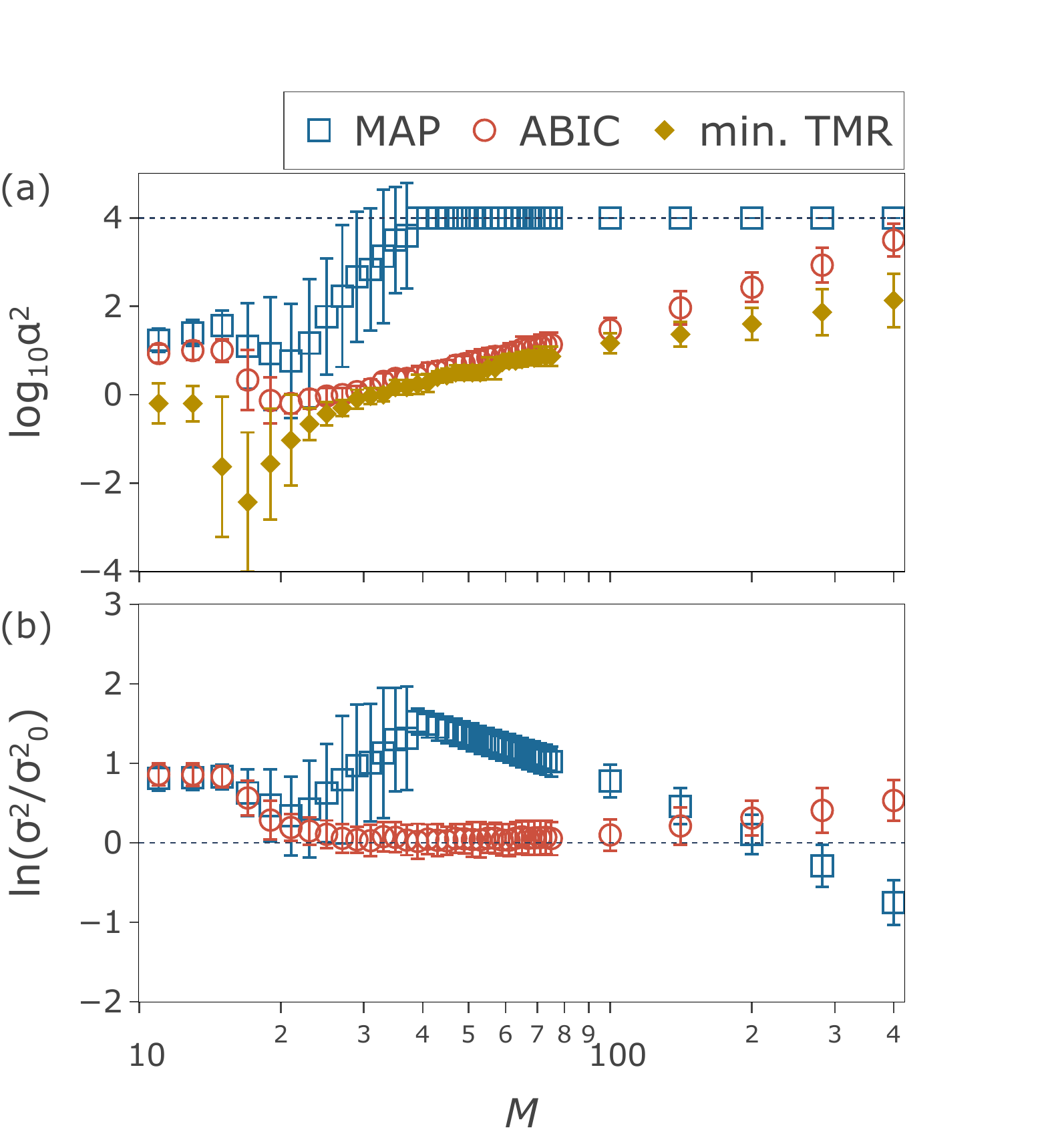}
 \caption{
Inverted $\alpha^2$ (a) and $\sigma^2$ (b) values on the log scale, plotted as functions of the number of model parameters $M$. 
The solved problem and visualisation method are the same as Fig.~\ref{fig:misfit_cosexp}. 
The broken lines indicate the upper bound of the search interval for $\alpha^2=10^4$ in (a) and the true value of $\sigma^2$ in (b). 
}
\label{fig:alphasigma_cosexp}
\end{figure}

Figure~\ref{fig:misfit_cosexp} shows the measures of misfit (DM, TMR and TMS,  normalised as in Fig.~\ref{fig:misfit}). For relatively small numbers of model parameters $M(\lesssim20)$, the misfits decline as $M$ increases for both the MAP and ABIC estimates, although such a trend is less significant in the MAP estimate. 
This accuracy improvement corresponds to the model resolution enhancement accompanying the increase in the number of model parameters. 
However, the misfits of the MAP estimate sharply retrograde for $M \gtrsim 20$ and affirm the oversmoothing tendency. 
The misfits of the ABIC estimate are regulated even for larger $M$. The TMR of the ABIC estimate is nearly identical to the min.~TMR within the whole plotted range of $M$. Incidentally, the misfit decrease in the MAP estimate is observed for very large $M$, but explicable by the same reason as in Fig.~\ref{fig:misfit}: the limitation of the search interval of $\alpha^2$ (see Fig.~\ref{fig:alphasigma_cosexp}).

Figure~\ref{fig:alphasigma_cosexp} shows a trend of the hyperparameter estimates consistent with the measures of misfit (DM, TMR and TMS) in Fig.~\ref{fig:misfit_cosexp}. For the MAP estimate, both $\alpha^2$ and $\sigma^2$ decrease as $M$ increases within $M\lesssim20$, but sharply grows when $M\gtrsim20$. 
Finally $\alpha^2$ reaches the upper bound of the search interval around $M=40$. Afterward ($M\gtrsim40$), $\alpha^2$ always takes the upper limit value, and $\sigma^2$ falls as $M$ increases. The drop in $\sigma^2$ continues even after it crosses the true variance $\sigma^2_0$ $(M\gtrsim300)$. 
As mentioned in the previous subsection, if we allow larger $\alpha^2$ by widening its search interval, we see larger $\sigma^2$ as well as larger misfits (DM, TMR and TMS) in the MAP estimates. 
Meanwhile, $\alpha^2$ of the ABIC estimate coincides with that of the min.~TMR within the error bars in the whole plotted range of $M$, and $\sigma^2$ of the ABIC estimate is consistently the same order of the true value $\sigma^2_0$.

We have investigated the inversion of a mixed-wavelength field eq.~(\ref{eq:secondex}). 
The MAP estimate reproduced only the long-wavelength pattern with large misfits (Fig.~\ref{fig:demo2}). 
For relatively larger $M\gtrsim20$, the $\alpha^2$ estimate of the MAP was excessively larger than the benchmark value of the min.~TMR estimate (Fig.~\ref{fig:alphasigma_cosexp}a), indicating the underfitting tendency of the MAP estimate. 
For relatively small $M\lesssim20$, where such a problem is not seen in the $\alpha^2$ value of the MAP estimate, there seem no ways to reproduce the short-wavelength part, as suggested from a rapid misfit decrease of the min.~TMR estimate around $M\sim 20$ (Fig.~\ref{fig:misfit_cosexp}). 
These indicate difficulties in using MAP estimates to invert complex true model-parameter fields 
with a large number of discrete model parameters.
In contrast, the ABIC estimate stably and reasonably reproduced the short-wavelength sinusoid as well with large $M$.

\section{Application}
\label{sec:applicationsec}
Our synthetic tests suggest counter-intuitive behaviours of the joint posterior in the fully Bayesian inversion for the case of a large number of model parameters, which corresponds to a high-resolution inversion with fine discretisation grids. 
In this section, 
we conduct a similar analysis with actual observed data and address how the discovered issue appears in a practical inverse problem. 
%

We solve an inverse problem of estimating the displacement-velocity field from spatially discrete GNSS data with the Laplacian smoothing constraint. 
This problem is a two-dimensional generalisation of our synthetic problem [$x\in (0,L)\to (x_1,x_2)\in (0,L)^2$ in eq.~\ref{eq:examplecase} and $\xi\in (0,L)\to (\xi_1,\xi_2)\in (0,L)^2$ in eq.~\ref{eq:defofG_ex}]. 
The studied area is central Japan, 
which indicates a relatively rougher spatial deformation pattern~\citep{sagiya2000continuous}, suitable to investigate the smoothing tendency (prior reliance) of estimates we are concerned with. 
We analyse the GNSS data of the GEONET archived by the Geospatial Information Authority of Japan (GSI) in 136$^\circ$--141$^\circ$E and 33$^\circ$--37$^\circ$N with 286 stations. 
We use the daily coordinates of the F3 solution~\citep{nakagawa2009development} provided by GSI, based on
International Terrestrial Reference Frame (ITRF) 2005~\citep{altamimi2007itrf2005}. The data period is from January 2006 to December 2009 including relatively fewer large earthquakes. 
The daily coordinate series of each component at each station
is fitted in a conventional least-square routine~\citep[][]{sagiya2000continuous} by a linear trend, annual and semiannual sinusoidal components and offsets related to coseismic deformation of large earthquakes ($M_j\geq 6$) and equipment maintenance catalogued by GSI. 
\citet{nozue2022comparison} employed the same processed data though analysed a wider region data. 
The fitted $286\times2$ linear trends represent velocities at respective stations, which set the number of data as $N=572$. 

The model-parameter field is a two-dimensional displacement-velocity field discretised by the cubic B-spline function. 
We span the coordinate space $\boldsymbol\xi\in(0,L)^2$ of $L=480$km, to which the model-parameter field belongs, the centre of which is placed at the centre of the observational region 138.5$^\circ$E and 35$^\circ$N. 
The problem is solved by the MAP and ABIC semianalytically with varying the number of basis functions (the number of model parameters, $M$). 
The associated $\alpha^2$ search is performed within a closed interval $\alpha^2\in [10^{-5}:10^5]$ slightly wider than that of the synthetic test ($\alpha^2\in [10^{-4}:10^4]$), later generalised to the unbounded $\alpha^2$ range. 
Unlike our synthetic case of the rigid-boundary condition ($u=0$), the present application treats an unfixed-boundary problem by following~\citet{okazaki2021consistent}, which space the basis functions in $(\xi_1,\xi_2)\in (-3\Delta\xi,L+3\Delta\xi)^2$ with a grid size $\Delta \xi$ and truncate them to $(\xi_1,\xi_2)\in (0,L)^2$; 
the basis functions near the edges are expressed by the products of the B-spline and step functions. 
It defines the number of model parameters as $M:=2(L/\Delta \xi+3)^2$.

\begin{figure*}
    \includegraphics[width=132mm]{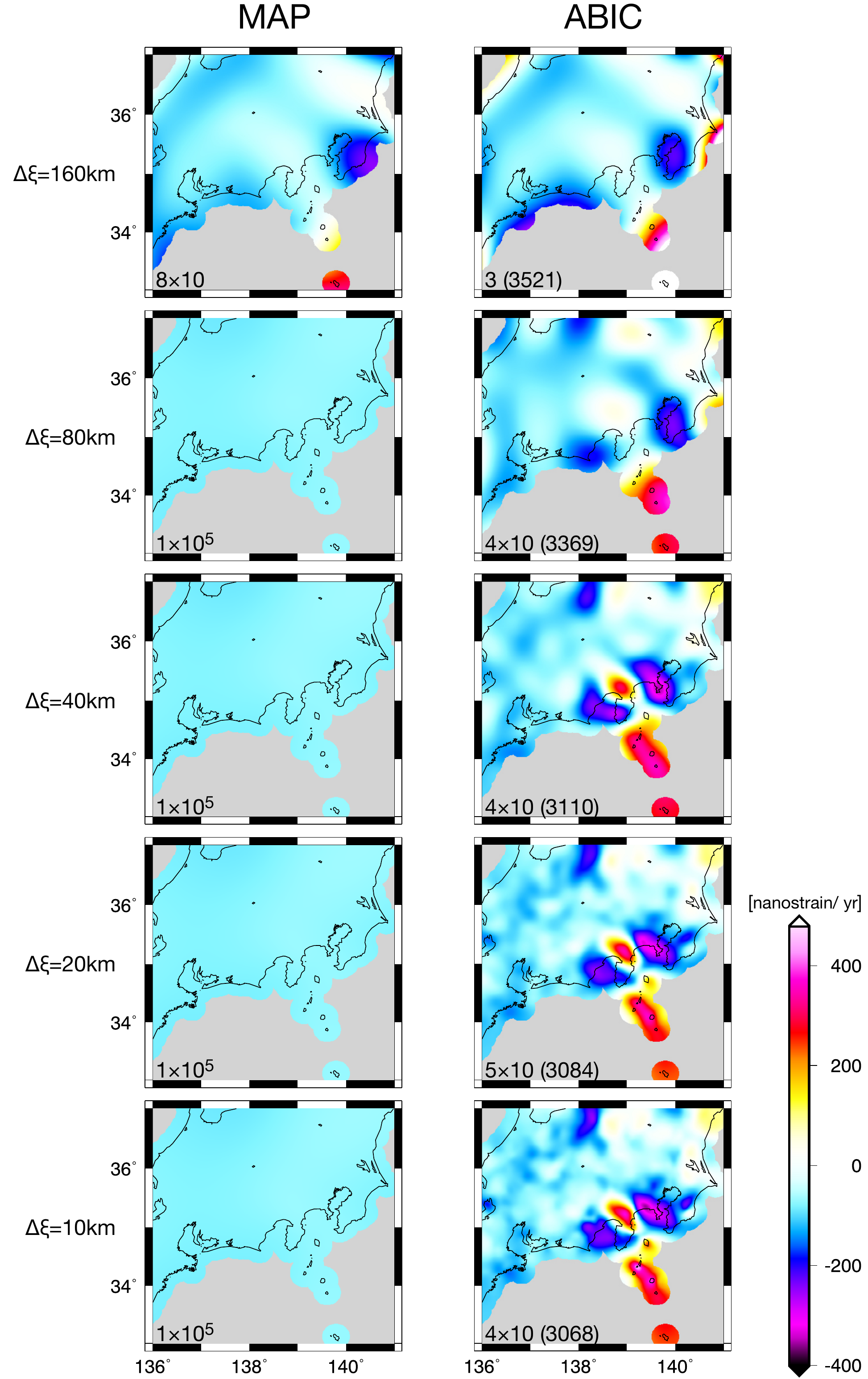} 
 \caption{
Estimated horizontal dilatation rate (the trace of the differentiated model-parameter field) in central Japan, obtained from GNSS data by the MAP (left) and ABIC (right). 
The grid size $\Delta \xi$ of the discretised model-parameter field varies as 
160, 80, 40, 20 and 10 in kilometer. 
The associated numbers of model parameters $M$ are 
72, 162, 450, 1458 and 5202, respectively. 
The data comprises 572 components that express the average north--south and east--west displacement rates of the 286 stations within 136--141$^\circ$E and 33--37$^\circ$N from January 2006 to December 2009. 
The number in each panel represents the selected $\alpha^2$ value. The parentheses for the ABIC estimates include the values of the minimisation function of ABIC regarding $M$ as another hyperparameter. 
}
\label{fig:App_dilatationrate}
\end{figure*}

Figure~\ref{fig:App_dilatationrate} compares the MAP and ABIC estimates for 
the grid size $\Delta \xi=160,80,40,20,10$km ($M=72,162,450,1458,5202$). 
We plot the trace of the first derivative of the estimated displacement rate, which is the estimated horizontal dilatation rate. 
The number in each panel is the selected $\alpha^2$ value, 
and the parentheses in the right column include the values of the minimisation function of ABIC in eq.~\ref{eq:ABICalpha} plus $\ln|\boldsymbol \Lambda_G|$, 
which is the minimisation function of ABIC when the number of model parameters $M$ is also a hyperparameter of a uniform hyperprior [obtained from eq.~(\ref{eq:lnmarginalposofhyperforlinearinverse}), considering the $M$-dependence of the normalisation function in the prior of the model parameters]. 
The MAP estimate for $\Delta \xi=160$km generates a long-wavelength pattern, similar to the ABIC estimates of the same grid size. 
However, the estimated dilatation rate fields are entirely smooth for the MAP with smaller $\Delta \xi$($=80, 40, 20, 10$km), 
despite that the estimates are generally expected to resolve more details with denser basis functions. 
It is followed by a sudden increase in the $\alpha^2$ MAP estimate, from $\alpha^2=10^{23/12}(\sim 8\times 10)$ ($\Delta \xi=160$km) to $\alpha^2=10^5$ ($\Delta \xi=80, 40, 20, 10$km), 
which is the upper limit of the search range. 
Hence, we observe the same characteristics as in the synthetic tests: the oversmoothness of the MAP estimates for large $M$. 
Since the boundary values are not forced to be zero in this problem setting, 
the oversmooth solution is here a linearly varying field with a constant first derivative and the zero second derivative, resulting in a constant dilatation-rate field. 
In contrast, the ABIC estimate generates shorter-wavelength patterns on finer grids with larger numbers of model parameters. The spatial pattern of the ABIC estimate is totally rougher than that of the MAP estimate 
and resolves a high-strain-rate zone in the back arc, called the Niigata-Kobe tectonic zone~\citep{sagiya2000continuous}, a low-strain-rate zone in the fore arc~\citep{okazaki2021consistent}, a high-strain-rate area along the Pacific coast, that is related to the collision of the Izu-Bonin arc~\citep{matsuda1978collision} and subduction of the Philippine Sea Plate, and a high-expansion-rate
area in the Izu island chain~\citep{nishimura2011back}. 

We also observe convergent spatial patterns of ABIC accompanying grid-size decrease. 
The obtained strain-rate fields are almost identical between $\Delta \xi=20$ and $10$km. 
Accordingly, a convergent decrease is also seen in the minimisation function of ABIC regarding the number of model parameters $M$ as another hyperparameter (the parentheses of Fig.~\ref{fig:App_dilatationrate}). 
It is considered that the ABIC estimate with $\Delta \xi\sim 20$km almost reaches the upper bound of resolution determined by the data information. 

This actual data analysis exhibits contrastive natures of the MAP and ABIC in parallel with our synthetic tests, 
demonstrating the increase in the number of basis functions results in 
an excessively smooth field of the MAP, while provides more details by ABIC, with a plateau of the resolution implying the resolution upper bound specified by the data. 
The reduction requires special attention to gain a reasonable resolution of the model-parameter field for the best use of observational data.

\section{Statistical properties of the joint posterior and marginal posterior of hyperparameters in fully Bayesian inversions}
The synthetic tests and geophysical application clarified a systematic oversmoothing (underfitting) tendency of the MAP estimate for the case of a large number of model parameters. 
The same problem is asymptotically expected of other various estimates with finite probabilities in the marginal posterior of the model parameters, given our analytic result eq.~(\ref{eq:deltafunctionalconvergence_modelparameter}). 
The synthetic tests also suggest the true solution is well reproduced by the use of ABIC. 
In this section, theoretical backgrounds to these results are explored. 
In \S\ref{sec:erasingoutsiders}, we determine what in the joint posterior causes the identified problems of the MAP in the fully Bayesian inversion. In \S\ref{sec:intepretationofprob}, we examine how this pathology is resolved by ABIC.

\subsection{Pathology in fully Bayesian joint posteriors}
\label{sec:erasingoutsiders}

Several literatures have pointed out the problem in the MAP estimate 
taking the overfitted solution as its global maximum~\citep[e.g.][]{takane1987relationship}. 
It twins with the problematic global maximality of the underfitted solution encountered in this study, 
and the same logic explain their causes, from the functional form of the joint posterior in the present linear inverse problem (eq.~\ref{eq:posterior4uniformhyperprior}): 
\begin{equation}
P({\bf a},\sigma^2,\rho^2|{\bf d})\propto (\sigma^2)^{-N/2}(\rho^2)^{-P/2}
\exp\left[
-\frac{U}{\sigma^2}-\frac{V}{\rho^2} 
\right].
\label{eq:jointposteriorform}
\end{equation}
Equation~(\ref{eq:jointposteriorform}) expresses
$P({\bf a},\sigma^2,\rho^2|{\bf d})$
diverges to infinity at $\sigma^2=0$ in the limit of the overfit $U=0$ 
and at $\rho^2=0$ in that of the underfit $V=0$. 
Therefore, 
the overfitted or underfitted estimate is selected as the global maximum of the joint posterior if either of them exists. 
The overfitted solution does not exist unless in an underdetermined problem, while the underfitted solution exists as ${\bf a}={\bf 0}$ whenever using zero-mean Gaussian priors, $V({\bf a})={\bf a}^{\rm T}{\bf Ga}/2$ (eq.~\ref{eq:defofV}).

The global probability maximisation strategy is therefore inappropriate for such cases to obtain an appropriate estimate from the joint posterior. 
We then excluded the global maximum of the joint posterior by using the weak hyperprior $P(\sigma^2,\rho^2)=c\theta(4-|\log_{10} \alpha^2|)$ in our synthetic tests, where $\theta(\cdot)$ is the Heaviside step function. 
It corresponds to defining the MAP estimate as a local maximum of the joint posterior because all its multimodalities arise along the $\alpha^2$ axis, or equivalently, the joint posterior is unimodal when $\alpha^2$ is fixed; as shown in \S\ref{sec:analyticsolution}, the joint posterior is Gaussian in terms of ${\bf a}$ with the mean ${\bf a}_*(\alpha^2)$ and has the unique extremum
$\tilde \sigma^2_{\rm MAP}({\bf a},\alpha^2)$ in terms of $\sigma^2$ (eq.~\ref{eq:sigmaatsaddleofposterior}). 
Substituting $\sigma^2=\tilde\sigma^2_{\rm MAP}$ and ${\bf a}={\bf a}_*$ into the joint posterior while fixing $\alpha^2$, 
we have 
\begin{equation}
\begin{aligned}
&P({\bf a},\sigma^2,\rho^2|{\bf d})|_{{\bf a}={\bf a}_*(\alpha^2),\sigma^2=\tilde \sigma^2_{\rm MAP}(\alpha^2),\rho^2=\tilde \sigma^2_{\rm MAP}(\alpha^2)/\alpha^2}
\\
&=c|\boldsymbol \Lambda_G|^{1/2}|{\bf E}|^{-1/2}
e^{-(N+P)/2}
\\&\times
\left[ \frac{2\pi s({\bf a}_*(\alpha^2),\alpha^2)}{N+P}\right]^{-(N+P)/2}
(\alpha^2)^{P/2},
\end{aligned}
\label{eq:saddlejointposterior}
\end{equation}
where $|_{...}$ denotes the substitution.
We abbreviated $\tilde\sigma^2_{\rm MAP}({\bf a}_*(\alpha^2),\alpha^2)$ as $\tilde\sigma^2_{\rm MAP}(\alpha^2)$ for brevity. 

Even defined as such a local maximum using the weak hyperprior, however, 
the MAP estimate still indicated the oversmooth tendency in the synthetic tests. 
This suggests there is another problem in the joint posterior, regarding its local maxima. 
To investigate it, we here analyse the transient behaviour of the joint posterior in terms of the multimodality with increasing $M$. 

The following is the comparative study that also evaluates
the profile of the marginal posterior of the hyperparameters, the multimodality of which is similarly allowed to appear only along the $\alpha^2$ axis; 
substituting
$\tilde\sigma^2_{\rm ABIC}(\alpha^2)$  [the unique extremum with respect to $\sigma^2$ given $\alpha^2$, shown in eq.~(\ref{eq:sigmaatsaddleofposterior})] into the marginal posterior of the hyperparameters while fixing $\alpha^2$, 
we have
\begin{equation}
\begin{aligned}
&
\left.
P(\sigma^2,\rho^2|{\bf d})
\right|_{\sigma^2=\tilde\sigma^2_{\rm ABIC},\rho^2=\tilde\sigma^2_{\rm ABIC}/\alpha^2}
\\&
=c
|\boldsymbol \Lambda_G|^{1/2}|{\bf E}|^{-1/2}e^{-(N+P-M)/2}| {\bf H}^{\rm T}{\bf E}^{-1}{\bf H}+\alpha^2 {\bf G}|^{-1/2}
\\&\times
\left[\frac{2\pi s({\bf a}_*(\alpha^2),\alpha^2)}{N+P-M}\right]^{-(N+P-M)/2}
(\alpha^2)^{P/2}.
\end{aligned}
\label{eq:marginalizedposforhyper}
\end{equation}
Different $\alpha^2$-dependencies arise from
$|{\bf H}^{\rm T}{\bf E}^{-1}{\bf H}+\alpha^2 {\bf G}|^{1/2}$ and $[s({\bf a}_*)]^{M/2}$ in eqs.~(\ref{eq:saddlejointposterior}) and (\ref{eq:marginalizedposforhyper}). 

\begin{figure*}
	\includegraphics[width=158mm]{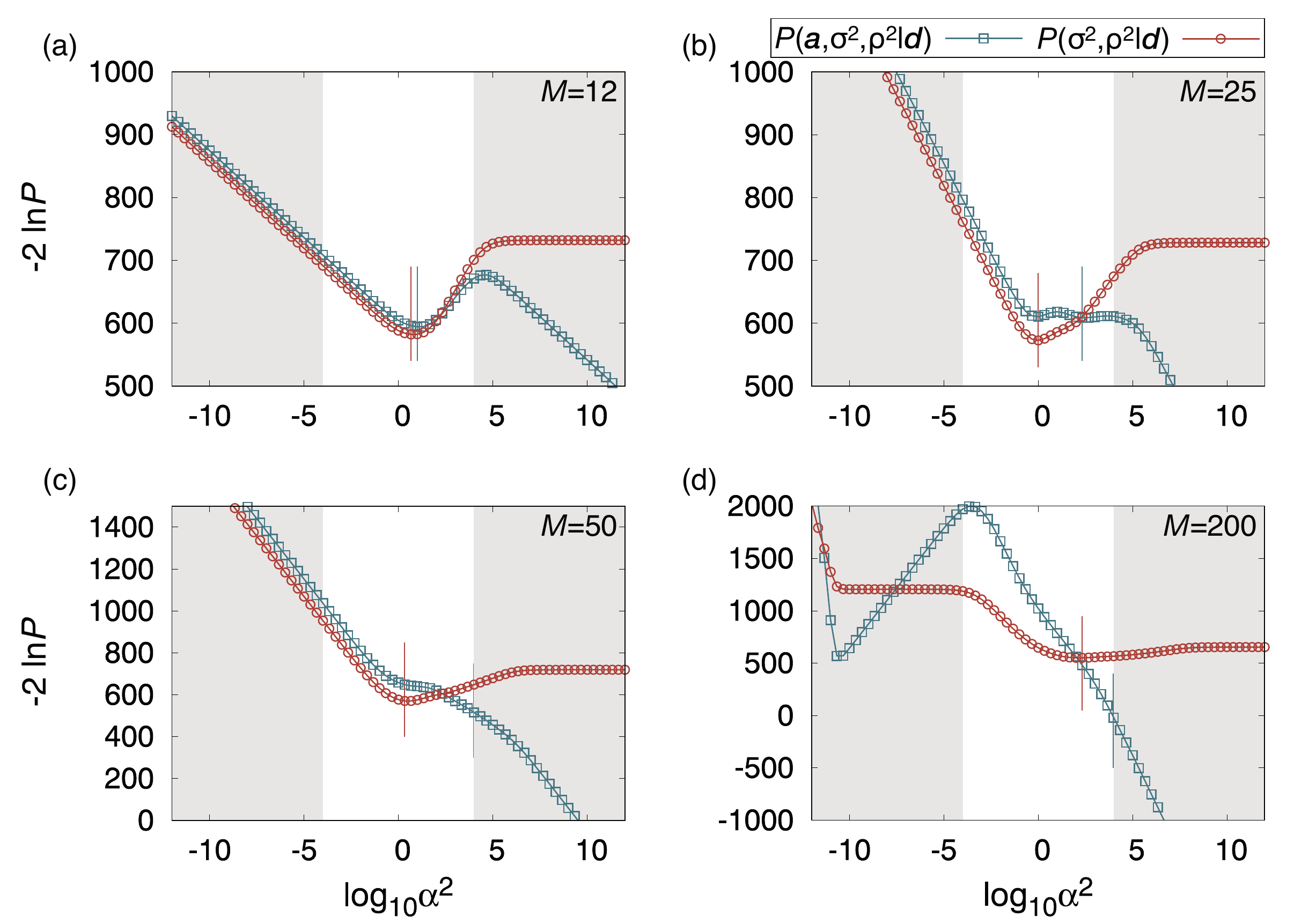}
\caption{
Joint posterior (MAP; blue) and marginal posterior of the hyperparameters (ABIC; red) on the log scale, maximised for given $\alpha^2$ values, shown for the second synthetic test eq.~(\ref{eq:secondex}) with varying the number of model parameters $M$ as 12 (a), 25 (b), 50 (c) and 200 (d). The number of data $N$ is fixed to 100. 
The log probabilities are offset and multiplied by $-2$ following the evaluation functions of the MAP and ABIC [eqs.~(\ref{eq:MAPalphadeterminationcondition}) and (\ref{eq:ABICalpha}), respectively]. 
The outside of the search interval for $\alpha^2$ in the synthetic test, $\alpha^2\in[10^{-4},10^4]$, is masked in grey. 
The selected $\alpha^2$ values are marked with vertical lines of the same colours as the probability profiles. 
}
\label{fig:lnpMAPABIC}
\end{figure*}

Figure~\ref{fig:lnpMAPABIC} measures eqs.~(\ref{eq:saddlejointposterior}) and (\ref{eq:marginalizedposforhyper}) for the mixed-wavelength field eq.~(\ref{eq:secondex}). 
The results are for a particular data set, but similar behaviours followed others. 
The vertical axis shows the non-constant part of the log probability multiplied by $-2$ as in the evaluation functions of the MAP (eq.~\ref{eq:MAPalphadeterminationcondition}) and ABIC (eq.~\ref{eq:ABICalpha}). 
We masked the outside of the grid search area in grey.

All the panels in Fig.~\ref{fig:lnpMAPABIC} indicate the joint posterior [blue, eq.~(\ref{eq:saddlejointposterior})] increases with $\alpha^2$ around the right end ($\alpha^2\to\infty$), corresponding to the aforementioned global maximality of the underfitted solution in the joint posterior. In contrast, the marginal posterior [red, eq.~(\ref{eq:marginalizedposforhyper})] of the hyperparameters flattens for $\alpha^2\to\infty$ and stably locates
its global maximum within a reasonable range of $\alpha^2$, 
consistent with the reasonable fits of the ABIC estimate to synthetic data in Fig.~\ref{fig:demo2}. 

The local maximum of the joint posterior in the given $\alpha^2$ interval is close to the global maximum of the marginal posterior of the hyperparameters for $M=12$, but it becomes obscure at $M=25$ (then making $\alpha^2$ of the MAP estimate jump to a larger value as one of branched maxima) and disappears at $M = 50$. This behaviour of the local maxima is consistent with the fitting results shown in Fig.~\ref{fig:demo2}, in which the MAP and ABIC estimates almost overlap at $M = 12$, but the MAP estimate is oversmoothed at $M = 25$ and $50$. 
The vanishment of the local maximum in the joint posterior also 
explains why the MAP estimate of $\alpha^2$ stuck to the upper bound of its search range [$M\gtrsim40$ in Fig.~\ref{fig:alphasigma_cosexp}(a)]. 
We also observe an overfitted solution emerges in the joint posterior at $M = 200$ as known previously (mentioned earlier), although it is outside the search range of $\alpha^2$, then now secondary.

\begin{figure}
	\includegraphics[width=75mm]{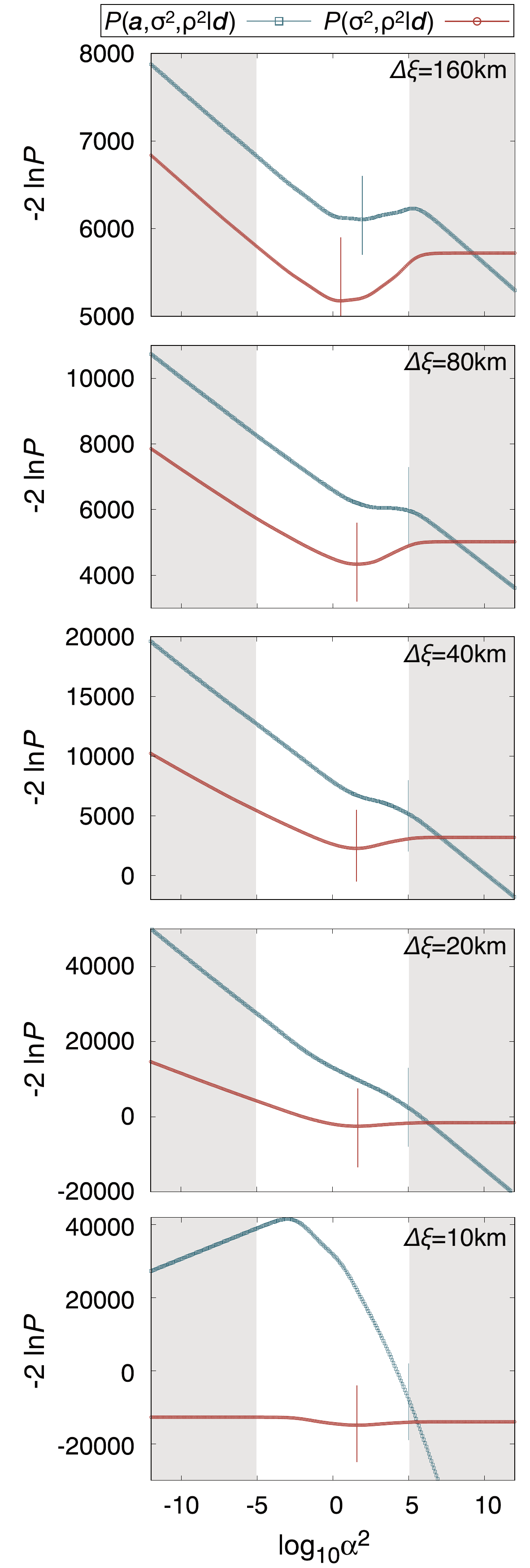}
\caption{
Joint posterior (MAP; blue) and marginal posterior of the hyperparameters (ABIC; red) on the log scale, maximised for given $\alpha^2$ values. 
Results of the data analysis in \S\ref{sec:applicationsec} are shown by the same visualisation scheme as that of Fig.~\ref{fig:lnpMAPABIC}, with varying
the grid size $\Delta \xi$ of the discretised model-parameter field as 
160, 80, 40, 20 and 10 in kilometer. 
The associated numbers of model parameters $M$ are 
72, 162, 450, 1458 and 5202, respectively. 
The number of data $N$ is fixed to 572. 
}
\label{fig:app_lnpMAPABIC}
\end{figure}

Figure~\ref{fig:app_lnpMAPABIC} plots eqs.~(\ref{eq:saddlejointposterior}) and (\ref{eq:marginalizedposforhyper}) 
for the actual data analysis conducted in \S\ref{sec:applicationsec}. 
The joint posterior (blue) has a local maximum only for $\Delta \xi=160$km and is unimodal for finer $\Delta \xi=80,40,20,10$km of larger $M$. 
Its global maximum is steadily the underfitted estimate of $\alpha^2=\infty$
as in the synthetic test, consistent with the above-mentioned theoretical consideration. 
These explain the resolution degradation of the MAP estimate accompanying the $M$ increase in Fig.~\ref{fig:App_dilatationrate}.
In brief, the problem of the MAP observed in the example of geophysical application (Fig.~\ref{fig:App_dilatationrate}) has the same structure as that in the synthetic test (Fig.~\ref{fig:demo2}). 
Again as in the synthetic test, 
the marginal posterior of the hyperparameters (red) locates the global maximum within a reasonable range of $\alpha^2$ for all plotted cases. 
Even though the true model-parameter field is unknown in the real data inversion, the marginal posterior of the hyperparameters is shown saved from the strange features of the joint-posterior profile. 

As shown in Fig.~\ref{fig:lnpMAPABIC}, it is the vanishment of the appropriate local maximum from the joint posterior for relatively large $M$ that results in the oversmooth tendency of the MAP estimate. 
Another underlying cause is the global maximality of the underfitted/overfitted solution in the joint posterior. 
The joint posterior value of the ABIC estimate is nearly zero,
for example, around $\exp(-50)\sim10^{-22}$ times the joint posterior peak of the MAP estimate for $M=50$ in this synthetic test (Fig.~\ref{fig:lnpMAPABIC}c). 
Figure~\ref{fig:app_lnpMAPABIC} exemplifies the same difficulty arises in a real data analysis. 
Recalling the $\alpha^2$ estimate of ABIC is close to the ideal $\alpha^2$ value [the min.~TMR, Fig.~\ref{fig:alphasigma_cosexp}(a)], 
an ideal estimate is also anomalously rare in the joint posterior, that is hard to win by numerical methods. 


\subsection{Appropriate dimensionality reduction and posterior averaging}
\label{sec:intepretationofprob}

The joint posterior was in substance zero around the appropriate estimates in Fig.~\ref{fig:lnpMAPABIC} for large $M$ because of the global maximality of the underfitted estimate and the asymptotic vanishment of the well-behaved local maxima. 
The pathology also follows the marginal posterior of the model parameters that concentrates on the MAP estimate for large $M$. 
These mean only the marginal posterior of the hyperparameters is the well-behaved distribution for a large number of model parameters in the reduction categories (1)--(3). 

We explore this change of the posterior profile, based on the reduction invariance of the posterior mean [EAP; eq.~(\ref{eq:defofmodelEAP})]: 
$
\langle 
{\bf a}
\rangle_{{\bf a},\sigma^2,\rho^2|{\bf d}}
=
\langle 
{\bf a}
\rangle_{{\bf a}|{\bf d}}
=
\langle \langle
{\bf a}
\rangle_{{\bf a}|\sigma^2,\rho^2,{\bf d}}
\rangle_{\sigma^2,\rho^2|{\bf d}}
$, 
coming from eqs.~(\ref{eq:twostage}) and (\ref{eq:elementarymath}). 
The reduction invariance is a special property of the cumulants, and the probability peaks we have investigated are not the invariants under the reduction, similarly to (Monte-Calro) model-parameter samples depending on generating distributions. 
The EAP can be a reference to measure the posterior peak shifts due to the reduction. 
We note the EAP minimises the posterior mean of the squared misfit $\langle |{\bf a}-\hat {\bf a}|^2\rangle_{{\bf a}|{\bf d}}$ of ${\bf a}$ from the estimate 
$\hat {\bf a}$~\citep[the squared error loss;][p.313]{carlin2008bayesian}, 
but its minimisation is not equivalent to minimising the misfit from the true solution (e.g. the TMR and TMS). 

We first derive an analytic form of the EAP estimate. 
We rewrite the EAP $\hat {\bf a}_{\rm EAP}$ of the model parameters with the mean ${\bf a}_*(\alpha^2)$ (eq.~\ref{eq:leastsquarewithpenalty}) of $P({\bf a}|\sigma^2,\rho^2,{\bf d})$ 
as $\hat {\bf a}_{\rm EAP}=\langle {\bf a}_*(\alpha^2) \rangle_{\sigma^2,\rho^2|{\bf d}}$ 
and expand 
$\langle({\bf H}^{\rm T}{\bf E}^{-1}{\bf H}+\alpha^2{\bf G})^{-1}\rangle_{\sigma^2,\rho^2|{\bf d}}$ in $\langle {\bf a}_*(\alpha^2) \rangle_{\sigma^2,\rho^2|{\bf d}}$ 
around the peak of the marginal posterior of $\ln\alpha^2$, which is quite steep as shown in the ABIC profile of Fig.~\ref{fig:lnpMAPABIC}. It yields the following series (Appendix~\ref{sec:FBEAP}):
\begin{equation}
\begin{aligned}
\hat {\bf a}_{\rm EAP}=&
\left[
{\bf I}
+
\left(\frac 1 2\alpha^2{\bf J} -(\alpha^2 {\bf J})^2 \right)
\left(
\frac{\partial^2\ln P(\ln\alpha^2|{\bf d})}{\partial (\ln\alpha^2)^2}
\right)^{-1}
\right.\\&\left.
+\mathcal O
\left(
\left(
\frac{\partial^2\ln P(\ln\alpha^2|{\bf d})}{\partial (\ln\alpha^2)^2}
\right)^{-2}
\right)
\right]_{\alpha^2=\hat\alpha^2_{\rm ABIC}}
\hat {\bf a}_{\rm ABIC},
\end{aligned}
\label{eq:EAPvsABIC_2ndorder}
\end{equation}
where ${\bf J}(\alpha^2):=({\bf H}^{\rm T}{\bf E}^{-1}{\bf H}+\alpha^2{\bf G})^{-1}{\bf G}$, and 
the second derivative of $P(\ln\alpha^2|{\bf d})$ is given as
\begin{equation}
\begin{aligned}
&\frac{\partial^2\ln P(\ln\alpha^2|{\bf d})}{\partial (\ln\alpha^2)^2}
=
\\&-\frac 1 2 \mbox{Tr}[\alpha^2
({\bf I}-\alpha^2{\bf J})
{\bf J}]
-\frac{N+P-M-4}{2}
\\&\times
\left[
-\left(\frac{\alpha^2{\bf a}_*^{\rm T}{\bf G}{\bf a}_*}{s({\bf a}_*(\alpha^2),\alpha^2)}\right)^2
+\frac{\alpha^2{\bf a}_*^{\rm T}{\bf G}({\bf I}-2\alpha^2 {\bf J}){\bf a}_*}{s({\bf a}_*(\alpha^2),\alpha^2)}
\right]. 
\end{aligned}
\label{eq:secondderivofmarginalposoflnalpha}
\end{equation}
The inverse of the second derivative of $P(\ln\alpha^2|{\bf d})$ expresses the second-order moment around the peak of $P(\ln\alpha^2|{\bf d})$. 
Equation~(\ref{eq:secondderivofmarginalposoflnalpha}) shows
the second derivative of $P(\ln\alpha^2|{\bf d})$ is $\mathcal O(P)$ (see Appendix~\ref{sec:FBEAP} for order estimation details).  
Then, the second-order moment around the peak of $P(\ln\alpha^2|{\bf d})$ is $O(1/P)$, and consequently, the law of large numbers of $P(\ln\alpha^2|{\bf d})$ brings the EAP estimate close to the ABIC estimate accompanying the increase in $P$ $(M)$: 
\begin{equation}
\hat {\bf a}_{\rm EAP}\approx\hat {\bf a}_{\rm ABIC}.
\label{eq:EAPsimeqABIC}
\end{equation} 
The EAP estimates 
($\hat \sigma^2_{\rm EAP}$, $\hat \rho^2_{\rm EAP}$) $:=$($\langle\sigma^2\rangle_{{\bf a},\sigma^2,\rho^2|{\bf d}}$, 
$\langle\rho^2\rangle_{{\bf a},\sigma^2,\rho^2|{\bf d}}$) 
of the hyperparameters are evaluated as 
$ 
\hat \sigma^2_{\rm EAP}
\approx
\hat \sigma^2_{\rm ABIC}
$ and $
\hat \rho^2_{\rm EAP}
\approx
\hat \rho^2_{\rm ABIC}
$ 
through eq.~(\ref{eq:deltafunctionalconvergence_hyperparameter}) for large $P$. 
The EAP estimate is as above asymptotically consistent with the ABIC estimate for both the model parameters and hyperparameters, and hence Fig.~\ref{fig:lnpMAPABIC} also means the EAP estimate takes an almost negligible probability value in the joint posterior for a large number of model parameters. 
The same applies to the marginal posterior of the model parameters; 
when $\lim_{M\to\infty}[\hat {\bf a}_{\rm ABIC}-\hat {\bf a}_{\rm MAP}]\neq 0$, 
eqs.~(\ref{eq:MMAPsimMAP}),  (\ref{eq:deltafunctionalconvergence_modelparameter}) and 
(\ref{eq:EAPsimeqABIC}) lead to  
\begin{equation}
    \lim_{M\to\infty}P(\hat {\bf a}_{\rm EAP}|{\bf d})=0.
    \label{eq:asymptoticalzeromeasureofEAP}
\end{equation}

\begin{figure}
  \includegraphics[width=90mm]{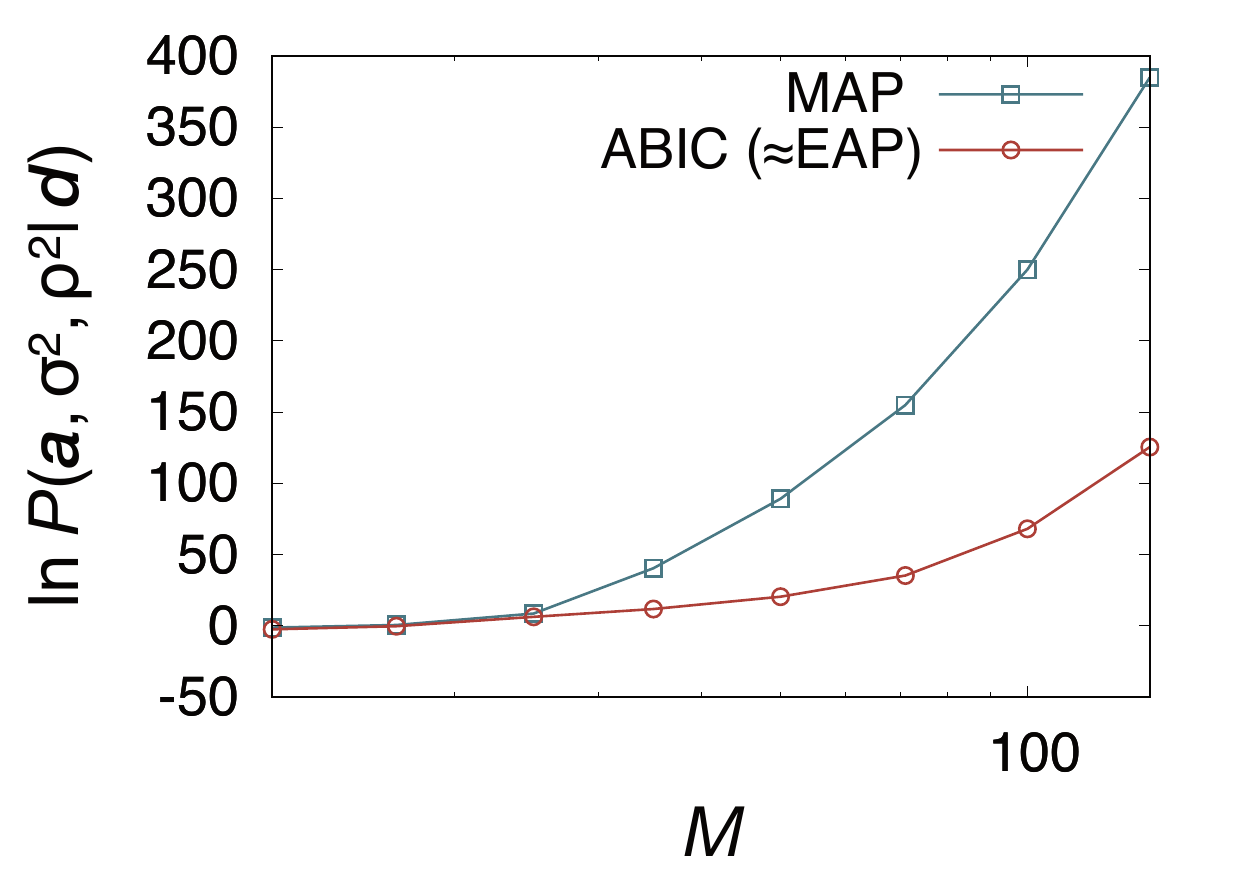}
 \caption{
Log joint posterior values taken by the estimates of the model parameters and hyperparameters, as a function of the number of model parameters $M$, 
comparing the MAP (blue) and ABIC (red), the latter of which is asymptotically the EAP. The plotted results are
for a particular data set of the second synthetic test eq.~(\ref{eq:secondex}). 
}
\label{fig:comparison_jointprobabilityvalue_ABICMAP}
\end{figure}

These analytic results suggest the marked differences between the mean and modes of the concentrating distributions induce the aforementioned pathologies. 
Although both $P({\bf a}|{\bf d})$ and $P(\sigma^2,\rho^2|{\bf d})$ are shown to concentrate on the peaks [eqs.~(\ref{eq:deltafunctionalconvergence_modelparameter}) and (\ref{eq:deltafunctionalconvergence_hyperparameter}), respectively], as above for a large number of model parameters, only the latter satisfies the law of large numbers, which states the asymptotic concentration of a distribution on its mean value. 
We then consider the joint posterior values of the mode (the MAP) and mean (the EAP, or given their asymptotic proximity, ABIC) in an intermediate $M$ range. 
Figure~\ref{fig:comparison_jointprobabilityvalue_ABICMAP} shows 
the probability values of the joint posterior at the MAP and ABIC estimates of the model parameters and hyperparameters for finite $M$ 
under the same setting as in Fig.~\ref{fig:lnpMAPABIC}. 
The ratio of the two probability values grows as the number of model parameters increases, and remarkably, their gap is widened exponentially: in an asymptotic sense, given eq.~(\ref{eq:EAPsimeqABIC}),
\begin{equation}
\ln \frac
{P(\hat{\bf a}_{\rm MAP},\sigma^2_{\rm MAP},\rho^2_{\rm MAP}|{\bf d})}
{P(\hat{\bf a}_{\rm EAP},\sigma^2_{\rm EAP},\rho^2_{\rm EAP}|{\bf d})}
%
=\mathcal O(M).
\end{equation}
We here ignore the trivial log order while the figure suggests this ratio may precisely be  $\mathcal O(M\ln M)$. 

The posterior mean of ${\bf a}$ ($\approx \hat {\bf a}_{\rm ABIC}$) takes exponentially smaller joint posterior values in terms of the dimension $M$ of the model-parameter space. 
It deduces the dominance of the posterior mean resides in 
exponentially many but exponentially rare
${\bf a}$ values, or a small number of exponentially rare yet exponentially large (i.e. unstable) inappropriate ${\bf a}$. 
The latter is obviously an improbable scenario, indeed,
one relation validates the former picture. 
We focus on the minimisation-function difference between the MAP and ABIC, the log marginal posterior of the model parameters (eq.~\ref{eq:twostage}; times $-2$): 
\begin{equation}
    \ln P({\bf a},\sigma^2,\rho^2|{\bf d})-\ln P(\sigma^2,\rho^2|{\bf d})
    =
    \ln P({\bf a}|\sigma^2,\rho^2,{\bf d}).
    \label{eq:logprobdecomposition}
\end{equation}
The minus mean of a log probability is generally called Shannon entropy (Gibbs entropy), which represents the concept of the number of states on the log scale (i.e. Boltzmann's entropy) extended to a probability space~\citep{gibbs1878equilibrium,shannon1948mathematical}. The Shannon entropy of a Gibbs distribution, the statistical thermodynamic entropy, is commonly an extensive variable proportional to the dimension of its probability space; it is true also for $P({\bf a}|\sigma^2,\rho^2,{\bf d})$ with $M$ degrees of freedom: 
\begin{equation}
\langle 
-\ln P({\bf a}|\sigma^2,\rho^2,{\bf d})
\rangle_{{\bf a}|\sigma^2,\rho^2,{\bf d}}
=
\mathcal O(M).
\end{equation}
Recalling a known form of the Shannon entropy for a Gaussian, 
we have 
$
\langle -\ln P({\bf a}|\sigma^2,\rho^2,{\bf d})
\rangle_{{\bf a}|\sigma^2,\rho^2,{\bf d}}=(M/2)\ln(2\pi e\sigma^2)-(1/2)\ln|{\bf H}^{\rm T}{\bf E}^{-1}{\bf H}+\alpha^2{\bf G}|$ in the present linear inverse problem, which is actually extensive. 
Furthermore, the conditional posterior cumulants of 
the cost functions are also commonly extensive in the Gibbs distributions~\citep[also see Supplement 3]{landau1994statistical}, including those of $U+\alpha^2 V$ in $P({\bf a}|\sigma^2,\rho^2,{\bf d})$; 
then for the suite of ${\bf a}\sim P({\bf a}|\sigma^2,\rho^2,{\bf d})$, $\sigma^2$ and $\rho^2$, 
\begin{equation}
-\ln P({\bf a}|\sigma^2,\rho^2,{\bf d})
=
    \langle-\ln P({\bf a}|\sigma^2,\rho^2,{\bf d})
\rangle_{{\bf a}|\sigma^2,\rho^2,{\bf d}}
+\mathcal O(\sqrt{M}), 
\label{eq:determinicacyofempiricalshannon}
\end{equation}
where $U+\alpha^2 V$ in $P({\bf a}|\sigma^2,\rho^2,{\bf d})$ (eq.~\ref{eq:posterior4uniformhyperprior}) is replaced with its conditional posterior mean given $\sigma^2$ and $\rho^2$ within $\mathcal O(\sqrt{M})$ accuracy. 
Equation~(\ref{eq:determinicacyofempiricalshannon}) means the empirical Shannon entropy is almost deterministic for the Gibbs distribution. 
For the present linear-inverse case, 
$-\ln P({\bf a}|\sigma^2,\rho^2,{\bf d})-
\langle 
-\ln P({\bf a}|\sigma^2,\rho^2,{\bf d})
\rangle_{{\bf a}|\sigma^2,\rho^2,{\bf d}}$ 
is $({\bf a}-{\bf a}_*)^{\rm T}{\bf C}_{{\bf a}_*}({\bf a}-{\bf a}_*)/2-M/2$ that becomes $\chi^2_{M}/2-M/2$ for ${\bf a}\sim P({\bf a}|\sigma^2,\rho^2,{\bf d})$, where $\chi^2_{M}$ denotes the $\chi^2$-distribution with $M$ degrees of freedom; the cumulants of $\chi^2_{M}/2$ are all $\mathcal O(M)$, and eq.~(\ref{eq:determinicacyofempiricalshannon}) follows asymptotics $\chi^2_{M}/2-M/2\approx\mathcal N(0,M/2)$ under the central limit theorem. 
Finally, eqs.~(\ref{eq:logprobdecomposition}) and
(\ref{eq:determinicacyofempiricalshannon}) lead to a key relation: 
\begin{equation}
\begin{aligned}
\ln P(\sigma^2,\rho^2|{\bf d})=&
\ln P({\bf a},\sigma^2,\rho^2|{\bf d})
\\&
+\langle 
-\ln P({\bf a}|\sigma^2,\rho^2,{\bf d})
\rangle_{{\bf a}|\sigma^2,\rho^2,{\bf d}}
+\mathcal O(\sqrt{M}). 
\end{aligned}
\label{eq:jointprobvsmarginalizedposofhyper}
\end{equation}

Equation~(\ref{eq:jointprobvsmarginalizedposofhyper}) 
indicates the $\mathcal O(M)$ entropy term
is the cause for the difference between the peaks of the joint posterior and marginal posterior of the hyperparameters, that is, the MAP and ABIC, and given eq.~(\ref{eq:EAPsimeqABIC}), the mode (the MAP) and mean (the EAP) of the joint posterior. 
An exponential representation of eq.~(\ref{eq:jointprobvsmarginalizedposofhyper}) is more intuitive: 
\begin{equation}
P(\sigma^2,\rho^2|{\bf d})\simeq
P({\bf a},\sigma^2,\rho^2|{\bf d})
\times 
W_{\sigma^2,\rho^2},
\label{eq:entropicexpressionofmarginalizedvsjoint}
\end{equation}
where $W_{\sigma^2,\rho^2}:=\exp[\langle -\ln P({\bf a}|\sigma^2,\rho^2,{\bf d})\rangle_{{\bf a}|\sigma^2,\rho^2,{\bf d}}]$, and $\simeq$ here denotes the asymptotic equality on the log scale of the leading order. 
Equation~(\ref{eq:entropicexpressionofmarginalizedvsjoint}) states the marginal posterior of the hyperparameters, and considering eq.~(\ref{eq:EAPsimeqABIC}), the EAP
count almost infinite $W_{\sigma^2,\rho^2}=\exp[\mathcal O(M)]$ states with almost zero probabilities, $P({\bf a},\sigma^2,\rho^2|{\bf d})=1/\exp[\mathcal O(M)]$.
The huge $\exp[\mathcal O(M)]$ number of states balances with unnaturally low $1/\exp[\mathcal O(M)]$ probabilities, 
and then the peak shifts from the inappropriate MAP to actually appropriate ABIC, subjected to the transform from the joint posterior to the marginal posterior of the hyperparameters. 
The same order balance also allows the mean (the EAP) to differ significantly from the peak (the MAP), 
and besides as the marginal posterior of the hyperparameters already integrates the joint posterior over the state space of the model parameters, its peak (ABIC) can be close to the mean (the EAP) all the possible model-parameter states participate. 
Equation~(\ref{eq:entropicexpressionofmarginalizedvsjoint}) signifies the pathology in the joint posterior is resolved by counting a literally infinite number of model-parameter states in the marginal posterior of the hyperparameters. 

\section{Discussion and conclusions}
\label{sec:pracuseofFB}

We have investigated the reduction in extracting useful information on the model parameters from the joint posterior. 
Our classification of reduction methods directs attention toward the marginalisation involved in the reduction process. 
The reduction issue then results in a study on the behaviours of the following three distributions: 
(1) the joint posterior itself, (2) the marginal posterior of the model parameters, and (3) the marginal posterior of the hyperparameters (plus the conditional posterior of the model parameters). 
We characterised distributions (1)--(3) by the analytic representations of their peaks and asymptotic shapes (with the second-order moments around the peaks and delta-functions). 
These distributions are certainly identical with respect to the model parameters as they are connected by the transformation formulae [integration eq.~(\ref{eq:defofmarginalized}) and association eq.~(\ref{eq:twostage})], so must include the same information on the model parameters, but their shapes are dissimilar, as their peaks are. 
This study elucidates only the two-stage inference of category (3) provides a well-behaved distribution for a large number of model parameters. 
Profiling the posterior in a well-behaved manner could be considered one mathematical aspect of what extracting useful information represents.

Parameters outside an estimation target are called nuisance parameters, and their elimination in the inference is a long-standing issue in statistics~\citep{dey2005bayesian}. 
Our classification of the reduction is generally in line with it. 
Category (1) using $P({\bf a},\sigma^2,\rho^2|{\bf d})$ to estimate ${\bf a}$ is similar to the profile likelihood method~\citep{murphy2000profile} that evaluates the joint posterior maximised given the parameters of interest [i.e. $\max_{\sigma^2,\rho^2|{\bf a},{\bf d}}P({\bf a},\sigma^2,\rho^2|{\bf d})$ for ${\bf a}$]. 
Category (2) of $P({\bf a}|{\bf d})$ corresponds to the marginal likelihood method sometimes recommended~\citep[p.209]{carlin2008bayesian,gelman2013bayesian}. 
Category (3) based on $P(\sigma^2,\rho^2|{\bf d})$ includes ABIC~\citep{akaike1980likelihood} and Type II likelihood~\citep{good1965estimation}, common for hyperparameter point estimations~\citep{bishop2006pattern}. 
In statistical terms, the scope of our study may be this elimination of the nuisance parameters in the model-parameter estimations.  
It would be surprising even in this context that the appropriate reduction is the empirical-Bayesian two-stage inference (ABIC), which is ordinarily regarded as a point-estimation technique of the hyperparameters in the fully Bayesian analysis~\citep{gelman2013bayesian}. 
Given their esteem as non-approximated styles of the empirical Bayes, more unexpected may be the pathologies in the joint posterior and marginal posterior of the model parameters. 

Several interesting properties were obtained in the asymptotic analysis.
The marginal posteriors of the model parameters and hyperparameters concentrate on the MAP and ABIC estimates, respectively. 
As confirmed, the latter is the law of large numbers, but the former is not. 
The model-parameter space expands at the same speed as the growth of the probability peak, and various model-parameter states then emerge [i.e. $c<1$ for $P({\bf a}|{\bf d})\to c\delta({\bf a}-\hat{\bf a}_{\rm MAP})$] with exponentially diminishing probabilities and finitely contribute to the posterior mean (the EAP). 
This effect is represented by the entropy term in the marginal posterior of the hyperparameters (eq.~\ref{eq:jointprobvsmarginalizedposofhyper}), and thus also in ABIC. 
The existence of the entropic effect in the marginal posterior of the hyperparameters itself appears to have been recognised in the statistical literature~\citep[e.g.][]{takane1987relationship,iba1989bayesian,iba1996learning}, while eq.~(\ref{eq:entropicexpressionofmarginalizedvsjoint}) provides an arguably novel relation: the number of states $W$ (multiplicity) is exactly the relative difference between the joint posterior and the marginal posterior of the hyperparameters. 
Equation~(\ref{eq:entropicexpressionofmarginalizedvsjoint}) is valid also for the marginalisation of the conditional likelihood $P({\bf d}|{\bf a},\sigma^2)$ of the model parameters ${\bf a}$ given the hyperparameter $\sigma^2$ (corresponding to $P=0$ in our analysis), and 
the marginalisation is widely the operation to include the combinatorial effect in the probability profile. 
The found entropic effect for a fully-Bayesian, or more specifically, multi-canonical~\citep{berg1992multicanonical} framework is analogous to that in the density of states~\citep{kittel1976introduction} of the Gibbs distributions with fixed hyperparameters. 
There are several entropy-oriented criteria~\citep{akaike1980likelihood,shore1980axiomatic,jaynes1982rationale}, which would be worthy of further investigation as in \citet{ulrych2001bayes}, to know the role of multiplicity in Bayesian inferences. 
Regarding the MAP, the global maximum of the joint posterior is an inappropriate underfitted or overfitted estimate whenever it exists, and local maxima may also be asymptotically inappropriate or vanish as observed in the synthetic tests, despite those are often chief candidates of good estimates in optimisation strategies. 
Given the asymptotic consistency between ABIC and the EAP, 
we intrinsically have only the MAP and EAP, which are typical states in the single- and second-stage model-parameter inferences, respectively. 
Zero probability of the EAP estimate should be regarded as an asymptotic pathological nature of the one-stage inferences [categories (1) and (2)] based on the joint posterior. 
The identified problem is essentially for a large $M$ setting, where the prior plays a major role in the inference and the fully Bayesian framework should have advantages over the likelihood-based frequentist approaches; in this setting, a serious defect becomes evident. 
Most distributions are well approximated by Gaussians around their extrema, so nonlinear problems would present with the same pathologies, 
including various geophysical analyses with linearisation techniques~\citep{tarantola2005inverse}. 
Besides, the problem in the global maximum of the joint posterior documented in \S\ref{sec:erasingoutsiders} is not limited to the two-hyperparamter formulation, because similar expressions to eq.~(\ref{eq:jointposteriorform}) are obtained also in multiple-hyperparameter cases, both for the joint posterior and marginal posterior of the hyperparameters~\citep{fukahata2004geodetic,malinverno2004expanded,fukuda2010mixed}. 
It is also evidenced by our application that the problem in the reduction could happen in an actual inversion analysis. 

The previous fully Bayesian numerical approaches may be affected by the above pathology in the joint posterior. 
The EAP close to ABIC is appropriate but exponentially rare in the joint posterior, so sampling low-probability events is essential for the MCMC joint-posterior samplers.  
Besides, counting an exponential number of model-parameter states  that contribute to the posterior mean is required to numerically average the model-parameter states in the joint posterior probability space. 
Although the MCMCs can sample large model-parameter subsets~\citep{malinverno2002parsimonious}, 
the accessible number of samples is frequently the polynomial order in numerics, so it is another matter to count an exponential number of samples with respect to the model-parameter dimension, $M$. 
The exponential rarity of the appropriate EAP and ABIC would become an issue even in the optimisation approaches. 
We usually naively believe a good solution can be constructed from a finite number of events with high probabilities, but the high joint-posterior domain could be inappropriate, and exponentially many samples become necessary for generating the posterior mean from the high-dimensional joint posterior sampling appropriately.

This study possibly cautions ordinary Monte Carlo methods could require exponential time to compute the EAP for large $M$ $(\gtrsim N)$. 
Here we examine it in detail for sample means generated by $P({\bf a}|{\bf d})$ or $P({\bf a},\sigma^2,\rho^2|{\bf d})$. 
The sample mean is an unbiased estimate of the posterior mean (the EAP) and hence converges to the EAP estimate in the limit of an infinite number of samples. 
At the same moment, the most frequent value of the samples is the MAP estimate, despite the significant difference between the MAP and EAP estimates. 
As shown earlier, these two propositions are compatible because after the summation over an exponential number of events, exponentially rare events take a finite probability in total, which drives the sample mean to the EAP estimate of an asymptotically zero probability (eq.~\ref{eq:asymptoticalzeromeasureofEAP}).
This reasoning parallels the aforementioned structure of how the EAP can be close to ABIC considerably shifted from the MAP. 
The requirement of an exponential number of samples ought to be called a sampling difficulty in the fully Bayesian techniques using the joint posterior. 
We may avoid that sampling problem for small $M$, by suitably adjusting the $\alpha^2$ search interval and using sufficient computational powers. 
However, the required computational effort to converge the sample mean to the EAP is an exponential of $M$ and easily surpasses available numerical resources as $M$ increases. 
Besides, our numerical experiments and actual data analysis show the 
ABIC estimate may not even be the local maximum of the joint posterior for large $M$ (Fig.~\ref{fig:lnpMAPABIC}c and Fig.~\ref{fig:app_lnpMAPABIC}), questioning the above presupposition that we can set an $\alpha^2$ range appropriately. 
Another fundamental problem in setting a finite search interval for $\alpha^2$ is that the sample mean may be biased due to dropping entropic contributions from almost-zero-probability events. 
If the posterior mean is simply rare, its sampling is within the realm of ordinary rare event sampling~\citep{swendsen1986replica,hukushima1996exchange,wang2001efficient}, but as it is affected by the combinatorial effect, the entropy, it would require another sampling scheme to reduce the required number of samples, probably similar to thermodynamic integral techniques for marginalisation~\citep{kirkwood1935statistical}.

These difficulties are all solved in the two-stage inference of category (3) (e.g. ABIC), with the analytic closed-form expression of the marginal posterior of the hyperparameters. 
However, analytic marginalisation of the model parameters, a successful strategy in the linear inverse problems~\citep{yabuki1992geodetic}, is often hard to accomplish in nonlinear problems~\citep[e.g. transdimensional inversions;][]{sambridge2013transdimensional,tomita2020development}. 
An approximation to Gaussian mixtures will be one practical way of evaluating ABIC with non-Gaussian posteriors~\citep{ishiguro1983bayesian,ogata1988likelihood}, but not obvious whether generally fast enough. 
Naively computationally marginalising out the model parameters ${\bf a}$ from the joint posterior requires counting an exponential number of events to evaluate the influence of the multiplicity, as in the posterior-mean evaluation of the model parameters. 
Computational use of ABIC is then also impracticable for large $M$ in a brute-force manner. A versatile ABIC evaluation method may be available with an advanced Monte Carlo approach~\citep[e.g.][]{ogata1990monte}. 
We will discuss numerical methods with regard to them elsewhere.

The present discussion rests on the relative smallness of the number of hyperparameters to the numbers of data and model parameters. 
The relevant entropy effect comes asymptotically solely from the model-parameter space in such cases, 
and for this reason, 
integration over the hyperparameters is not enough to elude the pathology in the marginal posterior of the model parameters. 
One may actually find the entropic effect distinguishes ABIC (eq.~\ref{eq:ABICalpha_explicitextremal}) from 
the MAP (and the MMPM, eq.~\ref{eq:MAPalpha}) in the hyperparameter estimation for this linear inverse problem, as supplemented in Appendix~\ref{sec:ABICextremal}. 
Considering the present analysis assumes a relatively small number of hyperparameters, 
appropriate reductions may have different properties in the inversions involving a large number of hyperparameters~\citep{minson2013bayesian,livermore2014core}. 
Note in this study the hyperparameter refers to the scale of variance ($\sigma^2$ and $\rho^2$, or widely, the parameter of the Gibbs ensembles, including some parts of the normalised coefficients of variances ${\bf E}^{-1}$ and ${\bf G}$ for a joint inversion; cf. Supplement 3). Another literature identifies the hyperparameters with the model parameters~\citep[e.g.][]{minson2013bayesian,livermore2014core}. 
Their position is the original fully Bayesian thought that regards both of them equally as unknowns~\citep{fukuda2008fully}. 
The Bayesian hierarchical model defines the stage I (the model parameters) and stage II (the hyperparameters) from given priors~\citep{gelman2013bayesian}, then deriving
polysemy of $\sigma^2$ and $\rho^2$ mentioned above: directly setting $P({\bf a},\sigma^2,\rho^2)$ regards all the unknowns as the model parameters while separate $P({\bf a}|\rho^2)$ and $P(\sigma^2,\rho^2)$ distinguish the model parameters and hyperparameters. There is the same terminological mixture in statistics~\citep{akaike1980likelihood,takane1987relationship}. 
Terms apart, the essence of this study is classifying posterior decompositions into pathological and well-behaved ones when the means and (scales of) variances are unknowns. The shown statistical structure thus holds regardless of whether they are the stage I or II unknowns. 
For the same reason, we foresee property changes in the marginal posterior of the hyperparameters for other hyperparameter designs~\citep[e.g. a dip angle hyperparameter in a finite fault inversion;][]{fukahata2008non}. 

As shown in this study, the posterior distribution possesses quite atypical properties in the fully Bayesian inversion. 
Our example application suggests this can be a significant issue in the actual inverse problems. 
Meanwhile, the multistage model-parameter estimation with the marginal posterior of the hyperparameters 
detaches the problem in profiling the joint posterior, and we can expect a simple feeling that events with high probabilities are close to the appropriate values, even in the fully Bayesian inference, as is often the case for the Bayesian non-hierarchical inference. 
Along with it, the preceding difficulty in sampling from the joint posterior naturally vanishes. 
The empirical-Bayesian multistage inference of ABIC, which has been underrated by various researchers as an approximation, perhaps because of its historical background of being introduced together with the point estimation of the hyperparameters, seems to have been an appropriate prescription for pathology in reduction of the joint posterior of the model parameters and hyperparameters in the fully Bayesian inversions. 

\begin{acknowledgments}
We appreciate the helpful comments of Dr Takaki Iwata, Dr Yuji Yagi, Dr Andrew Hooper, Dr Phil Livermore and Dr Ruth Amey. 
The author D.S. is also deeply grateful to Dr Tim Wright for accepting his stay at Leeds University, which led him to receive a large part of these comments. 
We also thank Dr Andrew Curtis and an anonymous reviewer for their insightful comments. 
This study was partly supported by MEXT KAKENHI Grant Numbers JP15K21755 and JP19K04030 and JSPS KAKENHI Grant Number 21J01694. 
\end{acknowledgments}

\section*{AUTHOR CONTRIBUTION STATEMENT}
DS derived the analytical results, coded the synthetic tests, conducted the synthetic and actual data analyses, and took the lead in writing. 
YF motivated the authors towards the present research project, designed synthetic test frameworks and data analysis and contributed to discussions and manuscript writing. 
YN provided the processed data with an application code and contributed to discussions. 
All authors read and approved the final manuscript.

\section*{DATA AVAILABILITY}
GNSS data used in this study is available from the Geospatial Information Authority of Japan. 

\bibliographystyle{gji}


\begin{appendix}
\renewcommand{\theequation}{\Alph{section}.\arabic{equation}}


\section{Differential forms of the ABIC minimisation}
\setcounter{equation}{0} 
\label{sec:ABICextremal}
In this section, we 
obtain
eq.~(\ref{eq:ABICalpha_explicitextremal}) as the extremum conditions with respect to $\alpha^2$, 
the formal representation of which is also supplemented. 

We 
rewrite eq.~(\ref{eq:ABICalpha}) and obtain the extremum condition eq.~(\ref{eq:ABICalpha_explicitextremal}). 
Equation~(\ref{eq:relofsatastar}) yields
\begin{flalign}
\frac{\partial}{\partial (\alpha^2)}s({\bf a}_*(\alpha^2),\alpha^2)
=&
\left.
\frac{\partial s({\bf a},\alpha^2)}{\partial {\bf a}}\right|_{{\bf a}={\bf a}_*(\alpha^2)}^{\rm T} \frac{\partial {\bf a}_*(\alpha^2)}{\partial (\alpha^2)}
\nonumber\\&
+
\left.
\frac{\partial s({\bf a},\alpha^2)}{\partial (\alpha^2)}\right|_{{\bf a}={\bf a}_*(\alpha^2)}
\\=&
{\bf a}_*^{\rm T}(\alpha^2){\bf G}{\bf a}_*(\alpha^2).
\label{eq:diffofsstaralpha}
\end{flalign}
Using this and eq.~(\ref{eq:diffoflndetnormalisedcovariance}) and
denoting the minimisation function in eq.~(\ref{eq:ABICalpha}) as ${\rm ABIC}(\alpha^2)$, 
we obtain the extremum condition for the minimiser $\alpha^2=\hat \alpha^2$ of ABIC as 
\begin{flalign}
0=&\frac{\partial}{\partial (\alpha^2)} {\rm ABIC}(\alpha^2)
\\=&
\mbox{Tr}[({\bf H}^{\rm T}{\bf E}^{-1}{\bf H}+\alpha^2  {\bf G})^{-1}{\bf G}]-\frac{P}{\alpha^2}
\nonumber\\&
+ {\bf a}_*^{\rm T}(\alpha^2){\bf G}{\bf a}_*(\alpha^2)\frac{N+P-M}{s({\bf a}_*(\alpha^2),\alpha^2)},
\end{flalign}
that is,
\begin{equation}
P=
\alpha^2 \mbox{Tr}[({\bf H}^{\rm T}{\bf E}^{-1}{\bf H}+\alpha^2  {\bf G})^{-1}{\bf G}]+\frac{\alpha^2 {\bf a}_*^{\rm T}(\alpha^2){\bf G}{\bf a}_*(\alpha^2)}{\tilde \sigma^2_{\rm ABIC}(\alpha^2)}, 
\end{equation}
where $\tilde \sigma^2_{\rm ABIC}(\alpha^2)$ is defined by eq.~(\ref{eq:sigmaatsaddleofmarginalposterior}).

The formal representations of the above results may be noteworthy in the context of the entropic effect appearing in the text.
Denoting only the non-constant part of the joint posterior as
\begin{equation}
P({\bf a},\sigma^2,\rho^2|{\bf d})\propto
(\sigma^2)^{-N/2}(\rho^2)^{-P/2}e^{-U/\sigma^2}e^{-V/\rho^2},
\end{equation}
we marginalise the model parameters from the joint posterior and obtain
\begin{equation}
P(\sigma^2,\rho^2|{\bf d})
=c(\sigma^2)^{-(N+P)/2}(\alpha^2)^{P/2}\int d{\bf a}e^{-(U+\alpha^2 V)/\sigma^2}.
\end{equation}
Partial differentiation of $\ln P(\sigma^2,\rho^2|{\bf d})$ with respect to $\sigma^2$ with fixing $\alpha^2$ yields the following from the extremum condition:
\begin{equation}
0=-\frac{N+P}2 \frac{1}{\sigma^2}+\frac{\int d{\bf a}(U+\alpha^2 V)/(\sigma^2)^2 e^{-(U+\alpha^2 V)/\sigma^2}}{\int d{\bf a}e^{-(U+\alpha^2 V)/\sigma^2}},
\end{equation}
or equivalently,
\begin{equation}
\sigma^2=\frac{2\langle U+\alpha^2 V \rangle_{{\bf a}|{\bf d},\sigma^2,\rho^2}}{N+P}=\frac{\langle s \rangle_{{\bf a}|{\bf d},\sigma^2,\rho^2}}{N+P}. 
\label{eq:ABICsigma_MAPlikeform}
\end{equation}
The set of $\sigma^2=\hat \sigma^2_{\rm ABIC}$ and $\rho^2=\hat\sigma^2_{\rm ABIC}/\hat\alpha^2_{\rm ABIC}$($=\hat \rho^2_{\rm ABIC}$) satisfies this relation. 
Likewise, partially differentiating the marginal posterior of the hyperparameters with respect to $\alpha^2$ with fixing $\sigma^2$, 
the extremum condition yields
\begin{equation}
\alpha^2=\frac{P}{2}\frac{\sigma^2}{\langle V\rangle_{{\bf a}|{\bf d},\sigma^2,\rho^2}}.
\label{eq:alphaforABIC_extremum}
\end{equation}
Substituting $\sigma^2=\hat \sigma^2_{\rm ABIC}$ into eq.~(\ref{eq:alphaforABIC_extremum}), 
we obtain the condition for $\alpha^2=\hat\alpha^2_{\rm ABIC}$: 
\begin{equation}
\alpha^2=
\left.
\frac{\langle U\rangle_{{\bf a}|{\bf d},\sigma^2,\rho^2}/N}{\langle V\rangle_{{\bf a}|{\bf d},\sigma^2,\rho^2}/P}.
\right|_{\sigma^2=\hat\sigma^2_{\rm ABIC},\rho^2=\hat\sigma^2_{\rm ABIC}/\alpha^2}
\label{eq:ABICalpha_MAPlikeform}
\end{equation}
Equations~(\ref{eq:ABICsigma_MAPlikeform}) and
(\ref{eq:ABICalpha_MAPlikeform})
are the counterparts of the MAP estimates (eqs.~\ref{eq:sigmaatsaddleofposterior} and \ref{eq:MAPalpha}) with respect to $\sigma^2$ and $\alpha^2$. 
We then also have their equivalents: 
\begin{flalign}
\sigma^2&=2\langle U\rangle_{{\bf a}|{\bf d},\sigma^2,\rho^2}/N
\label{eq:ABICsigma_equipartition_condition}
\\ 
\rho^2&=2\langle V\rangle_{{\bf a}|{\bf d},\sigma^2,\rho^2}/P.
\label{eq:ABICrho_equipartition_condition}
\end{flalign}
The posterior means of the cost functions per degrees of freedom, $U/N$ and $V/P$, are exactly halves of the associated hyperparameter estimates, $\sigma^2/2$ and $\rho^2/2$, respectively, in ABIC, analogously to the equipartition theorem~\citep{landau1994statistical} in statistical physics. 
One may notice the equivalence of 
eqs.~(\ref{eq:sigmaatsaddleofposterior}) and (\ref{eq:MAPalpha}) [rewritten as $\sigma^2=2U(\hat {\bf a}_*)/N$ and  $\rho^2=2V(\hat {\bf a}_*)/P$] to the mean field approximation of the equipartition relation, eqs.~(\ref{eq:ABICsigma_equipartition_condition}) and (\ref{eq:ABICrho_equipartition_condition}), dropping the fluctuations around the probability mean. These relations again manifest the MAP neglects the entropic (multiplicity) effects, the effect of fluctuations, counted in ABIC.

\section{Covariance of the marginal posterior of the hyperparameters}
\setcounter{equation}{0} 
\label{sec:ABIChyperparametervariance}
We herein calculate the variance of the hyperparameters in the ABIC estimate, for which we show the list of the first and second derivatives of the marginal likelihood (the marginal posterior times a constant, for the case of the uniform hyperprior) of the hyperparameters. 

Differentiating $P({\bf d}|\sigma^2,\rho^2)$ of the hyperparameters~\citep{akaike1980likelihood,yabuki1992geodetic}, [$P(\sigma^2,\rho^2|{\bf d})$ for the uniform hyperprior, eq.~(\ref{eq:lnmarginalposofhyperforlinearinverse}), times a constant], 
\begin{equation}
\begin{aligned}
\ln P({\bf d}|\sigma^2,\rho^2)=&c
-\frac{N-M}{2}\ln \sigma^2 -\frac P 2 \ln \rho^2
\\&
-\frac 1 2 \ln |{\bf H}^{\rm T}{\bf E}^{-1}{\bf H}+\alpha^2{\bf G}|
-\frac{s({\bf a}_*(\alpha^2),\alpha^2)}{2\sigma^2},
\end{aligned}
\end{equation}
we have its first derivatives as
\begin{flalign}
\left(
\frac{\partial}{\partial \sigma^2}
\ln P({\bf d}|\sigma^2,\rho^2)
\right)_{\rho^2}
&=\nonumber\\
-\frac{N-M}{2}\frac 1 {\sigma^2}
-\frac 1 {2\rho^2}
\mbox{Tr}[{\bf J}(\alpha^2)]
+\frac{U({\bf a}_*(\alpha^2))}{(\sigma^2)^2}
&\\
\left(
\frac{\partial}{\partial \rho^2}
\ln P({\bf d}|\sigma^2,\rho^2)
\right)_{\sigma^2}
&=\nonumber\\
-\frac{P}{2}\frac 1 {\rho^2}
+\frac {\alpha^2} {2\rho^2}
\mbox{Tr}[{\bf J}(\alpha^2)]
+\frac{V({\bf a}_*(\alpha^2))}{(\rho^2)^2}
&
\end{flalign}
and its second derivatives as
\begin{flalign}
\left(
\frac{\partial^2}{\partial (\sigma^2)^2}
\ln P({\bf d}|\sigma^2,\rho^2)
\right)_{\rho^2}
&=\nonumber\\
\frac{N-M}{2}\frac 1 {(\sigma^2)^2}
+\frac 1 {2(\rho^2)^2}
\mbox{Tr}\{[{\bf J}(\alpha^2)]^2\}
&\nonumber\\
-2\frac{U({\bf a}_*(\alpha^2))}{(\sigma^2)^3}-\frac{1}{\alpha^2(\rho^2)^3}\frac{dV({\bf a}_*(\alpha^2))}{d\alpha^2}
\label{eq:marginalposdelsigsig}
&\\
\left(
\frac{\partial}{\partial \sigma^2}
\left(
\frac{\partial}{\partial \rho^2}
\ln P({\bf d}|\sigma^2,\rho^2)
\right)_{\sigma^2}
\right)_{\rho^2}
&=\nonumber\\
\frac 1 {2(\rho^2)^2}
\mbox{Tr}[{\bf J}(\alpha^2)]
-
\frac {\alpha^2} {2(\rho^2)^2}
\mbox{Tr}\{[{\bf J}(\alpha^2)]^2\}&
\nonumber\\
+\frac{1}{(\rho^2)^3}\frac{dV({\bf a}_*(\alpha^2))}{d\alpha^2}
\label{eq:marginalposdelsigrho}
&\\
\left(
\frac{\partial^2}{\partial (\rho^2)^2}
\ln P({\bf d}|\sigma^2,\rho^2)
\right)_{\sigma^2}
&=\nonumber\\
\frac{P}{2}\frac 1 {(\rho^2)^2}
-\frac {\alpha^2} {(\rho^2)^2}
\mbox{Tr}[{\bf J}(\alpha^2)]
+
\frac {(\alpha^2)^2} {2(\rho^2)^2}
\mbox{Tr}\{[{\bf J}(\alpha^2)]^2\}
&\nonumber\\
-2\frac{V({\bf a}_*(\alpha^2))}{(\rho^2)^3}-\frac{\alpha^2}{(\rho^2)^3}\frac{dV({\bf a}_*(\alpha^2))}{d\alpha^2}
&
\label{eq:marginalposdelrhorho}
\end{flalign}
where ${\bf J}(\alpha^2):=({\bf H}^{\rm T}{\bf E}^{-1}{\bf H}+\alpha^2{\bf G})^{-1}{\bf G}$ [defined below eq.~(\ref{eq:EAPvsABIC_2ndorder})], and
\begin{equation}
\frac{dV({\bf a}_*(\alpha^2))}{d\alpha^2}=-{\bf a}_*^{\rm T}(\alpha^2){\bf G}
({\bf H}^{\rm T}{\bf E}^{-1}{\bf H}+\alpha^2{\bf G})^{-1}
{\bf G}{\bf a}_*(\alpha^2);
\label{eq:1stderivofVstar}
\end{equation}
the subscripts of partial derivatives represent fixed variables in the partial differentiation. 
The followings are useful in obtaining the above results:
\begin{flalign}
\left. 
\frac{\partial }{\partial {\bf a}} s({\bf a},\alpha^2)
\right|_{{\bf a}={\bf a}_*(\alpha^2)}&=
0
\label{eq:relofsatastar}
\\
\frac{\partial}{\partial (\alpha^2)} \ln |{\bf H}^{\rm T}{\bf E}^{-1}{\bf H}+\alpha^2  {\bf G}|
&=
\mbox{Tr}[{\bf J}(\alpha^2)].
\label{eq:diffoflndetnormalisedcovariance}
\\
\frac{\partial}{\partial \alpha^2} {\bf J}(\alpha^2)
&=
- [{\bf J}(\alpha^2)]^2
\label{eq:diffoflengthytr}
\\
\frac{\partial}{\partial \alpha^2}
{\bf a}_*^{\rm T}(\alpha^2){\bf G} {\bf a}_*(\alpha^2)
&=
-2 {\bf a}_*^{\rm T}(\alpha^2){\bf G} {\bf J}(\alpha^2) {\bf a}_*(\alpha^2)
\label{eq:diffofaGa}
\end{flalign}
Besides, we used ${\bf G}^{\rm T}={\bf G}$ and the following matrix calculus rules:
\begin{flalign}
\frac{\partial }{\partial x}\ln\det {\bf A}(x)
&=
\mbox{Tr}
\left [{\bf A}^{-1}(x) \frac{\partial }{\partial x}{\bf A}(x)
\right]
\label{eq:diffoflndet}
\\
\frac{\partial}{\partial x}\mbox{Tr}[{\bf A}(x){\bf B}] 
&=\mbox{Tr}\left[\frac{\partial {\bf A}(x)}{\partial x}{\bf B}\right]
\label{eq:diffofmattr}
\\
\frac{\partial}{\partial x}{\bf A}^{-1}(x)
&=-{\bf A}^{-1}\frac{\partial {\bf A}(x)}{\partial x}{\bf A}^{-1} 
\label{eq:diffofinvmat}
\end{flalign}
We note that eq.~(\ref{eq:relofsatastar}) gives $dU({\bf a}_*(\alpha^2))/d \alpha^2=-\alpha^2d V({\bf a}_*(\alpha^2))/d \alpha^2$; then, we can substitute $dU({\bf a}_*(\alpha^2))/d \alpha^2$ for $(-\alpha^2)d V({\bf a}_*(\alpha^2))/d \alpha^2$ in the above expressions. 
It may also be noteworthy for double checks the followings hold for ${\bf a}_*$ that satisfies the mode condition eq.~(\ref{eq:relofsatastar}) of the Gibbsian conditional posterior of ${\bf a}$, $P({\bf a}|\sigma^2,\rho^2,{\bf d})\propto \exp[-s/(2\sigma^2)]$: 
\begin{flalign}
\left(\frac{\partial [s({\bf a}_*(\alpha^2),\alpha^2)/(2\sigma^2)]}{\partial (1/\sigma^2)}\right)_{\rho^2}
&=U({\bf a}_*(\alpha^2))
\\
\left(\frac{\partial [s({\bf a}_*(\alpha^2),\alpha^2)/(2\sigma^2)]}{\partial (1/\rho^2)}\right)_{\sigma^2}
&=V({\bf a}_*(\alpha^2))
\end{flalign}

Using the expressions eqs.~(\ref{eq:marginalposdelsigsig})--(\ref{eq:marginalposdelrhorho}) of the second derivatives, for the uniform hyperprior, we 
expand the log marginal posterior $\ln P(\sigma^2,\rho^2|{\bf d})$ of $\sigma^2$ and $\rho^2$ up to the second order of $\delta\sigma^2=\sigma^2-\hat\sigma^2$ and $\delta\rho^2=\rho^2-\hat\rho^2$: 
\begin{equation}
\ln P(\sigma^2,\rho^2|{\bf d})=
c
-\frac 1 2 
(\delta\sigma^2\hspace{5pt} \delta\rho^2)
{\bf C}_{\hat {\bf h}_{\rm ABIC}}^{-1}
(\delta\sigma^2\hspace{5pt} \delta\rho^2)^{\rm T}
+...
\end{equation}
with
\begin{equation}
\begin{aligned}
&{\bf C}_{\hat {\bf h}_{\rm ABIC}}^{-1}
=
\\
&\left.
-
\left(
\begin{array}{cc}
\frac{\partial^2\ln P({\bf d}|\sigma^2,\rho^2)}{\partial (\sigma^2)^2}
&
\frac{\partial^2\ln P({\bf d}|\sigma^2,\rho^2)}{\partial (\sigma^2)\partial (\rho^2)}
\\
\frac{\partial^2\ln P({\bf d}|\sigma^2,\rho^2)}{\partial (\sigma^2)\partial (\rho^2)}
& 
\frac{\partial^2\ln P({\bf d}|\sigma^2,\rho^2)}{\partial (\rho^2)^2}
\end{array}
\right)
\right|_{\sigma^2=\hat\sigma^2_{\rm ABIC},\rho^2=\hat\rho^2_{\rm ABIC}}
\end{aligned}
\end{equation}
The 0th order is treated as constant in the above expression. 
Hereafter, brackets for the partial derivatives are omitted.
In terms of the diagonal part of the covariance, we have
\begin{flalign}
\frac{\partial^2 \ln P({\bf d}|\sigma^2,\rho^2)}{\partial (\sigma^2)^2}
=&
\frac{N-M}{2(\sigma^2)^2}-\frac{2U({\bf a}_*(\alpha^2))}{(\sigma^2)^3}
+
\frac {\mbox{Tr}[{\bf J}(\alpha^2)]} {2\alpha^2(\rho^2)^2}
\nonumber\\&
-\frac{1}{\alpha^2}
\frac{\partial^2 \ln P({\bf d}|\sigma^2,\rho^2)}{\partial \sigma^2\partial \rho^2}
\\
\frac{\partial^2\ln P({\bf d}|\sigma^2,\rho^2)}{\partial (\rho^2)^2}
=&
\frac{P}{2(\rho^2)^2}
-\frac{2V({\bf a}_*(\alpha^2))}{(\rho^2)^3}
-\frac {\alpha^2\mbox{Tr}[{\bf J}(\alpha^2)]} {2(\rho^2)^2}
\nonumber\\&
-
\alpha^2
\frac{\partial^2 \ln P({\bf d}|\sigma^2,\rho^2)}{\partial \sigma^2\partial \rho^2}
\end{flalign}
which are simplified as follows at the extrema $(\sigma^2,\rho^2)=(\hat\sigma^2_{\rm ABIC},\hat\rho^2_{\rm ABIC})$ of the marginal posterior of the hyperparameters that satisfy eqs.~(\ref{eq:sigmaatsaddleofmarginalposterior}) and (\ref{eq:ABICalpha_explicitextremal}):
\begin{flalign}
\frac{\partial^2 \ln P({\bf d}|\sigma^2,\rho^2)}{\partial (\sigma^2)^2}
=&
-\frac{U({\bf a}_*(\alpha^2))}{(\sigma^2)^3}
-\frac{1}{\alpha^2}
\frac{\partial^2 \ln P({\bf d}|\sigma^2,\rho^2)}{\partial \sigma^2\partial \rho^2}
\\
\frac{\partial^2\ln P({\bf d}|\sigma^2,\rho^2)}{\partial (\rho^2)^2}
=&
-\frac{V({\bf a}_*(\alpha^2))}{(\rho^2)^3}
-
\alpha^2
\frac{\partial^2 \ln P({\bf d}|\sigma^2,\rho^2)}{\partial \sigma^2\partial \rho^2}.
\end{flalign}
We then arrive at
\begin{equation}
\begin{aligned}
|{\bf C}_{\hat {\bf h}_{\rm ABIC}}^{-1}|
=
&\left[
\frac{s({\bf a}_*(\alpha^2),\alpha^2)}{2\sigma^2(\sigma^2\rho^2)}
\frac{\partial^2 \ln P({\bf d}|\sigma^2,\rho^2)}{\partial \sigma^2\partial \rho^2}
\right.
\\
&+\left.
\frac{U({\bf a}_*(\alpha^2))V({\bf a}_*(\alpha^2))}{(\sigma^2)^3(\rho^2)^3}
\right]_{\sigma^2=\hat\sigma^2_{\rm ABIC},\rho^2=\hat\rho^2_{\rm ABIC}}
\end{aligned}    
\end{equation}
Finally, we have a closed-form expression of the covariance ${\bf C}_{\hat {\bf h}_{\rm ABIC}}$:
\begin{flalign}
&{\bf C}_{\hat {\bf h}_{\rm ABIC}}
=
|{\bf C}_{\hat {\bf h}_{\rm ABIC}}^{-1}|^{-1}
\times\nonumber\\
&\left.
\left(
\begin{array}{cc}
-
\frac{\partial^2\ln P({\bf d}|\sigma^2,\rho^2)}{\partial (\rho^2)^2}
&
\frac{\partial^2\ln P({\bf d}|\sigma^2,\rho^2)}{\partial (\sigma^2)\partial (\rho^2)}
\\
\frac{\partial^2\ln P({\bf d}|\sigma^2,\rho^2)}{\partial (\sigma^2)\partial (\rho^2)}
& 
-
\frac{\partial^2\ln P({\bf d}|\sigma^2,\rho^2)}{\partial (\sigma^2)^2}
\end{array}
\right)
\right|_{\sigma^2=\hat\sigma^2_{\rm ABIC},\rho^2=\hat\rho^2_{\rm ABIC}}
\label{eq:covarianceand2ndderivatives}
\end{flalign}

The order estimate of $|{\bf C}_{\hat {\bf h}_{\rm ABIC}}^{-1}|$ and the second derivatives of the marginal posterior of the hyperparameters 
yield the following through eq.~(\ref{eq:covarianceand2ndderivatives}): 
\begin{equation}
{\bf C}_{\hat {\bf h}_{\rm ABIC}}=
\frac{1}{\mathcal O(NP)+\mathcal O(P^2)}
\left[
\mathcal O(P)+\left(\begin{array}{cc}0&0\\0&\mathcal O(N)\end{array}\right)
\right],
\end{equation}
or equivalently, 
\begin{equation}
\begin{aligned}
{\bf C}_{\hat {\bf h}_{\rm ABIC}}=&
\mathcal O[\min(1/N,1/P)]
\\
&+\left(\begin{array}{cc}0&0\\0&\mathcal O[(N/P)\min(1/N,1/P)]\end{array}\right).
\end{aligned}
\end{equation}
We consider $\mathcal O[\mbox{Tr}(A)]=\mathcal O[\ln \det (A)]=\mathcal O[\mbox{rk}(A)]$ for a matrix $A$ of rank $\mbox{rk}(A)$ with $U=\mathcal O(N)$, $V=\mathcal O(P)$ and $-\alpha^2 dV({\bf a}_*(\alpha^2))/d \alpha^2=dU({\bf a}_*(\alpha^2))/d \alpha^2=\mathcal O[\min(N,P)]$; throughout the paper, $f=\mathcal O(x)$ in calculations implies the asymptotic realisation of $c_-x<f<c_+x$ with constants $c_\pm$, which may be written as $\Theta(\cdot)$, rather than simply indicating $f<c_+x$ in an asymptotic sense. 
The first term vanishes for large $N$ or large $P$, and the second term remains as an $\mathcal O(1/P)$ term for large $N$ while vanishes for large $P$. 
For large $N$, the role of the prior is small from the beginning [$\mathcal O(P/N)$ in the regularised least-square solution], and probably for this reason, the constraint on $\rho^2$ (or equivalently, on $\alpha^2$) is relatively weaker for $N\gg P$. On the other hand, $\sigma^2$ variations and the cross-correlation of the hyperparameters are well regulated both for large $N$ and for large $P$. 
The standard deviations of the hyperparameters in the ABIC estimate are $\mathcal O[({\bf C}_{\hat {\bf h}_{\rm ABIC}})^{1/2}]$, which approaches to 0 for large $P$.

The log marginal posterior $\ln P({\bf d}|\sigma^2,\rho^2)$ of the hyperparameters comprises the terms proportional to $N$ and $P$ (and $N-M$), 
and hence with a similar argument to in \S\ref{sec:deltaposteriorintermsofmodelparameters}, 
$P(\sigma^2,\rho^2|{\bf d})$ approaches to $0$ as
$P$ increases, except for at the probability peak $\sigma^2=\hat\sigma^2$, $\rho^2=\hat\rho^2$, 
where the probability diverges to infinity. Then, we have the following relation for large $P$: 
\begin{equation}
P(\sigma^2,\rho^2|{\bf d})\to \delta(\sigma^2-\hat\sigma^2)\delta(\rho^2-\hat\rho^2),
\label{eq:uniquemodeofPsigrho}
\end{equation}
where the unique mode (the unique maximum) of $P(\sigma^2,\rho^2|{\bf d})$ is presumed, and we exclude an ill-posed case $N+P<M$ by considering $P=\mathcal O(M)$.
Equation~(\ref{eq:uniquemodeofPsigrho}) shows that the estimates of $\sigma^2$ and $\rho^2$ with finite probabilities, which include the posterior mean and median in this case, are asymptotically consistent with their ABIC estimates. 
The asymptotic form for large $N$ is rather complicated, as 
the increase in $N$ only affects partial derivatives of $P(\sigma^2,\rho^2|{\bf d})$ with respect to $\sigma^2$; it is noticed from the order estimate of the first derivatives, giving the higher-orders as their differentials. 
Then for large $N$, considering $N\geq M$ and excluding an ill-posed case $N+P<M$, we have
\begin{equation}
P(\sigma^2,\rho^2|{\bf d})\propto \delta(\sigma^2-\hat\sigma^2).
\end{equation}

\section{Reduction using the marginal posterior of the model parameters}
Some semianalytic results are obtained here for the marginal posterior of the model parameters, $P({\bf a}|{\bf d})$. We assume $N\gg1$ and $P\gg1$ in this section. 
\setcounter{equation}{0} 
\label{sec:FBmarginilisation}

\subsection{The second-order moment of the marginal posterior of the model parameters around the fully Bayesian MMPM estimate}

We evaluate the second-order moment of the marginal posterior around the extremum(s) below. 
The subscript ${}_{\rm MMPM}$ for the MMPM estimate $\hat\cdot_{\rm MMPM}$ is omitted throughout this subsection. 
Defining
$
\hat U:=U(\hat {\bf a})
$, $
\hat V:=V(\hat {\bf a})
$, $
\delta U({\bf a}):=U({\bf a})-\hat U
$ and $
\delta V({\bf a}):=V({\bf a})-\hat V
$, 
we rewrite the marginal posterior eq.~(\ref{eq:marginalizedpostprob}) as
\begin{flalign}
P({\bf a}|{\bf d})
&\propto [U^{-N/2+1}V^{-P/2+1}]/[\hat U^{-N/2+1}\hat V^{-P/2+1}] 
\\
&= \left (1+\frac{\delta U} {\hat U}\right)^{-N/2+1}
\left (1+\frac{\delta V}{\hat V}\right)^{-P/2+1},
\end{flalign}
where $\hat U$ and $\hat V$ are treated as constants. 
Considering the terms up to the second-order of 
$\delta U/\hat U$ and $\delta V/\hat V$ (which fully contain the deviation of ${\bf a}$ from $\hat {\bf a}$ up to its second-order), 
and using
$
1+x=e^{x-x^2/2}[1+\mathcal O(x^3)]
$, 
we have
\begin{equation}
\begin{aligned}
P({\bf a}|{\bf d})
\approx 
c\exp &
\left\{
\left(-\frac N 2+1\right)
\left[
\frac{\delta U}{\hat U}
-\frac 1 2 \left(\frac{\delta U}{\hat U}\right)^2
\right]
\right.\\&\left.
+\left(-\frac P 2 +1\right)
\left[
\frac{\delta V}{\hat V}
-\frac 1 2 \left(\frac{\delta V}{\hat V}\right)^2
\right]
\right\}.
\end{aligned}
\label{eq:varianceevaluationofPad_1st2nd}
\end{equation}
In the exponential of eq.~(\ref{eq:varianceevaluationofPad_1st2nd}), 
the first order of $\delta U/\hat U$ and $\delta V/\hat V$ is evaluated as
\begin{equation}
\begin{aligned}
&
\exp\left\{\left(-\frac N 2+1\right)
\frac{\delta U}{\hat U}
+\left(-\frac P 2 +1\right)
\frac{\delta V}{\hat V}
\right\}
=
\\&
\exp\left[
-\frac{1}{2\check \sigma^2}
\left(
s({\bf a},\check \alpha^2)
-s(\hat {\bf a},\check \alpha^2)
\right)\right]
\label{eq:firstorderofdelUV}
\end{aligned}
\end{equation}
with  
\begin{equation}
\check \sigma^2:= \frac{2\hat U}{N-2}=\frac {({\bf d}-{\bf H}\hat {\bf a})^{\rm T} {\bf E}^{-1} ({\bf d}-{\bf H}\hat {\bf a}) }{N-2}.
\label{eq:checksigmamode}
\end{equation}
We note
$s({\bf a},\check \alpha^2)
-s(\hat {\bf a},\check \alpha^2)=({\bf a}-\hat {\bf a})^{\rm T}
({\bf H}^{\rm T}{\bf E}^{-1}{\bf H}+\check\alpha^2 {\bf G})
({\bf a}-\hat {\bf a})$.
We also obtain a parallel relation for $\hat V$ from eqs.~(\ref{eq:checkalphamode}) and (\ref{eq:checksigmamode}):
\begin{equation}
\check \rho^2=\frac{\check \sigma^2}{\check\alpha^2}= \frac{2\hat V}{P-2}.
\label{eq:checkrhomode}
\end{equation}
Equations~(\ref{eq:checkalphamode}) and (\ref{eq:checksigmamode}) also give 
another form of $\check \sigma^2$: 
\begin{equation}
\check \sigma^2= \frac{s(\hat {\bf a},\check \alpha^2)}{N+P-4}.
\end{equation}
The virtually appearing hyperparameters $\check \sigma^2$ and $\check \rho^2$  ($\check\sigma^2_{\rm MMPM}$ and $\check\rho^2_{\rm MMPM}$) given by the same functional forms as $\hat\sigma^2_{\rm MAP}$ and $\hat\rho^2_{\rm MAP}$ in the MAP estimate, after replacing the associated ${\bf a}$ and $\alpha^2$ values
and excluding $\mathcal O[\min(1/N,1/P)]$ factors. 
Note the difference between
$\hat\sigma^2_{\rm MAP}$ and $\check\sigma^2_{\rm MMPM}$ is $\mathcal O(1/N)+\mathcal O(1/P)$ 
because of 
the $\mathcal O(1/N)+\mathcal O(1/P)$
difference between
$\hat \alpha^2_{\rm MAP}$ and $\check\alpha^2_{\rm MMPM}$. 
In the exponential of eq.~(\ref{eq:varianceevaluationofPad_1st2nd}), the second-order of 
$\delta U/\hat U$ and $\delta V/\hat V$ is expressed as follows up to the second-order of 
$({\bf a}-\hat{\bf a})$:
\begin{equation}
\begin{aligned}
&
\left(\frac N 2 -1\right)
\left(\frac {\delta U}{\hat U} \right)^2
+
\left(\frac P 2 -1\right)
\left(\frac {\delta U}{\hat U} \right)^2
\\
&\approx
\frac {4}{2\check\sigma^2}
({\bf a}-\hat {\bf a})^{\rm T}
\\&
\left\{
\frac 1 {(N-2)\check \sigma^2}
{\bf H}^{\rm T}{\bf E}^{-1}({\bf H}\hat{\bf a}-{\bf d})({\bf H}\hat{\bf a}-{\bf d})^{\rm T}{\bf E}^{-1}{\bf H}
\right.\\&\left.
+
\frac {(\check\alpha^2)^2} {(P-2)\check \sigma^2}
{\bf G}\hat{\bf a}\hat{\bf a}^{\rm T}{\bf G}
\right\}
({\bf a}-\hat {\bf a}), 
\end{aligned}
\label{eq:secondorderofdelUV}
\end{equation}
where we used eqs.~(\ref{eq:checksigmamode}) and (\ref{eq:checkrhomode}). 
Combining the first and second orders of 
$\delta U/\hat U$ and $\delta V/\hat V$ (eqs.~\ref{eq:firstorderofdelUV} and \ref{eq:secondorderofdelUV}), 
we have the following expression of eq.~(\ref{eq:varianceevaluationofPad_1st2nd}) 
up to the second-order of 
$({\bf a}-\hat{\bf a})$: 
\begin{equation}
P({\bf a}|{\bf d})
\propto 
\exp\left[
-\frac{1}{2\check\sigma^2}({\bf a}-\hat {\bf a})^{\rm T}({\bf C}^\prime_{\hat{\bf a}})^{-1}({\bf a}-\hat {\bf a})
+...
\right]
\end{equation}
with 
\begin{equation}
\begin{aligned}
({\bf C}^\prime_{\hat{\bf a}})^{-1}
&=
({\bf H}^{\rm T}{\bf E}^{-1}{\bf H}+\check\alpha^2 {\bf G})
\\
&+4\left[
\frac 1 {(N-2)\check \sigma^2}
{\bf H}^{\rm T}{\bf E}^{-1}({\bf H}\hat{\bf a}-{\bf d})({\bf H}\hat{\bf a}-{\bf d})^{\rm T}{\bf E}^{-1}{\bf H}
\right.
\\&
\left.
+
\frac {(\check\alpha^2)^2} {(P-2)\check \sigma^2}
{\bf G}\hat{\bf a}\hat{\bf a}^{\rm T}{\bf G}
\right]. 
\end{aligned}
\end{equation}
The second-order moment ${\bf C}_{\hat {\bf a}}$ (${\bf C}_{\hat {\bf a}_{\rm MMPM}}$) around $\hat{\bf a}$ ($\hat{\bf a}_{\rm MMPM}$) is then expressed as 
\begin{equation}
{\bf C}_{\hat {\bf a}}=\check\sigma^2{\bf C}^\prime_{\hat {\bf a}}.
\end{equation}

\subsection{An asymptotic form of the marginal posterior of model parameters}
\label{sec:deltaposteriorintermsofmodelparameters}
The marginal posterior of the model parameters (eq.~\ref{eq:marginalizedpostprob}), which can be written as
\begin{equation}
P({\bf a}|{\bf d})= c\left(UV^{\frac{P-2}{N-2}}\right)^{-\frac N 2 +1},
\end{equation}
takes large values only around the minima of $UV^{\frac{P-2}{N-2}}$, and $P({\bf a}|{\bf d})$ values for the other cases become negligible for large $N$ or $P$. 
Indeed, 
$P({\bf a}|{\bf d})$ converges to a delta function in the limit of infinite $N$ or $P$, as shown below. 
This means that a model-parameter estimate approaches to the MMPM estimate or has zero probability asymptotically for large $N$ or $P$. 
The following presumes the mode (the maximum) of $P({\bf a}|{\bf d})$ is unique. 

Hereafter, we proceed with the calculation by using the order estimate of $U$ and $V$: $U=\mathcal O(N)$ and $V=\mathcal O(P)$. 
For brevity, we introduce the following function:
\begin{equation}
x({\bf a}):=\frac{UV^{\frac{P-2}{N-2}}-\min_{\bf a}[UV^{\frac{P-2}{N-2}}]}{NP^{\frac{P-2}{N-2}}}. 
\end{equation}
The function $x({\bf a})$ takes a nonnegative number of $\mathcal O(N^0P^0)$ and is zero only when $UV^{\frac{P-2}{N-2}}$ takes its minimum value. 
Using $x({\bf a})$, we can express $P({\bf a}|{\bf d})$ as
\begin{equation}
P({\bf a}|{\bf d})= c[1+x({\bf a})]^{-N/2+1}.
\end{equation}
First, we consider the limit of $N,P\to\infty$ with keeping $N/P$ finite. The case of taking only $N\to\infty$ or only $P\to\infty$ is mentioned later.

We first consider a function
$[(N-4)/2] (1+x)^{-N/2+1}$, which is an increasing function of $N$ only at $x=0$, and it approaches to zero otherwise in the limit of $N\to\infty$ for fixed finite $N/P$: 
\begin{equation}
\frac {N-4} 2 (1+x)^{-N/2+1}
\to
\left\{
\begin{array}{ll}
\infty& x=0
\\
0& x> 0
\end{array}
\right.
\end{equation}
Besides, it satisfies the normalisation condition when $N>4$:
\begin{equation}
\int_0^\infty dx\frac {N-4} 2 (1+x)^{-N/2+1}=1.
\end{equation}
Because of these characteristics, $[(N-4)/2] (1+x)^{-N/2+1}$ approaches to the delta function of $x-0$ in the limit of $N\to\infty$: 
\begin{equation}
\frac {N-4} 2 (1+x)^{-N/2+1}\to \delta(x-0).
\end{equation} 
Note $\int^\infty_0 dx\delta (x-0)=1$. 
This asymptotic function shows that 
the ${\bf a}$-dependent part of 
$P({\bf a}|{\bf d})$ converges to a delta function with an appropriate constant multiplication. 

The remaining constant part is determined by the normalisation condition of $P({\bf a}|{\bf d})$:
\begin{equation}
\int d{\bf a}P({\bf a}|{\bf d})=1.
\end{equation}
This normalisation condition and the preceding asymptotic form of $[(N-4)/2](1+x)^{-N/2+1}$ deduce
\begin{equation}
P({\bf a}|{\bf d})\to
c_\infty\delta(x-0)
\end{equation}
and 
\begin{equation}
c_\infty\leq
\frac{1}
{\int d{\bf a}^\prime\delta(x({\bf a}^\prime)-0)},
\end{equation}
where constant $c_\infty$ does not necessarily satisfy the equality (supplemented below). 
Using $X({\bf a})$ defined in eq.~(\ref{eq:univariableofPad}) as $x({\bf a})=X({\bf a})-\min_{\bf a} X({\bf a})$, 
we obtain the desired asymptotic expression, eq.~(\ref{eq:deltafunctionalconvergence_modelparameter}) for $N/P\geq 1$, 
which is also applicable to the limit of increasing only $N$. 
Repeating the above calculation with converting 
$V\to U$ and $P \to N$ (and also $U\to V$ and $N\to P$) 
yields eq.~(\ref{eq:deltafunctionalconvergence_modelparameter}) for $N/P< 1$, which covers $P/N\to\infty$. 

As the marginal posterior approaches to the delta function, 
a model-parameter state with a finite probability 
approaches to the mode of $P({\bf a}|{\bf d})$ (the MMPM estimate). 
Here we should emphasise $c_\infty$ does not necessarily satisfy 
$c_\infty=1/\int d{\bf a}^\prime\delta(x({\bf a}^\prime)-0)$, a relation that means the population mean (the EAP estimate) coincides with the MMPM estimate. 
As explicated in the discussion, due to the balance between the exponential increase in the integration volume ($\int d{\bf a}$) of the probability space and the exponential decrease in the probability values [$P({\bf a}|{\bf d})$], the integral of the probability distribution over the asymptotically measure-zero domain can be finite [i.e. $c_\infty\leq 1/\int d{\bf a}^\prime\delta(x({\bf a}^\prime)-0)$] even in the asymptotic limit, especially for a large number $M\gg1$ ($P\gg1$) of model parameters.
Then, the posterior median is also not necessarily the mode, 
although the posterior median is neither necessarily consistent with the posterior mean for such a case.

The above results are further generalised for the cases of the hyperpriors given by
$P(\sigma^2)=(\sigma^2)^{n_{\sigma^2}}$ and
$P(\rho^2)=(\rho^2)^{n_{\rho^2}}$ treated in the text, 
such as the logarithmically uniform hyperpriors, 
considering the asymptotic limit of
$N-2n_{\sigma^2},P-2n_{\rho^2}\to\infty
$.

\section{Fully Bayesian EAP estimate}
\setcounter{equation}{0} 
\label{sec:FBEAP}
The EAP estimate of the model parameters ${\bf a}$ is calculated here for the uniform hyperprior of $\sigma^2$ and $\rho^2$. 
Decomposing the joint posterior of the model parameters ${\bf a}$ and the hyperparameters ${\bf h}$ as $P({\bf a},{\bf h}|{\bf d})=P({\bf a}|{\bf h},{\bf d})P({\bf h}|{\bf d})$ 
yields the following identity of the EAP estimate of ${\bf a}$:
\begin{equation}
\hat {\bf a}_{\rm EAP}
=
\langle
{\bf a}_*(\alpha^2)
\rangle_{{\bf h}|{\bf d}},
\end{equation}
where ${\bf a}_*(\alpha^2)=
\langle
{\bf a}
\rangle_{{\bf a}|{\bf h},{\bf d}}
$ is the conditional mean of ${\bf a}$ given the hyperparameters.
For the present case, 
\begin{equation}
\hat {\bf a}_{\rm EAP}=
\langle
({\bf H}^{\rm T}{\bf E}^{-1}{\bf H}+\alpha^2{\bf G})^{-1}
\rangle_{\alpha^2|{\bf d}}
{\bf H}^{\rm T}{\bf E}^{-1}{\bf d}.
\label{eq:formalEAPexpression}
\end{equation}
We obtain a closed-form expression of eq.~(\ref{eq:formalEAPexpression}) in this section. 

We start the calculation by expanding 
\begin{equation}
{\bf C}_{{\bf a}_*}^\prime(\alpha^2):=
({\bf H}^{\rm T}{\bf E}^{-1}{\bf H}+\alpha^2{\bf G})^{-1}
\end{equation}
as an infinite series in $\delta \alpha^2:=\alpha^2-\hat \alpha^2_{\rm ABIC}$: 
\begin{equation}
{\bf C}_{{\bf a}_*}^\prime(\alpha^2)=
\left[\sum^\infty_{m=0}(-\hat{\bf J})^m(\delta \alpha^2)^m\right]
{\bf C}_{{\bf a}_*}^\prime(\hat \alpha^2_{\rm ABIC}),
\label{eq:expanding_invcovmat}
\end{equation}
which is a matrix version of $(1+x)^{-1}=\sum_{n=0}^\infty (-x)^n$, 
where
$ \hat{\bf J}:=
{\bf J}(\hat \alpha^2_{\rm ABIC})
$, and 
${\bf J}(\alpha^2)$ is defined below eq.~(\ref{eq:EAPvsABIC_2ndorder}).
Substituting eq.~(\ref{eq:expanding_invcovmat}) into 
eq.~(\ref{eq:formalEAPexpression}), we have
\begin{equation}
\hat {\bf a}_{\rm EAP}=
\left\langle
\sum^\infty_{m=0}(-\hat{\bf J})^m(\delta \alpha^2)^m
\right\rangle_{\alpha^2|{\bf d}}
\hat {\bf a}_{\rm ABIC}.
\label{eq:expandingEAPfromABIC}
\end{equation}
Equation~(\ref{eq:expandingEAPfromABIC}) sets a correction factor for obtaining the EAP estimate from the ABIC estimate.

We further rewrite eq.~(\ref{eq:expandingEAPfromABIC}) 
as another series in $\delta \ln \alpha^2:=\ln \alpha^2-\ln \hat \alpha^2_{\rm ABIC}$, that is a fast convergent series for the case of $\delta\alpha^2/\hat \alpha^2_{\rm ABIC}\ll1$. 
Using 
\begin{equation}
\delta \alpha^2=\hat \alpha^2_{\rm ABIC}[\exp(\delta \ln\alpha^2)-1],
\end{equation}
we obtain the following: 
\begin{equation}
\begin{aligned}
&
\sum^\infty_{m=0}(-\hat{\bf J})^m(\delta \alpha^2)^m
\\
&=
{\bf I}-\hat \alpha^2_{\rm ABIC}\hat{\bf J}\delta \ln \alpha^2
-\left(\frac 1 2\hat \alpha^2_{\rm ABIC}\hat{\bf J} -(\hat \alpha^2_{\rm ABIC} \hat{\bf J})^2 \right)(\delta \ln\alpha^2)^2
\\&+...
\end{aligned}
\label{eq:transform_correctionfactorfromABICtoEAP}
\end{equation}

In the left part, we calculate the posterior mean of the right hand side in eq.~(\ref{eq:transform_correctionfactorfromABICtoEAP}) and evaluate eq.~(\ref{eq:expandingEAPfromABIC}). 
The marginal posterior of $\ln\alpha^2$ is obtained with that of $\alpha^2$
as follows using the variable transforms 
$P(\alpha^2,\sigma^2|{\bf d})
=\sigma^2/(\alpha^2)^2 P(\sigma^2,\rho^2|{\bf d})$ and 
$P(\ln\alpha^2|{\bf d})=\alpha^2P(\alpha^2|{\bf d})$ [note an equality $P(y)dy=P(x)dx$ for a bijective function $y(x)$ of $x$]: 
\begin{flalign}
P(\alpha^2|{\bf d})
=&\int d\sigma^2 P(\sigma^2,\alpha^2|{\bf d})
\\=&
c_\alpha
[s({\bf a}_*(\alpha^2),\alpha^2)]^{-(N+P-M-4)/2}
\nonumber\\&\times
(\alpha^2)^{(P-4)/2}|[{\bf C}^\prime_{{\bf a}_*}(\alpha^2)]^{-1}|^{-1/2}
\\
P(\ln\alpha^2|{\bf d})
=&
c_\alpha
[s({\bf a}_*(\alpha^2),\alpha^2)]^{-(N+P-M-4)/2}
\nonumber\\&\times
(\alpha^2)^{(P-2)/2}
|[{\bf C}^\prime_{{\bf a}_*}(\alpha^2)]^{-1}|^{-1/2} 
\end{flalign}
where a coefficient $c_\alpha=4^{-1}c\pi^{-(N+P-M)/2}\Gamma(-(N+P-M-4)/2)
|{\bf E}|^{-1/2}|\boldsymbol\Lambda_G|^{1/2}/P({\bf d})$ denotes the constant part independent of $\alpha^2$ and $\ln\alpha^2$ [here using the constant $c$ for normalising the uniform $P(\sigma^2,\rho^2)$, $P({\bf d})$ and $c$ in the prior of the model parameters]. 

Applying Laplace's method to the marginal posterior of
$\ln\alpha^2$ in eq.~(\ref{eq:transform_correctionfactorfromABICtoEAP}), we acquire
\begin{equation}
\begin{aligned}
&\left\langle
\sum^\infty_{m=0}(-\hat{\bf J})^m(\delta \alpha^2)^m
\right\rangle_{\alpha^2|{\bf d}}
\\
&=
\left[
{\bf I}
+
\left.
\left(\frac 1 2 \alpha^2{\bf J} -(\alpha^2 {\bf J})^2 \right)
\left(
\frac{\partial^2\ln P(\ln\alpha^2|{\bf d})}{\partial (\ln\alpha^2)^2}
\right)^{-1}
\right|_{\alpha^2=\hat\alpha^2_{\rm ABIC}}
\right.\\&\left.
+\mathcal O
\left(
\left.
\left(
\frac{\partial^2\ln P(\ln\alpha^2|{\bf d})}{\partial (\ln\alpha^2)^2}
\right|_{\alpha^2=\hat\alpha^2_{\rm ABIC}}
\right)^{-2}
\right)
\right].
\end{aligned}
\label{eq:lap_EAP_lnalpha_Cov}
\end{equation}
We further utilised the fact that the peak of $P(\ln\alpha^2|{\bf d})$ coincides with $\hat\alpha^2_{\rm ABIC}$, located at the peak of $P(\sigma^2,\rho^2|{\bf d})$, excluding
an $\mathcal O(1/P)$ shift [obtained from the comparison between the maximisation function of ABIC eq.~(\ref{eq:ABICalpha}), corresponding to eq.~(\ref{eq:marginalizedposforhyper}), and $\ln P(\ln \alpha^2|{\bf d})$]; the error due to the $\mathcal O(1/P)$ peak shift is on the same order as the third term mentioned later, so collectively expressed in it. 
Substituting eq.~(\ref{eq:lap_EAP_lnalpha_Cov}) into eq.~(\ref{eq:expandingEAPfromABIC}), we find
\begin{equation}
\begin{aligned}
\hat {\bf a}_{\rm EAP}&=
\left[
{\bf I}
+\left(\frac 1 2 \alpha^2{\bf J} -( \alpha^2 {\bf J})^2 \right)
\left(
\frac{\partial^2\ln P(\ln\alpha^2|{\bf d})}{\partial (\ln\alpha^2)^2}
\right)^{-1}
\right.\\&\left.
+\mathcal O
\left(
\left(
\frac{\partial^2\ln P(\ln\alpha^2|{\bf d})}{\partial (\ln\alpha^2)^2}
\right)^{-2}
\right)
\right]_{\alpha^2=\hat\alpha^2_{\rm ABIC}}
\hat {\bf a}_{\rm ABIC}.
\end{aligned}
\label{eq:lap_EAP_lnalpha}
\end{equation}
The following rewritten forms of eqs.~(\ref{eq:diffofsstaralpha}) and (\ref{eq:diffoflndetnormalisedcovariance}), 
\begin{flalign}
\frac{\partial}{\partial \ln\alpha^2}\log s({\bf a}_*(\alpha^2),\alpha^2)
&=
\frac{\alpha^2{\bf a}_*^{\rm T}{\bf G}{\bf a}_*}{s({\bf a}_*(\alpha^2),\alpha^2)}
\\
\frac{\partial}{\partial \ln\alpha^2}\log
|[{\bf C}^\prime_{{\bf a}_*}(\alpha^2)]^{-1}|
&=
\mbox{Tr}(\alpha^2{\bf J}),
\end{flalign}
and $\partial {\bf J}/\partial \alpha^2=-{\bf J}^2$
(eq.~\ref{eq:diffoflengthytr}) yield 
\begin{flalign}
\frac{\partial^2}{\partial (\ln\alpha^2)^2}\log s({\bf a}_*(\alpha^2),\alpha^2)
&=
-\left(\frac{\alpha^2{\bf a}_*^{\rm T}{\bf G}{\bf a}_*}{s({\bf a}_*(\alpha^2),\alpha^2)}\right)^2
\nonumber\\&
+\frac{\alpha^2{\bf a}_*^{\rm T}{\bf G}({\bf I}-2\alpha^2 {\bf J}){\bf a}_*}{s({\bf a}_*(\alpha^2),\alpha^2)}
\\
\frac{\partial^2}{\partial (\ln\alpha^2)^2}\log
|[{\bf C}^\prime_{{\bf a}_*}(\alpha^2)]^{-1}|
&=
\mbox{Tr}[\alpha^2
({\bf I}-\alpha^2{\bf J})
{\bf J}]
\end{flalign}
and then the second derivative of the marginal posterior of $\ln\alpha^2$ is evaluated as
\begin{equation}
\begin{aligned}
&\frac{\partial^2\ln P(\ln\alpha^2|{\bf d})}{\partial (\ln\alpha^2)^2}=
\\
&-\frac 1 2 \mbox{Tr}[\alpha^2
({\bf I}-\alpha^2{\bf J})
{\bf J}]
-\frac{N+P-M-4}{2}
\\&\times
\left[
-\left(\frac{\alpha^2{\bf a}_*^{\rm T}{\bf G}{\bf a}_*}{s({\bf a}_*(\alpha^2),\alpha^2)}\right)^2
+\frac{\alpha^2{\bf a}_*^{\rm T}{\bf G}({\bf I}-2\alpha^2 {\bf J}){\bf a}_*}{s({\bf a}_*(\alpha^2),\alpha^2)}
\right].
\end{aligned}
\label{eq:secondderivoflnPlnalpha_app}
\end{equation}
We evaluate the derivatives of the determinant and trace as eqs.~(\ref{eq:diffoflndet}) and (\ref{eq:diffofmattr}), respectively, and the derivative of $s({\bf a}_*,\alpha^2)$ as eq.~(\ref{eq:diffofsstaralpha}). 
The first term is $\mathcal O(P)$, and the second term in eq.~(\ref{eq:secondderivoflnPlnalpha_app}) is $\mathcal O(N+P-M)\leq \mathcal O(N)$ for large $P$ [where $s=\mathcal O(V)=\mathcal O(P)$] and $\leq \mathcal O(P)$ for large $N$ [where $s=\mathcal O(U)=\mathcal O(N)\geq\mathcal O(N+P-M)$], where excluding an ill-posed case $N+P<M$ is presumed. 
The second derivative of the marginal posterior of $\ln\alpha^2$ is then $\mathcal O(P)$, so the posterior covariance of $\ln \alpha^2$ given by its inverse is $\mathcal O(1/P)$. 
Therefore, 
the difference between $\hat {\bf a}_{\rm EAP}$ and $\hat {\bf a}_{\rm ABIC}$ in eq.~(\ref{eq:lap_EAP_lnalpha}) is also $\mathcal O(1/P)$: 
\begin{equation}
\hat {\bf a}_{\rm EAP}=
[
{\bf I}+\mathcal O(1/P)
]
\hat {\bf a}_{\rm ABIC}.
\end{equation}

We note expanding the log posterior $\ln P(\alpha^2|{\bf d})$ of $\alpha^2$ around its maximum for $\delta \alpha^2\ll1$ leads to another series expansion, which may be more straightforward: 
\begin{equation}
\begin{aligned}
&\left\langle
\sum^\infty_{m=0}(-\hat{\bf J})^m(\delta \alpha^2)^m
\right\rangle_{\alpha^2|{\bf d}}
\\
=&
\sum^\infty_{n=0}({\bf J}^2)^n\frac{(2n)!}{2^nn!}
\left.
\left(
-
\frac{\partial^2\ln P(\alpha^2|{\bf d})}{\partial (\alpha^2)^2}
\right)^{-n}
\right|_{\alpha^2=\hat\alpha^2_{\rm ABIC}+\mathcal O(1/P)}
\\&
+\mathcal O
\left(
\left.
\left(
\frac{\partial^2\ln P(\alpha^2|{\bf d})}{\partial (\alpha^2)^2}
\right|_{\alpha^2=\hat\alpha^2_{\rm ABIC}+\mathcal O(1/P)}
\right)^{-2}
\right)
.
\end{aligned}
\label{eq:lap_EAP_alpha}
\end{equation}
We applied Laplace's method to the marginal posterior of $\alpha^2$
and utilised the fact that the peak of $P(\alpha^2|{\bf d})$ [the peak of $\ln P(\alpha^2|{\bf d})$] coincides with $\hat\alpha^2_{\rm ABIC}$, located at the peak of $P(\sigma^2,\rho^2|{\bf d})$, excluding
an $\mathcal O(1/P)$ shift. 
Equation~(\ref{eq:lap_EAP_alpha}) is applicable to $\delta \alpha^2\ll1$ [precisely, $\hat {\bf J}\delta \alpha^2\ll{\bf I}$, meaning a negligible prior] but not necessarily to $\delta \alpha^2/\hat \alpha^2_{\rm ABIC}\ll1$ [corresponding to the law of large numbers with respect to the hyperparameters] treated in the text, which requires the higher orders of $\delta \alpha^2$ not evaluable in the Gaussian approximation of the $\delta \alpha^2$ distribution assumed in eq.~(\ref{eq:lap_EAP_alpha}).

\end{appendix}
\setcounter{section}{0}
\setcounter{equation}{0}
\renewcommand{\thesection}{S.\arabic{section}}
\renewcommand{\theequation}{S.\arabic{section}.\arabic{equation}}
\section{Supplement 1: Propagation of uncertainty from hyperparameters to model-parameter estimates}
\label{sec:errorprop}

The joint posterior of model parameters ${\bf a}$ and hyperparameters ${\bf h}$ [${\bf h}=(\sigma^2,\rho^2)^{\rm T}$ in the text] is expressed as $P({\bf a},{\bf h}|{\bf d})=P({\bf a}|{\bf h},{\bf d})P({\bf h}|{\bf d})$ by the conditional posterior $P({\bf a}|{\bf h},{\bf d})$ of ${\bf a}$ given ${\bf h}$ and marginal posterior $P({\bf h}|{\bf d})$ of ${\bf h}$. 
When the joint posterior is approximated using the mode (the maximiser) $\hat {\bf h}$ of $P({\bf h}|{\bf d})$ as $P({\bf a},{\bf h}|{\bf d})\approx P({\bf a}|\hat {\bf h},{\bf d})$ [precisely, $P({\bf a}|{\bf d}):=\int d{\bf h} P({\bf a},{\bf h}|{\bf d})\approx P({\bf a}|\hat {\bf h},{\bf d})$] as in the point-estimation ABIC, 
the neglected effect is the propagation of uncertainty resulting from the stochastic fluctuation of the hyperparameters, as often criticised as a problem of the empirical Bayes method~\citep{gelman2013bayesian}. 
The following shows a brief summary of the lowest-order calculation method for this propagation of uncertainty from the hyperparameters to the model parameters.  Specific expressions for the ABIC estimate in the present linear inverse problem are also attached. 

\subsection{Propagation of uncertainty between stochastic variables}
We evaluate the second-order moment of ${\bf a}$ around a given model-parameter estimate $\hat {\bf a}$: 
\begin{equation}
{\bf C}_{\hat {\bf a},{\rm ep}}
:=
\langle
({\bf a}-\hat {\bf a})
({\bf a}-\hat {\bf a})^{\rm T}
\rangle_{{\bf a},{\bf h}|{\bf d}}.
\end{equation}
${\bf C}_{\hat {\bf a},{\rm ep}}$ corresponds to the covariance when $\hat {\bf a}$ is the posterior mean (the EAP estimate). 
This cross product of the deviation of ${\bf a}$ from $\hat{\bf a}$ contains uncertainty propagated from the hyperparameters, as shown in the following decomposition identity of ${\bf C}_{\hat {\bf a},{\rm ep}}$:
\begin{equation}
{\bf C}_{\hat {\bf a},{\rm ep}}
=
\langle
{\bf C}_{{\bf a}|{\bf h}}
\rangle_{{\bf h}|{\bf d}}
+
\left
\langle
(\hat {\bf a}-\langle {\bf a}\rangle_{{\bf a}|{\bf h},{\bf d}})
(\hat {\bf a}-\langle {\bf a}\rangle_{{\bf a}|{\bf h},{\bf d}})^{\rm T}
\right
\rangle_{{\bf h}|{\bf d}},
\label{eq:errorprop_gen}
\end{equation}
where 
${\bf C}_{{\bf a}|{\bf h}}
:=\langle
({\bf a}-\langle {\bf a}\rangle_{{\bf a}|{\bf h},{\bf d}})
({\bf a}-\langle {\bf a}\rangle_{{\bf a}|{\bf h},{\bf d}})^{\rm T}
\rangle_{{\bf a}|{\bf h},{\bf d}}
$ denotes the conditional posterior covariance of the model parameters given the hyperparameters. 
We use some equations for calculation:
\begin{flalign}
\langle
({\bf a}-\hat {\bf a})
({\bf a}-\hat {\bf a})^{\rm T}
\rangle_{{\bf a},{\bf h}|{\bf d}}
&=
\langle
{\bf aa}^{\rm T}
\rangle_{{\bf a},{\bf h}|{\bf d}}
-
\langle
{\bf a}
\rangle_{{\bf a},{\bf h}|{\bf d}}
\hat {\bf a}^{\rm T}
-
\hat
{\bf a}
\langle
{\bf a}
\rangle_{{\bf a},{\bf h}|{\bf d}}
^{\rm T}
+
\hat {\bf a}
\hat {\bf a}^{\rm T}
\\
\langle
{\bf a a}^{\rm T}
\rangle_{{\bf a},{\bf h}|{\bf d}}
&=
\langle
{\bf C}_{{\bf a}|{\bf h}}
\rangle_{{\bf h}|{\bf d}}
+
\langle
\langle {\bf a}\rangle_{{\bf a}|{\bf h},{\bf d}}
\langle {\bf a}\rangle_{{\bf a}|{\bf h},{\bf d}}^{\rm T}
\rangle_{{\bf h}|{\bf d}}
\end{flalign}
Regarding propagation of uncertainty, 
eq.~(\ref{eq:errorprop_gen}) indicates the second-order moment of ${\bf a}$ around $\hat {\bf a}$ (${\bf C}_{\hat {\bf a},{\rm ep}}$) 
comprises (1, the first term) the posterior mean of the conditional posterior covariance ${\bf C}_{{\bf a}|{\bf h}}$ of ${\bf a}$ given ${\bf h}$ 
and (2, the second term)
the second-order moment of the conditional posterior mean $\langle {\bf a}\rangle_{{\bf a}|{\bf h},{\bf d}}$ of ${\bf a}$ given ${\bf h}$, around $\hat {\bf a}$. 
The posterior mean of ${\bf C}_{{\bf a}|{\bf h}}$ (the first term) is not necessarily the same as a specific ${\bf C}_{{\bf a}|{\bf h}}$ value, which appears in the ABIC estimate with ${\bf h}=\hat {\bf h}$. 
Besides, the second term is nonzero (although can be negligibly small) unless the conditional posterior mean
$\langle {\bf a}\rangle_{{\bf a}|{\bf h},{\bf d}}$ of ${\bf a}$ (denoted by ${\bf a}_*$ in the text) takes the same value for all the possible ${\bf h}$ values and is equated to $\hat {\bf a}$. 

Further considering the following decomposition using the posterior mean $\hat{\bf a}_{\rm EAP}:=\langle
{\bf a}
\rangle_{{\bf a},{\bf h}|{\bf d}}$ of ${\bf a}$, 
\begin{flalign}
&
\left
\langle
(\hat {\bf a}-\langle {\bf a}\rangle_{{\bf a}|{\bf h},{\bf d}})
(\hat {\bf a}-\langle {\bf a}\rangle_{{\bf a}|{\bf h},{\bf d}})^{\rm T}
\right
\rangle_{{\bf h}|{\bf d}}
\nonumber
\\&=
\left
\langle 
(\hat {\bf a}_{\rm EAP}-\langle {\bf a}\rangle_{{\bf a}|{\bf h},{\bf d}})
(\hat {\bf a}_{\rm EAP}-\langle {\bf a}\rangle_{{\bf a}|{\bf h},{\bf d}})^{\rm T}
\right
\rangle_{{\bf h}|{\bf d}}
+
\left
\langle
(\hat {\bf a}-\hat {\bf a}_{\rm EAP})
(\hat {\bf a}-\hat {\bf a}_{\rm EAP})^{\rm T}
\right
\rangle_{{\bf h}|{\bf d}},
\end{flalign}
we have
\begin{equation}
{\bf C}_{\hat {\bf a},{\rm ep}}
=
{\bf C}_{\hat {\bf a}_{\rm EAP},{\rm ep}}
+
(\hat {\bf a}-\hat {\bf a}_{\rm EAP})
(\hat {\bf a}-\hat {\bf a}_{\rm EAP})^{\rm T}.
\label{eq:codeviationconversion}
\end{equation}
Equation~(\ref{eq:codeviationconversion}) is a simple conversion rule of ${\bf C}_{\hat {\bf a},{\rm ep}}$ for an estimate $\hat {\bf a}$ to the covariance of $\hat {\bf a}_{\rm EAP}$, or consequently to ${\bf C}_{\hat {\bf a},{\rm ep}}$ for arbitrary another $\hat {\bf a}$. 
Equation~(\ref{eq:codeviationconversion}) also leads to that the EAP estimate of the model parameters ``minimises'' the second-order moment ${\bf C}_{\hat {\bf a},{\rm ep}}$: 
\begin{equation}
{\bf C}_{\hat {\bf a},{\rm ep}}
\geq
{\bf C}_{\hat {\bf a}_{\rm EAP},{\rm ep}},
\label{eq:EAPtocovarianceminimizingestimator}
\end{equation}
where the inequality for a matrix denotes the difference of both sides is a positive-semidefinite matrix; 
a matrix ${\bf A}\in\mathbb R^{M\times M}$ is positive semidefinite if and only if ${\bf x}^{\rm T}{\bf Ax}\geq 0$ holds for any nonzero vector ${\bf x}\in\mathbb R^M$. 
The positive semidefiniteness holds for $(\hat {\bf a}-\hat {\bf a}_{\rm EAP})(\hat {\bf a}-\hat {\bf a}_{\rm EAP})^{\rm T}$, and then eq.~(\ref{eq:EAPtocovarianceminimizingestimator}) follows eq.~(\ref{eq:codeviationconversion}). 
Because a partitioning method between the model parameters and hyperparameters has been unspecified in the above calculation, 
eq.~(\ref{eq:EAPtocovarianceminimizingestimator}) intrinsically 
expresses that the EAP for an arbitrary subset of random variables minimises the second-order moments for that subset, which is a consequence of that the optimisation function of the EAP is the
squared error loss (the trace of ${\bf C}_{\hat {\bf a},{\rm ep}}$)~\citep[p.313]{carlin2008bayesian}.

A specific form of the uncertainty propagation (eq.~\ref{eq:errorprop_gen}) for $\hat {\bf a}_{\rm ABIC}=
\langle{\bf a}\rangle_{{\bf a}|\hat {\bf h}_{\rm ABIC},{\bf d}}$ describes
the second-order moment of ${\bf a}$ around the ABIC estimate $\hat {\bf a}_{\rm ABIC}=
\langle{\bf a}\rangle_{{\bf a}|\hat {\bf h}_{\rm ABIC},{\bf d}}$ of the model parameters ${\bf a}$, 
and approximately the covariance of ${\bf a}$ given the approximate coincidence between the ABIC and EAP estimate, which is as below analogous to the ordinary uncertainty propagation law. 
Expanding the right-hand side of eq.~(\ref{eq:errorprop_gen}) with respect to the hyperparameter fluctuations around the peak of $P({\bf h}|{\bf d})$, that is ${\bf h}=\hat{\bf h}_{\rm ABIC}$, we have the following up to the second-order of the hyperparameter fluctuations: 
\begin{equation}
{\bf C}_{\hat {\bf a}_{\rm ABIC},{\rm ep}}
=
\left.
\left[
{\bf C}_{\bf a|{\bf h}}
+
\frac{\partial \langle{\bf a}\rangle_{{\bf a}|{\bf h},{\bf d}}}{\partial {\bf h}^{\rm T}}
{\bf C}_{{\bf h}}
\frac{\partial \langle{\bf a}\rangle_{{\bf a}|{\bf h},{\bf d}}^{\rm T}}{\partial {\bf h}}
+
\frac 1 2
\mbox{Tr}\left({\bf C}_{\bf h}\frac{\partial^2}{\partial {\bf h}\partial {\bf h}^{\rm T}}\right){\bf C}_{{\bf a}|{\bf h}}
\right]
\right|_{{\bf h}=\hat{\bf h}_{\rm ABIC}},
\label{eq:ABIC_errorprop_2ndorder_gen}
\end{equation}
where ${\bf C}_{{\bf h}}
=\langle
({\bf h}-\langle {\bf h}\rangle_{{\bf h}|{\bf d}})
({\bf h}-\langle {\bf h}\rangle_{{\bf h}|{\bf d}})^{\rm T}
\rangle_{{\bf h}|{\bf d}}
$ denotes the posterior covariance of the hyperparameters. 
Equation~(\ref{eq:ABIC_errorprop_2ndorder_gen}) is the second-order approximation with respect to the hyperparameter fluctuations. 
The first term is the model-parameter covariance for the point estimation of the hyperparameters~\citep{yabuki1992geodetic}, thus the leading order of ${\bf C}_{\hat {\bf a}_{\rm ABIC},{\rm ep}}$. The others express the propagation of uncertainty as the lowest-order correction to ${\bf C}_{\hat {\bf a}_{\rm ABIC},{\rm ep}}$. 
The second term is identical to the ordinary second-order uncertainty propagation law.
The third term is the fourth-order of the fluctuations (the second-order for both the model-parameter and hyperparameter fluctuations), but is the lowest-order perturbation to ${\bf C}_{\hat {\bf a}_{\rm ABIC},{\rm ep}}$ in terms of the hyperparameter fluctuations, as the second term is. Namely, the second and third terms are the same order when the model-parameter fluctuations are not small, as in the linear inverse problem considered in the text. 


\subsection{Representations of the model-parameter second-order moments around the ABIC and EAP estimates in the linear inverse problems}
Using eq.~(\ref{eq:ABIC_errorprop_2ndorder_gen}), 
we obtain the second order moment ${\bf C}_{\hat {\bf a}_{\rm ABIC},{\rm ep}}$ of the model parameters ${\bf a}$ around their ABIC estimate $\hat {\bf a}_{\rm ABIC}$ for the linear inverse problem treated in the text,  
where
$\langle {\bf a}\rangle_{{\bf a}|{\bf h},{\bf d}}={\bf a}_*(\alpha^2)$ holds for ${\bf h}=(\sigma^2,\rho^2)^{\rm T}$. 
Given the approximate coincidence between $\hat {\bf a}_{\rm ABIC}$ and $\hat {\bf a}_{\rm EAP}$ (eq.~\ref{eq:EAPsimeqABIC}), 
${\bf C}_{\hat {\bf a}_{\rm ABIC},{\rm ep}}$ is approximately equated to the covariance ${\bf C}_{\hat {\bf a}_{\rm EAP},{\rm ep}}$ of the EAP estimate $\hat {\bf a}_{\rm EAP}$ 
as 
\begin{equation}
{\bf C}_{\hat {\bf a}_{\rm EAP},{\rm ep}}\approx{\bf C}_{\hat {\bf a}_{\rm ABIC},{\rm ep}}.
\end{equation}

A closed-form expression of ${\bf C}_{{\bf h}}$ is obtained in Appendix~\ref{sec:ABIChyperparametervariance} as ${\bf C}_{\hat {\bf h}_{\rm ABIC}}$, and below we calculate the remaining expansion coefficients. 
Twice differentiating the conditional covariance of the model parameters given the hyperparameters,
\begin{equation}
{\bf C}_{{\bf a}|{\bf h}}=\sigma^2({\bf H}^{\rm T}{\bf E}^{-1}{\bf H}+\alpha^2{\bf G})^{-1},
\end{equation}
we have
\begin{flalign}
&\mbox{Tr}\left({\bf C}_{\bf h}\frac{\partial^2}{\partial {\bf h}\partial {\bf h}^{\rm T}}\right){\bf C}_{{\bf a}|{\bf h}}
\nonumber
\\&=
2\alpha^2/\sigma^2
[
C_{\sigma^2\sigma^2}
-\alpha^2
(C_{\sigma^2\rho^2}+
C_{\rho^2\sigma^2})
+(\alpha^2)^2
C_{\rho^2\rho^2}
]
\nonumber\\
&\times
[
-
{\bf I}
+
\alpha^2({\bf H}^{\rm T}{\bf E}^{-1}{\bf H}+\alpha^2{\bf G})^{-1}{\bf G}
]
({\bf H}^{\rm T}{\bf E}^{-1}{\bf H}+\alpha^2{\bf G})^{-1}{\bf G}
({\bf H}^{\rm T}{\bf E}^{-1}{\bf H}+\alpha^2{\bf G})^{-1},
\end{flalign}
where 
$C_{\sigma^2\sigma^2}$ and $C_{\rho^2\rho^2}$ denote the autocorrelations of $\sigma^2$ 
and $\rho^2$, respectively, and $C_{\sigma^2\rho^2}$ denotes their cross-correlation, all contained in ${\bf C}_{{\bf h}}$.
Besides, by using 
\begin{equation}
\frac{\partial}{\partial \alpha^2}
({\bf H}^{\rm T}{\bf E}^{-1}{\bf H}+\alpha^2{\bf G})^{-1}
=
-({\bf H}^{\rm T}{\bf E}^{-1}{\bf H}+\alpha^2{\bf G})^{-1}{\bf G}
({\bf H}^{\rm T}{\bf E}^{-1}{\bf H}+\alpha^2{\bf G})^{-1},
\end{equation}
which can be obtained through eq.~(\ref{eq:diffofinvmat}), 
we have
\begin{equation}
\left(
\left(
\frac{\partial {\bf a}_*}{\partial \sigma^2}
\right)_{\rho^2}
,
\left(
\frac{\partial {\bf a}_*}{\partial \sigma^2}
\right)_{\rho^2}
\right)
=
(-1/\sigma^2,1/\rho^2)
\alpha^2
({\bf H}^{\rm T}{\bf E}^{-1}{\bf H}+\alpha^2{\bf G})^{-1}{\bf G}{\bf a}_*. 
\end{equation}
It gives
\begin{flalign}
\frac{\partial {\bf a}_*
}{\partial {\bf h}^{\rm T}}
{\bf C}_{{\bf h}}
\frac{\partial {\bf a}_*^{\rm T}}{\partial {\bf h}}
&=
(\alpha^2/\sigma^2)^2
[
C_{\sigma^2\sigma^2}
-\alpha^2
(C_{\sigma^2\rho^2}+
C_{\rho^2\sigma^2})
+(\alpha^2)^2
C_{\rho^2\rho^2}
]
\nonumber\\&\times
({\bf H}^{\rm T}{\bf E}^{-1}{\bf H}+\alpha^2{\bf G})^{-1}{\bf G}{\bf a}_*
{\bf a}_*^{\rm T}
{\bf G}
({\bf H}^{\rm T}{\bf E}^{-1}{\bf H}+\alpha^2{\bf G})^{-1}.
\end{flalign}

\section{Supplement 2: Relationships between the true misfit on receivers and statistical quantities}
\label{sec:statisticalmeaningoftruemisfit}

The true misfit on receivers (TMR) in the text is related to some statistical quantities. Examples include the Kullback-Leibler distance. 
For the linear inverse cases with true $\sigma^2_0$, the Kullback–Leibler distance from
$P({\bf d}^\prime|{\bf a}_0,\sigma^2_0)$ to $P({\bf d}^\prime|\hat{\bf a},\sigma^2_0)$ is given as follows~\citep{gelman2013bayesian}:
\begin{equation}
D[P({\bf d}^\prime|\hat{\bf a},\sigma^2_0)|P({\bf d}^\prime|{\bf a}_0,\sigma^2_0)]
:=-\int d{\bf d}^\prime 
P({\bf d}^\prime|{\bf a}_0,\sigma^2_0)
\ln
\frac
{
P({\bf d}^\prime|\hat{\bf a},\sigma^2_0)
}{
P({\bf d}^\prime|{\bf a}_0,\sigma^2_0)
}.
\end{equation}
The estimate $\hat {\bf a}$ is a function of ${\bf d}$ and does not depend on the other (out-of-sample) events ${\bf d}^\prime$ generated by ${\bf a}_0$ independently of ${\bf d}$. 
In the linear inverse problem, when the true $\sigma^2$ ($\sigma^2_0$) is known, the Kullback–Leibler distance is evaluated as 
\begin{equation}
D[P({\bf d}^\prime|\hat{\bf a},\sigma^2_0)|P({\bf d}^\prime|{\bf a}_0,\sigma^2_0)]
=
({\bf d}_0-{\bf H}\hat{\bf a})^{\rm T}
{\bf E}^{-1}
({\bf d}_0-{\bf H}\hat{\bf a})
/(2\sigma^2_0).
\end{equation}
This is the TMR divided by $2\sigma^2_0$. 
Then, the TMR is also related to the cross entropy (often adopted in bootstrapping and machine learning problems), which is the sum of the Kullback–Leibler distance plus the Shannon entropy of $P({\bf d}|\hat{\bf a},\sigma^2_0)$.

Another statistic related to the TMR is $\langle \log P({\bf d}^\prime|\hat {\bf a}) \rangle_{{\bf d}^\prime|{\bf a}_0}$, termed expected log pointwise predictive density for a new data set~\citep[elppd;][]{gelman2013bayesian} for the point estimate $\hat {\bf a}$. 
For the linear inverse cases with true $\sigma^2_0$, we have
\begin{equation}
    -2\sigma^2_0\times\mbox{elppd}=\langle
({\bf d}^\prime-{\bf H}\hat{\bf a})^{\rm T}
{\bf E}^{-1}
({\bf d}^\prime-{\bf H}\hat{\bf a})
\rangle_{{\bf d}^\prime|{\bf a}_0}={\rm TMR}+N\sigma^2_0.
\end{equation}
We see elppd multiplied by $-2\sigma^2_0$ equals the TMR plus the constant $N\sigma^2_0$.

\section{Supplement 3: Handling the Gibbs ensembles}
\label{sec:withrespectogibbens}
Here, we outline generic properties of the Gibbs distribution and utilise them to specifically obtain the means and variances (and an order estimate of the higher order cumulants) of Gibbsian cost functions in the fully Bayesian formulation. 
We first introduce the Gibbs distribution (\S\ref{sec:propofgibbens}). 
Refer to, say, \citet{landau1994statistical} for details.
Subsequently, we perform specific calculations for inverse problems of the Gibbsian likelihood and prior (\S\ref{sec:priposgibbens}-\S\ref{sec:cumulantofuv_posterior}).

\subsection{Relationships between the free energy and cumulant generating function of the cost function in the Gibbs ensemble}
\label{sec:propofgibbens}

The Gibbs ensemble is the following distribution, in which the probability value $P_\beta(x)$ of a state $x$ is given by an exponential of the product of the cost function $E(x)$ (energy) and the weighting hyperparameter $\beta$ (inverse temperature): 
\begin{equation}
P_\beta(x)=  e^{-\beta E(x)+F(\beta)},
\end{equation}
where $F(\beta)$ is a normalisation constant such that $\int dxP_\beta(x)=1$, called the free energy, which is the following function of $\beta$: 
\begin{equation}
F(\beta):=-\ln \int dx e^{-\beta E(x)}.
\label{eq:defofF}
\end{equation}
Although the free energy is originally the above definitional form (eq.~\ref{eq:defofF}) divided by $\beta$ in statistical thermodynamics, then having the same dimension as the energy $E$, herein we refer to this non-dimensional function $F$ (originally, the free entropy) as the (Bayesian) free energy, following the nomenclature of Bayesian statistics~\citep[e.g.][]{iba1996learning}. 

The free energy $F$ is related to the cumulant of energy $E$.
We define the cumulant generating function of $E$ as
\begin{equation}
K_{\beta,E}(\delta \beta):= \ln\left(\int dx P_\beta(x)e^{\delta \beta E}
\right),
\label{eq:defofK}
\end{equation}
where $\delta \beta$ is a parameter of this cumulant generating function. 
The integral in the logarithm represents the average of $\exp(\delta \beta E)$ weighted by the Gibbs ensemble.
Considering the Maclaurin series for the cumulant generating function in parameter $\delta \beta$, 
we define the $n$-th cumulant $\kappa_{\beta,E,n}$ such that
\begin{equation}
K_{\beta,E}(\delta \beta):=\sum_n \kappa_{\beta,E,n} \frac{(\delta \beta)^n}{n!}.
\end{equation}
That is, the $n$-th cumulant $\kappa_{\beta,E,n}$ is the 
$n$-th derivative of the cumulant generating function $K_{\beta,E}(\delta \beta)$ at $\delta \beta=0$: 
\begin{equation}
\kappa_{\beta,E,n} = \frac{\partial ^n}{\partial (\delta \beta)^n}K_{\beta,E}(\delta \beta)|_{\delta \beta=0}.
\label{eq:defofkappa}
\end{equation}
Then, by definition, the free energy $F$ and the $n$-th cumulant $\kappa_{\beta,E,n}$ of $E$ possess the following relationship: 
\begin{equation}
\kappa_{\beta,E,n} = -\frac{\partial ^n}{\partial (-\beta)^n} F(\beta)
\left(
= \frac{\partial ^n}{\partial (-\beta)^n}\ln \int dx e^{-\beta E(x)} 
\right).
\label{eq:tobeproved}
\end{equation}

We supplement the derivation of eq.~(\ref{eq:tobeproved}) below. 
First, we substitute the definitional identity of 
the cumulant generating function eq.~(\ref{eq:defofK})
into the differential expression of the cumulant eq.~(\ref{eq:defofkappa}): 
\begin{flalign}
\kappa_{\beta,E,n}
&= 
\frac{\partial ^n}{\partial (\delta\beta)^n} \left.\left[
\ln\int dx e^{-(\beta-\delta \beta)E(x)+ F(\beta)} 
\right]\right|_{\delta \beta=0}
\\&=
\frac{\partial ^n}{\partial (\delta\beta)^n} \left.\left[
\ln\int dx e^{-(\beta-\delta \beta)E(x)} +\ln  F(\beta)
\right]\right|_{\delta \beta=0}.
\end{flalign}
Differentiation by $\delta \beta$ in the first term is equivalent to that by $-\beta$, and the second term is exactly zero; 
\begin{equation}
\kappa_{\beta,E,n}
=
\frac{\partial ^n}{\partial (-\beta)^n}\ln
\left.
\int dx e^{-(\beta-\delta \beta)E(x)}\right|_{\delta \beta=0}.
\end{equation}
Substituting $\delta \beta=0$, we obtain eq.~(\ref{eq:tobeproved}): 
\begin{equation}
\kappa_{\beta,E,n}
=
\frac{\partial ^n}{\partial (-\beta)^n}\ln
\int dx e^{-\beta E(x)}
=-\frac{\partial ^n}{\partial (-\beta)^n}F(\beta).
\end{equation}
In statistical-mechanical literature, 
this cumulant-generating nature of the free energy is associated with that the free energy is the thermodynamic potential, the differentials of which give all the thermodynamic state variables, such as the energy in this context, and the differentials of the thermodynamic state variables, for example the specific heat~\citep{landau1994statistical}. 

As above, the $n$-th cumulant of the energy $E$ in the Gibbs ensemble is the $n$-th derivative of $-F(\beta)$ with respect to $-\beta$.
From this, when
\begin{equation}
F\propto N_x
\end{equation}
holds for the dimension $N_x$ of $x$, 
we have the same order estimate
\begin{equation}
\kappa_{\beta,E,n}\propto N_x.
\end{equation}
That is, 
when $F$ is a variable proportional to $N_x$ [precisely, an $\mathcal O(N_x)$ variable i.e. $\propto N_x$ in an asymptotic sense], the so-called extensive variable, 
the cumulant is also proportional to $N_x$ [$\mathcal O(N_x)$] and is namely an extensive variable. 

The relationship between the cumulants and free energy holds also for multi-hyperparameter cases:
\begin{equation}
P_{\boldsymbol\beta}(x) =e^{-\boldsymbol\beta^{\rm T} {\bf E}+ F(\boldsymbol\beta)}, 
\end{equation}
where the (intensive) inverse temperature $\boldsymbol\beta$ is a vector, and the associated (extensive) energy ${\bf E}$ is also. 
$F$ is still a $\boldsymbol\beta$-dependent normalisation factor. 
We define the multivariate cumulant generating function as
\begin{equation}
K_{\boldsymbol\beta,{\bf E}} =\ln \left(\int dx P_{\boldsymbol\beta}(x) \prod_i e^{\delta \beta_i E_i}
\right)
\end{equation}
and 
the ${\bf n}=(n_1,n_2,...)$-th cumulant  $\kappa_{\boldsymbol\beta,{\bf E},{\bf n}}$ of ${\bf E}$ such that 
\begin{equation}
K_{\boldsymbol\beta,{\bf E}}
=\sum_{\bf n}\left[\prod_i
\frac{(\delta \beta_i)^{n_i}}{n_i!}
\right]
\kappa_{\boldsymbol\beta,{\bf E},{\bf n}},
\end{equation}
which satisfies
\begin{equation}
\kappa_{\boldsymbol\beta,{\bf E},{\bf n}}
=
\left.
\left[\prod_i
\frac{\partial^{n_i}}{\partial (\delta \beta_i)^{n_i}}
\right]
K_{\boldsymbol\beta,{\bf E}}\right|_{\delta \beta_1=\delta\beta_2=...=0}.
\end{equation}
This contains cross cumulants, the cumulants that involve multiple subscripts $i$ such that $n_i\neq0$. 
Then, the cumulant and the free energy satisfy the following relationship:
\begin{equation}
\kappa_{\boldsymbol\beta,{\bf E},{\bf n}}
=
\left[\prod_i
\frac{\partial^{n_i}}{\partial (-\beta_i)^{n_i}}
\right]
[-F(\boldsymbol\beta)],
\label{eq:cumulantvsfreeen}
\end{equation}
which can be derived through the same procedure as the single cost function case. 
Meanwhile, the order-estimate discussion may be complicated when the cost-functions are not the same order, as in the conditional posterior of the model parameters treated in the text. 

Equation~(\ref{eq:cumulantvsfreeen}) shows 
the mean values of $E$ for adjacent $\beta$ values give any higher-order cumulants as the derivatives of the averaged $E$ values, 
and hence given the positivity of the second-order cumulant (variance), 
the first order cumulant (mean) is a monotonically-decreasing function of the conjugate intensive variable $\beta$, namely a bijective function of $\beta$; note as $\beta$ at least owns some value for the mean of $E$, then surjectivity of the mean $E$ is here obvious. 
In summary, specifying (the mean of) the extensive variable $E$ is equivalent to specifying the intensive variable $\beta$. 
Besides, when the first order cumulant is always positive, as in the present linear inverse problem giving $E(x)>0$ for any $x$, 
the free energy $F$ is also a bijective function of $\beta$.
For multiple intensive variables, 
we can repeat the same discussion with diagonalisation of the covariance matrix, 
as long as the covariance matrix is positive definite, 
and obtain the one-to-one correspondence between the averaged cost functions (and the free energy) and the hyperparameters. 

The above relation indicates, in a statistical sense, 
the mean of $E$ is the sufficient estimator of the hyperparameter $\beta$ in the Gibbs ensemble (presuming nonzero variance of $E$, for the above discussion of the bijectivity between $\beta$ and the mean of $E$). 
Besides, when the any-order cumulant of $E$ is $\mathcal O(N_x)$, then we can get the leading order of the mean of $E$ with just one sample asymptotically almost surely for large $N_x$. That is, the leading order of $E$ is asymptotically an almost surely constant variable, allowing us to calculate the mean of $E/N_x$, which is also a sufficient estimator of $\beta$. The cost function $E$ occupies a special place in the Gibbs ensemble. 

We note the situation complicates when considering a zero-variance cost function (corresponding to the thermodynamic phase transition), the mean of which is still a monotonic function of $\beta$, but not bijective (just surjective), so not the sufficient estimator.  

\subsection{Gibbsian expressions and characters of the observation equation, prior and conditional posterior of the model parameters given the hyperparameters}
\label{sec:priposgibbens}
Examples of the Gibbs ensembles are seen in the inverse problems considered in the text.
The observation and prior give a distribution $P({\bf a},{\bf d}|\sigma^2,\rho^2)$ of the model parameters ${\bf a}$ and data ${\bf d}$ given the hyperparameters, which corresponds to a probability for a forecast of ${\bf a}$ and ${\bf d}$ before the observation of  ${\bf d}$. 
We consider $P({\bf a},{\bf d}|\sigma^2,\rho^2)$ taking the following form: 
\begin{equation}
P({\bf a},{\bf d}|\sigma^2,\rho^2)=e^{-U({\bf d},{\bf a})/\sigma^2-V({\bf a})/\rho^2+F_{\rm pri+obs}(1/\sigma^2,1/\rho^2)}, 
\end{equation}
which is a Gibbs ensemble involving two extensive variables in the variable space $x=({\bf a},{\bf d})$ normalised by $F_{\rm pri+obs}$, the cost functions of which are $U$ and $V$ for the conjugate intensive variables $\sigma^2$ and $\rho^2$ ($1/\sigma^2$ and $1/\rho^2$). 
$P({\bf a},{\bf d}|\sigma^2,\rho^2)$ in the inverse problem treated in the text separates into two Gibbs ensembles (for the observation equation and prior) each of which involves a single cost function:
\begin{flalign}
P({\bf d}|{\bf a},\sigma^2)&=e^{-U({\bf d},{\bf a})/\sigma^2+F_{\rm obs}(1/\sigma^2)}
\\
P({\bf a}|\rho^2)&=e^{-V({\bf a})/\rho^2+F_{\rm pri}(1/\rho^2)}
\end{flalign}
$F_{\rm obs}$ and $F_{\rm pri}$ denote
the normalisation constants for $P({\bf d}|{\bf a},\sigma^2)$ (the observation equation) 
and 
$P({\bf a}|\rho^2)$ (the prior), respectively, 
such that
\begin{equation}
F_{\rm pri+obs}=F_{\rm pri}+F_{\rm obs};
\label{eq:separabilityofFpriobs}
\end{equation}
the Gibbs distribution for the observation equation is the probability for the variable space $x={\bf d}$, weighted by 
$E(x)=U({\bf d},{\bf a})$ [$U({\bf a})$ in the text] with $\beta=1/\sigma^2$, 
and for the prior, $\beta=1/\rho^2$, $x={\bf a}$, and $E(x)=V({\bf a})$. 
We note the arguments of $F_{\rm pri}$ and $F_{\rm obs}$ are now
$1/\sigma^2$ and $1/\rho^2$, instead of $\sigma^2$ and $\rho^2$ in the text, and this notation modification does not affect any following calculations. 
Likewise, the conditional prior of the model parameters given the hyperparameters takes the following form:
\begin{equation}
P({\bf a}|{\bf d},\sigma^2,\rho^2)=e^{-U/\sigma^2-V/\rho^2+F_{\rm pos}({\bf d},1/\sigma^2,1/\rho^2)},
\end{equation}
which is the Gibbs ensemble in the probability space $x={\bf a}$ with two cost functions $U$ and $V$,
or formally identically, with a single cost function $U+\alpha^2 V$ (or $U/\alpha^2+V$), where $F_{\rm pos}$ denotes a normalisation constant. 
We note the ${\bf a}$-independence of $F_{\rm pos}$ [or equivalently, that of $F_{\rm obs}$ when assuming separability eq.~(\ref{eq:separabilityofFpriobs}) of $F_{\rm pos}$] is intrinsic for the Gibbsianity of the conditional posterior of the model parameters.
Comparing $P({\bf a}|{\bf d},\sigma^2,\rho^2)$ with 
the product of
$P({\bf a},{\bf d}|\sigma^2,\rho^2)$ and $P({\bf a}|{\bf d},\sigma^2,\rho^2)$, 
we find
\begin{equation}
F_{\rm pos}=F_{\rm pri+obs}-\ln P({\bf d}|\sigma^2,\rho^2).
\label{eq:freeenergyvsmarginalizedprob}
\end{equation}

Equation~(\ref{eq:freeenergyvsmarginalizedprob}) indicates that 
for the uniform hyperprior, 
the maximisation criterion of ABIC
$ 
\max\ln P({\bf d}|\sigma^2,\rho^2)
$ 
is equivalent to the difference minimisation 
$
\min[
F_{\rm pos}
-F_{\rm pri+obs}
]
$
of the posterior free energy $F_{\rm pos}$ from the free energy $F_{\rm pri+obs}$ of the prior and observation equation: 
\begin{equation}
    \max\ln P({\bf d}|\sigma^2,\rho^2)\to \min[
F_{\rm pos}
-F_{\rm pri+obs}
].
\end{equation}
[When ${\bf d}$ takes discrete values (then making $P$ not a density), $\ln P({\bf d}|\sigma^2,\rho^2)\leq 0$ holds, then $F_{\rm pos}\geq F_{\rm pri+obs}$ follows, and the above difference minimisation of $F_{\rm pos}
$ from $F_{\rm pri+obs}$ results in their absolute difference minimisation.] 
Treating this minimisation condition of the free energy difference as the extremum condition 
and using the relation eq.~(\ref{eq:cumulantvsfreeen}) between the cost-function cumulant 
and free energy,
we have the following condition of ABIC requiring the 
same expectation values of the cost functions between $P({\bf a},{\bf d}|\sigma^2,\rho^2)$ 
and 
$P({\bf a}|\sigma^2,\rho^2,{\bf d})$:
\begin{flalign}
\langle
U
\rangle_{{\bf a},{\bf d}|\sigma^2,\rho^2}
&=
\langle
U
\rangle_{{\bf a}|{\bf d},\sigma^2,\rho^2}
\\\langle
V
\rangle_{{\bf a},{\bf d}|\sigma^2,\rho^2}
&=
\langle
V
\rangle_{{\bf a}|{\bf d},\sigma^2,\rho^2}.
\end{flalign}
These are the generalisations of the extremum expressions eqs.~(\ref{eq:ABICsigma_equipartition_condition}) and (\ref{eq:ABICrho_equipartition_condition}) for the linear inverse problem. 
As above, ABIC is a criterion giving the expectation values of the cost functions $U$ and $V$ independent of whether ${\bf d}$ is fixed or not, that is, invariant from whether we observe ${\bf d}$ or not~\citep{iba1996learning}. 
\citet{akaike1980likelihood} originally seems to dispute the prior selection affected by the arbitrariness of the fixed random-variable subsets (in this case, treating ${\bf d}$ as constant or not), supposedly including the difference between $P({\bf a},{\bf d}|\sigma^2,\rho^2)$ and $P({\bf a}|{\bf d},\sigma^2,\rho^2)$, so the identity 
$\langle U\rangle_{{\bf a},{\bf d}|\sigma^2,\rho^2}=\langle U\rangle_{{\bf a}|{\bf d},\sigma^2,\rho^2}$
would be a natural consequence of his proposition. 

Besides, commonly, $F_{\rm pri}$ and $F_{\rm obs}$ are extensive variables of order $N$ and $P$, respectively, 
and any order cumulants (e.g. the mean and variance) of their cost functions ($U$ and $V$) are correspondingly also extensive [i.e. $\mathcal O(N)$ and $\mathcal O(P)$, respectively]. Namely, their $\mathcal O(\sqrt N,\sqrt P)$ fluctuations are asymptotically negligible for their $\mathcal O(N,P)$ means. 
The same commonly applies to $F_{\rm pos}$ for the cost function $U+\alpha^2 V$, except its ground-state-energy part $\min_{{\bf a}|\alpha^2}(U+\alpha^2 V)$ [i.e. only the 0th and 1st partial derivatives of $F_{\rm pos}$ with respect to $\sigma^2$ given $\alpha^2$ include $\mathcal O(N)$ and $\mathcal O(P)$ terms, and the cumulants excluding these offsets are $\mathcal O(M)$]. 
Cumulants of $U$ and $V$ in the conditional posterior of $F_{\rm pos}$ are mixtures of $\mathcal O(N)$, $\mathcal O(M)$ and $\mathcal O(P)$ terms even in the linear problems as seen later, 
but $U$ and $V$ are commonly expected to be almost deterministic before and after the observation for large $N$ and $P$ (and simultaneously large $M\geq P$). 

\subsection{A formal expression of the hyperparameter covariance in ABIC}
\label{sec:varianceofhyperparameter}
The cumulants of $U$ and $V$ are related to the cumulants of the hyperparameters 
through the rewritten form of 
eq.~(\ref{eq:freeenergyvsmarginalizedprob}):
\begin{equation}
\ln P({\bf d}|\sigma^2,\rho^2)=F_{\rm pri+obs}- F_{\rm pos}.
\end{equation}

Further setting $\boldsymbol\beta=(1/\sigma^2,1/\rho^2)^{\rm T}$ 
and introducing 
$\delta \boldsymbol\beta
:=\boldsymbol\beta
-
\hat{\boldsymbol\beta}_{\rm ABIC}$, 
where 
$\hat{\boldsymbol\beta}_{\rm ABIC}
=(1/\hat\sigma^2_{\rm ABIC},1/\hat\rho^2_{\rm ABIC})^{\rm T}$, 
we have
\begin{flalign}
&\ln P(\sigma^2,\rho^2|{\bf d})
\nonumber
\\
&\propto
\ln P({\bf d}|\sigma^2,\rho^2)
\\
&=
\max_{\sigma^2,\rho^2}[\ln P(\sigma^2,\rho^2|{\bf d})]
-\frac 1 2 
\delta\boldsymbol\beta^{\rm T}
\left.
\left(
\frac{\partial^2 F_{\rm pos}}
{\partial \boldsymbol\beta \partial \boldsymbol\beta^{\rm T}}
-
\frac{\partial^2 F_{\rm pri+obs}}
{\partial \boldsymbol\beta \partial \boldsymbol\beta^{\rm T}}
\right)
\right|_{\boldsymbol\beta=\hat{\boldsymbol\beta}_{\rm ABIC}}
 \delta \boldsymbol\beta
+...,
\end{flalign}
or equivalently, for the uniform hyperprior, up to the second-order of
$\delta\sigma^2=\sigma^2-\hat\sigma^2_{\rm ABIC}$ and $\delta\rho^2=\rho^2-\hat\rho^2_{\rm ABIC}$, 
\begin{equation}
\ln P(\sigma^2,\rho^2|{\bf d})=
c
-\frac 1 2 
(\delta\sigma^2\hspace{5pt} \delta\rho^2)
\hat{\bf C}_{\hat {\bf h}_{\rm ABIC}}^{-1}
(\delta\sigma^2\hspace{5pt} \delta\rho^2)^{\rm T}
+...,
\end{equation}
with
\begin{equation}
\hat{\bf C}_{\hat {\bf h}_{\rm ABIC}}^{-1}
:=
\mbox{diag}(\hat{\boldsymbol\beta}_{\rm ABIC}^2)
\left.
\left(
\frac{\partial^2 F_{\rm pos}}
{\partial \boldsymbol\beta \partial \boldsymbol\beta^{\rm T}}
-
\frac{\partial^2 F_{\rm pri+obs}}
{\partial \boldsymbol\beta \partial \boldsymbol\beta^{\rm T}}
\right)
\right|_{\boldsymbol\beta=\hat{\boldsymbol\beta}_{\rm ABIC}}
\mbox{diag}(\hat{\boldsymbol\beta}_{\rm ABIC}^2)
.
\label{eq:formalexpofhypercov}
\end{equation}
where $\mbox{diag}(\cdot)$ denotes the diagonal matrix storing the bracketed variables $\cdot$ as diagonal entries. 
As seen in 
\S\ref{sec:propofgibbens}, 
the second derivatives of the free energy with respect to the inverse temperature constitute the covariance of the cost functions multiplied by $-1$, 
and 
hence from eq.~(\ref{eq:formalexpofhypercov}) and \S\ref{sec:propofgibbens},
the posterior covariance of the hyperparameters are noticed to be related to the covariance of the cost functions (exactly the inverse of the cost-function covariance decrement for the inverse temperature). 

Considering a sort of the inverse proportionality between the covariance matrices for the cost functions and hyperparameters,
one would expect the law of large numbers of the hyperparameters [i.e. $P(\sigma^2,\rho^2|{\bf d})$ approaches to zero except at the probability peak(s) $\sigma^2=\hat\sigma^2$, $\rho^2=\hat\rho^2$ in some limit], especially when $\ln P(\sigma^2,\rho^2|{\bf d})$ (which is the difference of $F_{\rm pri+obs}$ from $F_{\rm pos}$) is expressed by extensive variables of order $N$ and $P$ as actually does in the linear inverse problems (Appendix~\ref{sec:ABIChyperparametervariance}). 
It appears to immediately follow the extensive property of the free energies (or more plausibly, those of the cost functions), but as explicitly shown in the linear inverse cases, the free-energy landscapes [and the $P(\sigma^2,\rho^2|{\bf d})$ profile] are not necessarily sharpened when $N$ increases despite their extensive properties while are sharpened for large $P$ even with small $N$, in brief, rather complicated.

\subsection{Cumulants of $U$ and $V$ in the observation equation and prior given the hyperparameters}
\label{sec:cumulantofuv}
The contents in 
\S\ref{sec:priposgibbens} and
\S\ref{sec:varianceofhyperparameter} 
generally hold with applicability to nonlinear problems, as long as the 
conditional likelihood and prior of ${\bf a}$ given the hyperparameters are Gibbs ensembles in the forms of eqs.~(\ref{eq:obseqprob}) and (\ref{eq:conjpriorobs}). 
Below, we apply these to the present linear inverse problem, 
calculate the free energies $F_{\rm obs}$ and $F_{\rm pri}$ and obtain the cumulants of $U$ and $V$ from their derivatives.

Using
\begin{equation}
F_{\rm obs}:=
-\ln\int d{\bf d}\exp\left[-\frac {1}{2\sigma^2} ({\bf d}-{\bf Ha})^{\rm T}{\bf E}^{-1}({\bf d}-{\bf Ha})\right]
=-\frac{N}{2} \ln (2\pi\sigma^2)-\frac 1 2 \ln |{\bf E}|,
\end{equation}
we can calculate the cumulants $\kappa_{{\rm obs},U,n}$ of the cost function $U$ for the observation equation as
\begin{equation}
\kappa_{{\rm obs},U,n}=\frac{\partial^n}{\partial(-1/\sigma^2)^n} \frac{N}{2} \ln \sigma^2=\frac N 2 (n-1)! (\sigma^2)^{n}=\mathcal O(N).
\end{equation}
Through a similar procedure, 
differentiating
\begin{equation}
F_{\rm pri}:=
-\ln\int d{\bf a}\exp\left[-\frac {1}{2\rho^2} {\bf a}^{\rm T}{\bf G}{\bf a}\right]
=-\frac{P}{2} \ln (2\pi\rho^2)+\frac 1 2 \ln |\boldsymbol\Lambda_G|(+c),
\end{equation}
we evaluate the cumulants $\kappa_{{\rm pri},V,n}$ of the cost function $V$ for the prior as
\begin{equation}
\kappa_{{\rm pri},V,n}=\frac{\partial^n}{\partial(-1/\rho^2)^n}\frac{P}{2} \ln \rho^2=\frac P 2 (n-1)! (\rho^2)^{n}=\mathcal O(P).
\end{equation}
The extensive property holds for both the cumulants.
Specifically, the means are
\begin{flalign}
\kappa_{{\rm obs},U,1}=\langle U\rangle_{{\bf a},{\bf d}|\sigma^2,\rho^2}=\frac{N}{2} \sigma^2
\\
\kappa_{{\rm pri},V,1}=\langle V\rangle_{{\bf a},{\bf d}|\sigma^2,\rho^2}=\frac{P}{2} \rho^2,
\end{flalign}
and the covariance components are
\begin{flalign}
&\langle (U 
-
\langle U \rangle_{{\bf a},{\bf d}}|\sigma^2,\rho^2)^2
\rangle_{{\bf a},{\bf d}|\sigma^2,\rho^2}=\frac N 2 (\sigma^2)^2
\\
&\langle (V 
-
\langle V \rangle_{{\bf a},{\bf d}|\sigma^2,\rho^2})^2
\rangle_{{\bf a},{\bf d}|\sigma^2,\rho^2}=\frac P 2 (\rho^2)^2
\\&\langle 
(U -\langle U \rangle_{{\bf a},{\bf d}|\sigma^2,\rho^2})
(V -\langle V \rangle_{{\bf a},{\bf d}|\sigma^2,\rho^2})
\rangle_{{\bf a},{\bf d},\sigma^2,\rho^2}=0.
\end{flalign}
The representations of the first cumulants (the means) of quadratic $U$ and $V$ are known as the equipartition theorem~\citep{landau1994statistical} in statistical thermodynamics, and the representations of the $n$th cumulants are analogous to that of the exponential distribution ${\rm Exp}(x;\lambda)=\lambda \exp(-\lambda x)$ of a random variable $x$ with rate parameter $\lambda$ [evaluated as $\lambda^{-n}(n-1)!$]. 
We note the cross cumulants of any orders are identically zero between $U$ and $V$ in the distribution $P({\bf a},{\bf d}|\sigma^2,\rho^2)$ when $F_{\rm pri+obs}(1/\sigma^2,1/\rho^2)=F_{\rm pri}(1/\rho^2)+F_{\rm obs}(1/\sigma^2)$. 

\subsection{Conditional posterior cumulants of $U$ and $V$ given the hyperparameters}
\label{sec:cumulantofuv_posterior} 
We here consider similar calculations to those in \S\ref{sec:cumulantofuv} for the conditional posterior of the model parameters.

Using eq.~(\ref{eq:freeenergyvsmarginalizedprob}), $F_{\rm pos}$ is obtained as 
\begin{equation}
    F_{\rm pos}:=-\ln\int d{\bf a}\exp(-U/\sigma^2-V/\rho^2) =-\frac M 2 \ln (2\pi\sigma^2)+\frac 1 2 \ln |{\bf H}^{\rm T}{\bf E}^{-1}{\bf H}+\alpha^2{\bf G}|+\frac{s({\bf a}_*(\alpha^2),\alpha^2)}{2\sigma^2}.
\end{equation}
It is apparently the free energy for a quadratic cost function [$U+\alpha^2 V-\min_{\bf a} (U+\alpha^2 V)$ for this case] as in the observation equation and prior of the model parameters, except an additional term (the third term) for the ground-state value [$\min_{\bf a} (U+\alpha^2 V)$] of $U+\alpha^2 V$. 
The cumulants $\kappa_{{\rm pos},U+\alpha^2V,n}$ of the cost function $U+\alpha^2 V$ for the conditional posterior are then evaluated as 
\begin{equation}
    \kappa_{{\rm pos},U+\alpha^2V,n}=\frac M 2 (n-1)! (\sigma^2)^{n}+\frac{s({\bf a}_*(\alpha^2),\alpha^2)}{2}\delta_{n1}
\end{equation}
where $\delta_{ab}$ is the Heaviside delta function. 
Except for the $\mathcal O(N)+\mathcal O(P)$ offset of $U+\alpha^2V$, the extensive property holds for its cumulants:
\begin{equation}
    \kappa_{{\rm pos},U+\alpha^2V,n}-\frac{s({\bf a}_*(\alpha^2),\alpha^2)}{2}\delta_{n1}=\mathcal O(M).
\end{equation}

The conditional posterior of the model parameters is also a Gibbs ensemble for two cost functions $U$ and $V$. Then using eqs.~(\ref{eq:relofsatastar})--(\ref{eq:diffofinvmat}), 
the conditional posterior means and covariance of the cost functions $U$ and $V$ given the hyperparameters are evaluated as follows: 
\begin{flalign}
\langle U \rangle_{{\bf a}|{\bf d},\sigma^2,\rho^2}
&=\left(\frac{\partial F_{\rm pos}}{\partial (1/\sigma^2)}\right)_{\rho^2}
\nonumber
\\
&=\frac M 2 \sigma^2 -\frac{\alpha^2\sigma^2}2 \mbox{Tr}[({\bf H}^{\rm T}{\bf E}^{-1}{\bf H}+\alpha^2{\bf G})^{-1}{\bf G}]+U({\bf a}_*(\alpha^2))
\\
\langle V \rangle_{{\bf a}|{\bf d},\sigma^2,\rho^2}
&=\left(\frac{\partial F_{\rm pos}}{\partial (1/\rho^2)}\right)_{\sigma^2}
\nonumber
\\
&=\frac{\sigma^2}2 \mbox{Tr}[({\bf H}^{\rm T}{\bf E}^{-1}{\bf H}+\alpha^2{\bf G})^{-1}{\bf G}]+V({\bf a}_*(\alpha^2))
\end{flalign}
and 
\begin{flalign}
&\langle (U 
-
\langle U \rangle_{{\bf a}|{\bf d},\sigma^2,\rho^2})^2
\rangle_{{\bf a}|{\bf d},\sigma^2,\rho^2}
=-\left(\frac{\partial^2F_{\rm pos}}{\partial (1/\sigma^2)^2}\right)_{\rho^2}
\nonumber
\\
=&
\frac M 2 (\sigma^2)^2
-\alpha^2(\sigma^2)^2\mbox{Tr}[({\bf H}^{\rm T}{\bf E}^{-1}{\bf H}+\alpha^2{\bf G})^{-1}{\bf G}]
+(\alpha^2)^2\langle (V 
-
\langle V \rangle_{{\bf a}|{\bf d},\sigma^2,\rho^2})^2
\rangle_{{\bf a}|{\bf d},\sigma^2,\rho^2}
\\
&\langle (V 
-
\langle V \rangle_{{\bf a}|{\bf d},\sigma^2,\rho^2})^2
\rangle_{{\bf a}|{\bf d},\sigma^2,\rho^2}
=-\left(\frac{\partial^2 F_{\rm pos}}{\partial (1/\rho^2)^2}\right)_{\sigma^2}
\nonumber
\\
=&
\frac{(\sigma^2)^2}2 \mbox{Tr}\{[({\bf H}^{\rm T}{\bf E}^{-1}{\bf H}+\alpha^2{\bf G})^{-1}{\bf G}]^2\}-\sigma^2\frac{dV({\bf a}_*(\alpha^2))}{d\alpha^2}
\\
&\langle (U -
\langle U \rangle_{{\bf a}|{\bf d},\sigma^2,\rho^2})
(V -
\langle V \rangle_{{\bf a}|{\bf d},\sigma^2,\rho^2})
\rangle_{{\bf a}|{\bf d},\sigma^2,\rho^2}
=-
\left(
\frac{\partial }{\partial (1/\sigma^2)}
\left(\frac{\partial F_{\rm pos}}{\partial(1/\rho^2)} 
\right)_{\sigma^2}
\right)_{\rho^2}
\nonumber
\\
=&
\frac{(\sigma^2)^2}2\mbox{Tr}[({\bf H}^{\rm T}{\bf E}^{-1}{\bf H}+\alpha^2{\bf G})^{-1}{\bf G}]
-\alpha^2\langle (V 
-
\langle V \rangle_{{\bf a}|{\bf d},\sigma^2,\rho^2})^2
\rangle_{{\bf a}|{\bf d},\sigma^2,\rho^2}
\end{flalign}
where $d V({\bf a}_*(\alpha^2))/d\alpha^2$ is shown in eq.~(\ref{eq:1stderivofVstar}). 
We note $dU({\bf a}_*(\alpha^2))/d \alpha^2=-\alpha^2d V({\bf a}_*(\alpha^2))/d \alpha^2$ derived from eq.~(\ref{eq:relofsatastar}). 
Equation~(\ref{eq:freeenergyvsmarginalizedprob}) balances the shown first derivatives of $F_{\rm pos}$ with those of $\ln P({\bf d}|\sigma^2,\rho^2)$ (in Appendix~\ref{sec:ABIChyperparametervariance}) and $F_{\rm pri+obs}$ (=$F_{\rm pri}+F_{\rm obs}$, shown in \S\ref{sec:cumulantofuv}). 
Their second derivatives are shifted from an equality by the derivative of the Hessian for the map of the hyperparameters to their inverses, while the off-diagonal parts balance; 
denoting ${\bf h}=(\sigma^2,\rho^2)^{\rm T}$ 
and using
eq.~(\ref{eq:freeenergyvsmarginalizedprob}), 
we actually have the followings:
\begin{equation}
\frac{\partial F_{\rm pos}}
{\partial \boldsymbol\beta}
=
\frac{\partial F_{\rm pri+obs}}
{\partial \boldsymbol\beta}
+
\mbox{diag}({\bf h}^2)
\frac{\partial \ln P({\bf d}|\sigma^2,\rho^2)}{\partial {\bf h}}
\end{equation}
and
\begin{flalign}
\frac{\partial^2 F_{\rm pos}}
{\partial \boldsymbol\beta \partial \boldsymbol\beta^{\rm T}}
&=
\frac{\partial^2 F_{\rm pri+obs}}
{\partial \boldsymbol\beta \partial \boldsymbol\beta^{\rm T}}
-
\mbox{diag}({\bf h}^2)
\frac{\partial^2 \ln P({\bf d}|\sigma^2,\rho^2)}{\partial {\bf h}\partial {\bf h}^{\rm T}}
\mbox{diag}({\bf h}^2)
\nonumber\\&
-2
\mbox{diag}({\bf h}^3)
\mbox{diag}
\left(
\frac{\partial \ln P({\bf d}|\sigma^2,\rho^2)}{\partial {\bf h}}
\right).
\end{flalign}
Except for $U({\bf a}_*(\alpha^2))=\mathcal O(N)$ in the conditional posterior mean of $U$, the first and second cumulants of $U$ and $V$ are $\mathcal O(M)+\mathcal O(P)$, not exceeding the order of the dimension of the model-parameter space. Then excluding the offset of $U$, the extensive property loosely holds for the cumulants of $U$ and $V$ when $P=\mathcal O(M)$. 
The same applies to the higher-orders given by the derivatives of the covariance. 
$\mathcal O(M)$ terms are only contained in the cumulants of $U$ corresponding to the cumulants of the cost function in the least square estimation when $\alpha^2\to 0$. We have presumed the positive definite covariance of $P({\bf a}|\sigma^2,\rho^2,{\bf d})$ in the text, and then these $\mathcal O(M)$ terms arise in the cumulants of $U$; their more precise evaluation is $\mathcal O[\mbox{rk}({\bf C}_{{\bf a}_*})]$ using rank $\mbox{rk}({\bf C}_{{\bf a}_*})$ of ${\bf C}_{{\bf a}_*}$. 
The cumulants of $V$ and cross cumulants of $U$ and $V$ are all $\mathcal O(P)$ as they are necessarily expressed as the derivatives of the $\mathcal O(P)$ mean of $V$.

\label{lastpage}
\end{document}